\pdfoutput=1
\documentclass[
superscriptaddress,
preprint,
nofootinbib,
amsmath,amssymb,
aps,
prd,
]{revtex4-1}

\usepackage{graphicx}
\usepackage{bm}
\usepackage[colorlinks=true,hyperfootnotes=true,breaklinks=true,citecolor=PineGreen]{hyperref}
\usepackage[dvipsnames]{xcolor}
\usepackage{physics}
\usepackage{diagbox}

\begin{document}
\count\footins = 1000
\newcommand{\tbf}[1]{\textbf{#1}}
\newcommand{\p}{\partial}
\title{Dynamical Vacuum Compressibility of Space}
\author{Yu-Cun Xie}
\email{xyc@terpmail.umd.edu}
\affiliation{Department of Physics, University of Maryland, College Park, Maryland 20742, USA}
\author{Jen-Tsung Hsiang}
\email{cosmology@gmail.com}
\affiliation{College of Electric Engineering and Computer Science, National Taiwan University of Science and Technology, Taipei City 106, Taiwan, R.O.C.}
\affiliation{Center for High Energy and High Field Physics (CHiP), National Central University, Taoyuan 320317, Taiwan, R.O.C.} 
\author{Bei-Lok Hu}
\email{blhu@umd.edu}
\affiliation{Maryland Center for Fundamental Physics and Joint Quantum Institute, University of Maryland, College Park, Maryland 20742, USA}
\date{\small Dec 26, 2023 submitted to arXiv-v2 and PRD} 
\begin{abstract}
This paper continues the investigation initiated in Ref.~\cite{CHH1} into the quantum thermodynamic properties of space by deriving the vacuum compressibility of a variety of {\it dynamical} spacetimes containing massive and massless conformally coupled quantum fields. The quantum processes studied here include particle creation, Casimir effect, and the trace anomaly. The spaces include $S^2, S^3$, and $T^3$ with prescribed time evolution and $S^1$, where the temporal development is  backreaction determined. Vacuum compressibility belongs to the same group of quantum thermodynamic / mechanical response functions as {\it vacuum viscosity}, a concept first proposed in 1970 by Zel'dovich \cite{Zel70} for capturing the effects of vacuum particle production on the dynamics of the early universe, made precise by rigorous work of many authors in the following decade using quantum field theory in curved spacetime methodologies and semiclassical gravity theory for treating backreaction effects.  Various subtleties in understanding the behavior of the vacuum energies of quantum field origins, negative pressures and novel complicated features of dynamical compressibility are discussed. 
\end{abstract}
\maketitle
\parskip=10pt
\section{Introduction}

The title contains four key words in three areas of research: vacuum refers to the ground state of a quantum field, space conjures (general) relativity, and compressibility brings up the (nonequilibrium) thermodynamics of matter \cite{IrrTD}. The first two elements make up quantum field theory (QFT) in curved spacetime (CST), a well-established field since the 1970s \cite{BirDav,ParTom},  Hawking effect \cite{Haw75} being a well-known example. 
Quantum compressibility refers to the compressibility of space filled with a quantum field at zero or finite temperature. This bears on the emergent field of quantum thermodynamics (QTD) emphasizing how considerations of the quantum properties of matter entice us to  extend, even revise, our understanding of classical thermodynamics and explore new laws at low temperature for small objects. Studies on the mechanical response to quantum fields, in particular, electromagnetic and gravitational radiation,  and the reverse process of how quantum fields respond to mechanical actions,  make up an equally exciting field of quantum optomechanics (QOM), LIGO/VIRGO/KAGRA based on interferometry are its signature highlights. 

Can one apply quantum thermodynamic concepts to understand the fundamental properties of spacetime with the help of quantum field theory in curved spacetimes, in more general terms than the much studied and better understood black hole thermodynamics \cite{Bek72,Bek73,Haw76,Wald} in the class of spacetimes with event horizons? This was what motivated Cho and two of the present authors \cite{CHH1} (CHH) to investigate into the quantum thermodynamic properties of spacetimes by deriving the quantum heat capacity and vacuum compressibility of a variety of (static) spacetimes containing a quantum field at zero or finite temperatures.  This paper continues that line of pursuit extending it to dynamical conditions.

\paragraph{Quantum Thermodynamics of Dynamical Spacetimes}

Spacetime thermodynamics \cite{STDJac1,STDJac2,STDPad,STDLib} is a derivative and extension of black hole thermodynamics, also related to the view that gravity is of the nature of thermodynamics \cite{STDJac1,PaddyRPP,Verlinde,HuGravNEqTh} or hydrodynamics \cite{HuGRhydro,VolHe3,HuSTcond,GieSinCond,OritiCond}, and should be treated on the same footing as the  collective excitations of (spacetime) atoms as in condensed matter physics \cite{GuWen06,SinEmG,OritiQMB,VolCMP}. These are examples of emergent gravity, a fundamentally different vein from quantum gravity. (For a sample of discourses on conceptual issues of emergent gravity, see, e.g., \cite{HuE/QG,Carlip,Marolf,SinEmG,BarGaray}).  In this broader perspective, considerations of spacetime thermodynamics need not be limited to spacetimes with event horizons and thermodynamics should not be restricted to equilibrium conditions. This brings up the first key word of this paper: Dynamical.  

In the context of quantum fields in curved spacetimes,  a process of similar importance as the Hawking effect in black holes is cosmological particle creation (CPC), prominent at the Planck time.   The underlying mechanism of CPC is parametric amplification of vacuum fluctuations in a quantum field. The backreaction effects of CPC have been shown to be strongly responsible for the isotropization and homogenization of the early universe. Seeing the dissipative effects of these ubiquitous quantum processes,  Zel'dovich in a short essay in 1970 \cite{Zel70} called it ``vacuum viscosity".  It was a brilliant foresight, but showing it formally was nontrivial, as it involves three steps: a) a well-defined vacuum for quantum fields in an evolutionary spacetime, b) a viable procedure to regularize the UV divergences of the stress energy tensor to arrive at  a finite expression for the energy density of particles produced, and c) calculate the backreaction  of quantum processes (involving the Casimir energy, the trace anomaly and particle creation) by solving the semiclassical Einstein equation and the field equations self-consistently.  It wasn't till 1982 \cite{HuVacVis} that `vacuum viscosity' acquires its full meaning by rigorous derivations in the context of QFTCST and semiclassical gravity \cite{HuVer}.  Many interesting ideas emanated from this vein of thought on viscous cosmology, both of classical and quantum origins, including current speculations on the nature of dark matter and dark energy.  We mention a few representative works here \cite{Gron,MakHar,Zimdahl,ZimPav,SinKal,WarmInf,ChaSah,Barrow,VisDE,VisCos}. The two 2008 papers by Klinkhamer and Volovik \cite{KliVol08a,KliVol08b} are of special relevance to our present work since they also invoke a time changing vacuum compressibility in the vein of emergent gravity.  While their goal is to explain the dynamical behavior of the cosmological constant, our work at this stage only aims at deriving the vacuum energy and pressure of a dynamical space from the three unavoidable sources of vacuum processes, namely, the Casimir effect, the trace anomaly and particle creation. In a way we are applying a solid layer of prime over a rough and spotty wall, so others can later paint on it with colors of their choice for whatever purposes without much worry of anything underneath falling off. Only after we have done the same for thermal quantum fields \cite{XHH2} and then for nonequilibrium quantum fields \cite{CHH2} suitable for an evolutionary universe will we feel comfortable enough to begin thinking about how these sources of vacuum energy bear on the cosmological constant issue.

\paragraph{Goal of this work}

Understanding these QFT phenomena in quantum thermodynamics terms is a primary goal of CHH's work: exploring whether one can see them through the quantum capacity and vacuum compressibility of curved and dynamical spacetimes.  In this work we have the same goal in mind but  a restricted scope,  working with more tangible objects which may prove more useful in analogy gravity experiments such as a pulsating ring, a moving box or a bouncing ball.  Placing ourselves for a moment on mundane ground we can think of a simple question like, how can you tell what a balloon is filled with? gas or liquid? What kind of gas? The answer is known to even toddlers: bounce it on the floor. A water balloon will probably splash rather than bounce. A lighter gas will likely bounce higher. Compressibility often refers to the material of the ball, e.g., a volleyball is made of softer material and thus bouncier under the same pressure than a basketball. Here we are not concerned with the material which the ball is made of, rather, we consider different kinds of quantum fields filling a variety of spaces, like, on a 2-sphere $S^2$, in a 3-sphere $S^3$ and in a rectangular box (with topology 3-torus $T^3$). We are interested in how dynamics would alter the static values of vacuum compressibility, comparing a) nonadiabatic versus adiabatic conditions, (or sudden versus gradual changes in the geometry of the spaces), as well as comparing b) the contributions from the three quantum sources in various spaces: Casimir energy, particle creation and trace anomaly, the latter containing higher derivatives of the scale factor with time.

Nonadiabatic dynamics strongly amplifies the vacuum fluctuations parametrically, as testified by many cosmological problems studied before:  rapid expansion of the early universe produces copious amounts of particles and their backreaction effects are much stronger than those scantily produced under adiabatic expansions.  For the present problem we anticipate different responses from nonadiabatic versus adiabatic processes (hitting the ball hard versus slowly squeezing it) and a noticeable variation in its vacuum compressibility.   It would be interesting to see how such quantum effects show up in simpler setups like a pulsating ring, a moving box or a bouncing ball. 

\paragraph{Five cases studies}

Indeed, these are the five cases we shall investigate: 1) a massless conformal field in a (1+1)D ring with time-dependent radius, where the trace anomaly and Casimir effect is the only vacuum quantum effects, 2) a massless conformal field in a 2-sphere with changing radius, where the trace anomaly is identically zero (for even spatial dimensions \cite{TAevendim}),  but with a massive conformal field,  there is particle creation.  In (1+3)D we study two situations, particle creation of a massive conformal field in 3) a 3D conducting box with one moving side and 4) a 3-sphere in nonadiabatic expansion,  and finally, 5) trace anomaly associated with a massless conformal field in a 3-sphere.  See Table 1 for the layout. 

In Case 1 we shall make use of recent results \cite{XBH} from backreaction studies of the effect of trace anomaly acting on the ring. This result is derived with full-fledged QFT CST methods which involve a) deriving a formal expression for the trace anomaly and the energy density, b) using suitable regularization schemes to remove the ultraviolet divergences, and c) analyzing the backreaction\footnote{Even though these quantum vacuum effects outside of strong field conditions as in the early universe or near the black hole are small, when future experiments reach a high enough sensitivity level the results obtained here for these simple and easily accessible systems may prove useful. They could, in the spirit of analog gravity, serve as a guide to connect with the backreaction effects of conformal anomalies and cosmological particle creation.} of these effects on the rate of contraction of the ring (over and above that due to the static Casimir effects). 
We shall use the vacuum energy obtained there to calculate (numerically) the vacuum compressibility under dynamical conditions.  
% This could serve as a benchmark to compare with results for problems in the same setups but obtained under specific approximations.   

For the three new cases 2, 3, 4, and 5 studied here we shall ignore backreaction effects on the external drive and assume the dynamics is prescribed\footnote{Notice that in all realistic experimental setups, backreaction effects are innate, i.e., intrinsically included in the physical parameters. Conceptually this is similar to bare  versus dressed masses or charges in the context of renormalization, namely, quantum radiative corrections are already included in actual physical parameters or dressed quantities. To study the backreaction effects in analog gravity experiments one needs to add the dynamics of an external drive into the overall picture in order to study how quantum effects such as particle creation would alter its dynamics compared to the case where the drive undergoes a prescribed evolution. Quantum friction is a good example where a neutral atom moving at constant speed near a dielectric surface experiences a steady drag force from the dielectric-modulated quantum field. Compared to the simplistic case of an atom undergoing uniform motion driven by some external source, the drive in the case of quantum friction certainly experiences nontrivial backreaction from the dielectric-modulated quantum field through its action on the atom's dynamics.}

\paragraph{Approaches and Justifications} 

Our specific goals in the present problem is to derive the dynamical vacuum compressibility of model spacetimes. For this we need to calculate the vacuum energy due to all quantum field effects contributing to it, which includes the (static) Casimir effect, the dynamical Casimir effect, particle creation and trace anomaly. We can take advantage of the techniques developed for QFTCST and semiclassical gravity in the decade since 1977  \cite{Har77,HuPar77,HuPar78,FHH79,HarHu79,HarHu80,HarV,And85,CalHu87} and make reasonable  approximations/assumptions based on results obtained therefrom.  
% a) Since the dominant contributions to compressibility come from the low lying modes of excitation,  the ultraviolet divergences and the regularization issues, though one needs to account for them, are not of primary concern here. 
% b) Since particle creation is the strongest under nonadiabatic parametric amplification,  one can determine the range of normal modes which yields the most particle creation at a particular moment of time by the condition that the nonadiabaticity parameter $\bar\Omega_k \equiv \Omega_k'/\Omega_k^2$ (a prime denotes taking the derivative with respect to time) associated with mode $k$ is much greater than unity, namely\footnote{The number density of particles in the $kth$ mode expressed as a function of $\bar\Omega_k$ in the lowest two adiabatic orders is given in Eqs. (68-74) of \cite{Hu74}},  $\bar\Omega_k  \gg 1$.  Integrating the energy density of particles created over this dominant range of $k$ modes will give a good approximation to the total energy from particle creation.  This condition justifies the use of a sudden approximation \textcolor{red}{(eh, do we still use this?)} in the calculation of the energy density, as was explained in \cite{HuPar77,HuPar78}. This approximation is used in \cite{XBH} for studying the backreaction effects of particle creation in $\mathbf{R}\times T^3$. The results from \cite{XBH} are adopted here for the calculation of vacuum compressibility of particle creation in $\mathbf{R}\times T^3$ presented in Sec.~\ref{S:eotbsf}.  
a) Since the backreaction effect on the drive is not of our concern here we won't pursue this issue in the new cases studied in Sec.~\ref{S:esefsg} and~\ref{S:dbshdj}. Instead, we shall stipulate two types of temporal evolution which enable us to calculate the effects related to particle creation in $\mathbf{R}\times S^2$, $\mathbf{R}\times S^3$, and $\mathbf{R}\times T^3$ without any approximation. b) We do so by solving the three coupled first order differential equations governing the Bogoliubov coefficients: two complex or four real quantities minus one Wronskian condition (e.g., Eqs. (75-77) of \cite{Hu74}) numerically with the specified time-dependence of the scale factor. This new approach is an improvement over earlier treatments.  Once the vacuum energy from these two processes in these spacetimes are obtained it is easy in technical terms to derive the vacuum pressure and the compressibility. 

Here it is perhaps appropriate to add a comment on the value of our present work. We mention three aspects 1) An experienced researcher in cosmological particle creation and backreaction may find there is little fundamentally new in terms of principles. Granted, it is when these principles are applied to corresponding analog gravity setups that calculations found in the literature are either incomplete or incorrect\footnote{An example of the incompleteness is, despite an abundance of work on Casimir and dynamical Casimir effects,  we have not seen much discussion of the effect of trace anomaly in the corresponding setups,  even though both effects are on the same footing. In fact the latter's importance takes over the former as higher time derivatives of the scale factor contribute. An example of `incorrectness' is in Ref. \cite{2J} where presenting the dynamics is mis-conjured as including the backreaction.} We want to employ the best methodology established in the cosmological context and apply them to the problems under study of value to possible future analog gravity experiments.  2) To calculate the vacuum dynamical compressibility of a space we need to include all contributions of vacuum energy and pressure. One cannot use one kind of partial pressure due to one kind of vacuum process, such as Casimir or particle creation or trace anomaly, to calculate the partial dynamical compressibility, because they don't add up to give the correct expression of dynamical compressibility. It is the derivative with respect to the total pressure which counts. Because of this, and the fact that vacuum energies have been calculated in the literature for some but not all sources, we need to `fill in the gaps' and do it in as vigorous a way as possible. 3) Even though in this paper we have derived the behavior of vacuum dynamical compressibility for some typical spaces, because of its novelty, further work is needed to unlock its deeper physical meanings in our understanding of the quantum thermodynamics of spacetime. We view this work as only taking a first step in that direction.

\paragraph{Organization}

This paper is organized as follows: In Sec.~\ref{S:eotbsf} we present a summary of the recent results in \cite{XBH}  for the energy density of a conformal quantum field in a 1D ring with changing radius. We then work out the dynamical vacuum compressibility of this space, due to the trace anomaly and Casimir effect in the ring. In Sec.~\ref{S:esefsg} we present a calculation of the energy density of particle creation from the nonadiabatic parametric amplification of vacuum fluctuations of a massive conformal scalar field in a two-dimensional sphere with time varying radius. From this we obtain the dynamical vacuum compressibility of a two-sphere due to particle creation. In Sec.~\ref{S:dbshdj} we do the same for the 3-sphere, calculating the energy density first for particle creation from a massive conformal field, then for the trace anomaly of a massless conformal field, both under specified time variation of the radius. After integrating over the spatial volume we obtain the energy,  the pressure and the vacuum compressibility as functions of time. In Sec.~\ref{S:t3}, we present a careful analysis of the particle creation process of a conformal massless scalar field in a symmetric Bianchi type-I spacetime with $T^3$ topology, by allowing only one-dimension to expand. We work out the energy density, pressure and dynamical vacuum compressibility of this space,  neglecting all other less important effects. The results are shown in plots. This is a prelude to a more complete and in-depth treatment of this problem in the more realistic context of cosmology by the authors of~\cite{CHH1} in their second paper~\cite{CHH2}, by way of the nonequilibrium influence action for the free energy density.    
In Sec.~\ref{S:reiuss} we summarize the key points in the quantum field theoretical derivations and highlight the distinct features in the vacuum energy, pressure and compressibility in these two kinds of quantum field processes in the different spacetimes studied.

\begin{table}
\caption{\label{table}The red cross mark (left column) is used with the effect that is present for massless conformal field, while the blue check mark (right column) is for massive conformal field.}
\begin{ruledtabular}
\begin{tabular}{lllll}
 & $\mathbf{R}\times S^1 $&$\mathbf{R}\times S^2 $ &$\mathbf{R}\times S^3 $&$\mathbf{R}\times T^3 $\\
Casimir energy & \textcolor{Red}{$\checkmark$} \textcolor{Blue}{$\checkmark$}&\textcolor{red}{$\times$}\; \textcolor{Blue}{$\checkmark$}& \textcolor{Red}{$\checkmark$} \textcolor{Blue}{$\checkmark$}&\textcolor{Red}{$\checkmark$} \textcolor{Blue}{$\checkmark$}\\
Trace anomaly & \textcolor{Red}{$\checkmark$} \textcolor{Blue}{$\checkmark$} &\textcolor{red}{$\times$}\;\;\textcolor{Blue}{$\times$}\;\; &\textcolor{Red}{$\checkmark$} \textcolor{Blue}{$\checkmark$}& \textcolor{Red}{$\checkmark$} \textcolor{Blue}{$\checkmark$}\\
Particle production &\textcolor{red}{$\times$} \textcolor{Blue}{$\checkmark$}&\textcolor{red}{$\times$}\;\;\textcolor{Blue}{$\checkmark$} &\textcolor{red}{$\times$}\;\hspace{0.5mm}\textcolor{Blue}{$\checkmark$} & \textcolor{Red}{$\checkmark$} \textcolor{Blue}{$\checkmark$}\\
\end{tabular}
\end{ruledtabular}
\end{table}

\newpage

\section{\texorpdfstring{Vacuum compressibility of $\mathbf{R}\times S^1$ with back reaction}{}}\label{S:eotbsf}
In this section, we derive the dynamical vacuum compressibility from the vacuum energy density of a conformal quantum field by considering the one-dimensional moving mirror configuration. We impose periodic boundary conditions on the field at the surface of the mirror. Thus the configuration is equivalent to a one-dimensional ring with changing radius~\cite{XBH}. In so doing the original problem can be recast to that of finding the dynamical vacuum compressibility of a conformally-coupled quantum field in $\mathbf{R}\times S^1$ topology. Two different vacuum quantum processes need be considered: Casimir effect (static and dynamic) and trace anomaly. We include the key steps in the derivation for the convenience of those readers who may not be so familiar with the relevant quantum field theory in curved spacetime calculations.   In this  case the backreaction effects of these quantum processes on the system are included, so the scale factor $a(t)$ that account for the mirror's motion is dynamically determined, rather than prescribed a priori. They will provide a benchmark for comparison with cases treated in Sec.~\ref{S:esefsg},~\ref{S:dbshdj}, and \ref{S:t3} where the backreaction effects are ignored, that is, $a(t)$ following a given trajectory. 

% \subsection{\texorpdfstring{Vacuum compressibility of $\mathbf{R}\times S^1$}{} -- Casimir effect and trace anomaly}
% In this section, we study the simplest case: massless conformal field in (1+1)-dimension with topology of $\mathbf{R}^1\times S^1$. 

We shall calculate the vacuum energy, pressure, and compressibility of the quantum field associated with the Casimir effect and the trace anomaly. In this lower dimensional case, the expression for the trace anomaly is sufficiently simple, enabling one to dynamically determine the evolution of the scale factor including the backreaction effects of the quantum field.

Here we consider two mirrors in a one-dimensional space, one of which is fixed at the spatial origin and the other is allowed to move. The spatial location of the moving mirror varies with time $t$ as $a(t)\,L$. Here  $L$ is a constant measuring the `size'  and $a(t)$ is a scale factor which varies with time, describing the motion of the second mirror. 
We impose a periodic condition on the field at the surfaces of the two mirrors, so as far as the field is concerned, the space has a $S^1$ topology with time-dependent scale factor $a(t)$. 

Equivalently we can consider a two-dimensional spacetime with line element
\begin{align}
    ds^2 &=dt^2 - a(t)^2 \,dx^2\,,
    \label{metric}
\end{align}
Identifying $x$ and $x + L$, $L$ becomes the coordinate circumference. This is an Einstein cylinder (See, e.g., \cite{LCH_EinCyl}) with topology $ \mathbf{R}^1\times S^1$.  Introducing the conformal time $\eta$ defined by $d\eta=d t / a(t)$, the metric can be written in a conformally-flat form:
\begin{equation}\label{E:rtbfsd}
	ds^2 = a^2(\eta)\qty[d \eta^2 - d x^2]\,.
\end{equation}
A conformally-coupled, massless scalar field $\phi$ living in this spacetime defined by the action
\begin{equation}
    S = \frac{1}{2}\int d^2x \sqrt{-g}\;\qty[g^{\alpha\beta}\phi_{,\alpha}\phi_{,\beta}]\,,
\end{equation}
satisfies the Klein-Gordon equation, 
\begin{equation}\label{E:bggfgd}
    \Box \phi=0\,,
\end{equation}
where the Beltrami-d'Alembertian operator $\Box$ on a scalar is given by
\begin{equation}
    \Box=g^{\alpha\beta}\nabla_{\alpha}\nabla_{\beta}=\frac{1}{\sqrt{-g}}\,\partial_{\alpha}\Bigl(g^{\alpha\beta}\sqrt{-g}\,\partial_{\beta}\Bigr)
\end{equation}    
and $\nabla_{\alpha}$ is the covariant derivative operator. The partial differentiation of a tensor will be denoted by a comma in the subscript followed by the coordinate with which the differentiation is carried out, while the covariant derivative of a tensor is denoted by a semicolon.

Carrying out a normal mode expansion for this field,  we have
\begin{equation}
    \hat{\phi}(\eta,x)=\sum_{k_n}\left[\hat{a}^{\vphantom{*}}_{k_n}f^{\vphantom{*}}_{k_n}(\eta,x)+\hat{a}^\dagger_{k_n}f^*_{k_n}(\eta,x)\right]\,,
\label{mode expansion}
\end{equation}
where $f^{\vphantom{*}}_{k_n}(\eta,x)$ is the positive-frequency solution of \eqref{E:bggfgd}. Under the specified spatial topology, the $k_n$ form a discrete set $k_n=2\pi n/L$, $n\in\mathbb{Z}$. In a conformally-flat spacetime, the mode function $f_{k_n}$ is exactly solvable:
\begin{equation}
    f_{k_n}=\frac{1}{\sqrt{2\lvert k_n\rvert}}\,e^{i(k_nx - |k_n|\eta)}\,,\label{2d mode}
\end{equation}
with the normalization condition 
\begin{equation}
    i\,f^*_{k_n}(\eta,x)\,\overset{\text{\tiny$\leftrightarrow$}}{\partial}_{\eta} f^{\vphantom{*}}_{k_{n'}}(\eta,x)=\delta_{n,n'}\,.
\end{equation}    
Here $\delta_{n,n'}$ is the Kronecker delta.
For positive $n$, $f_{k_n}$ represent modes moving counterclockwise; when $n$ is negative, $f_{k_n}$ corresponds to modes moving clockwise.  

Taking the variation of the action with respect to the metric tensor, we obtain the {classical} energy-momentum tensor:
\begin{equation}
    T_{\mu\nu}=\frac{2}{\sqrt{|g|}}\frac{\delta S}{\delta g_{\mu\nu}}=(\partial_{\mu}\phi)(\partial_{\nu}\phi)-\frac{1}{2}g_{\mu\nu}g^{\rho\sigma}(\partial_{\rho}\phi)(\partial_{\sigma}\phi)\,,
    \label{1+1 stress tensor}
\end{equation}
which is traceless, $T_{\mu}{}^{\mu}=0$.

With the field variable promoted to be field operator we have the energy-momentum tensor operator. Using the mode expansion Eq.~\eqref{mode expansion} for Eq.~\eqref{1+1 stress tensor} the (unrenormalized) vacuum expectation values are obtained. From there, we obtain an expression for the energy density
\begin{align}
    \langle\hat{T}_{\mu\nu}\rangle U^\mu U^\nu&=\sum_{k_n}\qty[(\partial_{t}f^{\vphantom{*}}_{k_n})(\partial_{t}f_{k_n}^*)-\frac{1}{2}g_{tt}g^{\rho\sigma}(\partial_{\rho}f^{\vphantom{*}}_{k_n})(\partial_{\sigma}f_{k_n}^*)]=\frac{2\pi }{L^2}\sum_{n=0}^{\infty}n\,.\label{dT1}
\end{align}
where $U^\mu$ is the observer's four-velocity. In particular, $U^\mu$ has components $(1,0,\cdots,0)$ in the comoving frame.

This formal expression for the energy density in Eq.~\eqref{dT1} contains ultraviolet divergences which need to unambiguously identified and removed, we adopt the adiabatic regularization method, briefly explained in the following (see, e.g., \cite{XBH} for details). 

It is more instructive to work with a massive conformally-coupled field $\phi$. With a mode expansion in the form
\begin{equation}
    \hat{\phi}(\eta,x)=\sum_{k}\left[\hat{A}_{k}f_{k}(\eta,x)+\hat{A}^\dagger_{k}f^*_{k}(\eta,x)\right]\,,\label{adia mode expan}
\end{equation}
(we denote the annihilation operator of each mode by $\hat{A}_{k}$ to distinguish it from its counterpart in the massless field case) in a spatially homogeneous spacetime, the mode amplitude functions $f_{k}(\eta,x)$ are separable into
\begin{equation}
    f_{k}(\eta,x)=(2\pi)^{-1/2}e^{ikx}\chi_{k}(\eta).\label{mode}
\end{equation}
The normalization condition for $f_k$ leads to $\chi_{k}\partial_{\eta}\chi_{k}^*-\chi_{k}^*\partial_{\eta}\chi_{k}=i$. Substituting  Eq.~\eqref{mode} into the Klein-Gordon equation \eqref{E:bggfgd} yields  
\begin{equation}
    \chi''_{k}(\eta) + \omega_k^2(\eta)\chi_{k}(\eta)=0,\label{tm}
\end{equation}
with \(\omega_{k}=\sqrt{k^2+m^2a^2}\). This is the equation of motion for the amplitude functions of the normal modes of a parametric oscillator.

To identify the UV divergences in two-dimensional spacetime considered here, it is sufficient to find a (positive-frequency) WKB solution to Eq. \eqref{tm} up to the second adiabatic order, which should capture all the vacuum divergent terms
\begin{align}\label{wkb}
    \chi_k&=(2W_{k})^{-1/2}\exp{-i\int^{\eta}\!\dd\eta'\;W_k(\eta')}\,,\\
    W_k&=\omega_k(1+\epsilon_2)^{1/2}\approx\omega_k(1+\frac{1}{2}\epsilon_2)\,,&\epsilon_2&=-\omega_k^{-1/2}\frac{\dd^2}{\dd\eta^2} \omega_k^{1/2}\,.\notag
\end{align}
Substituting $f_k$ into Eq. \eqref{dT1} with $\chi_k$ given by \eqref{wkb}, then taking the massless limit, we find an expression for the energy density of the conformally coupled field evaluated in the second-order adiabatic vacuum, $\lvert0_A\rangle$,
\begin{align}
    \langle0_A\vert\hat{T}_{\mu\nu}\vert0_A\rangle U^\mu U^\nu&=\frac{1}{2\pi a} \int_{0}^{\infty}\!\dd k\;\omega_k + \frac{1}{24\pi }\frac{\dot{a}^2}{a^2}\,,
\end{align}
in the comoving frame, where the overhead dot denotes taking the derivative with respect to the comic time $t$. The renormalized energy  density of the conformal scalar field in its conformal vacuum is given by   
\begin{align}\label{E:nbgfs}
   \expval{T_{\mu\nu}}_{\text{reg}}U^\mu U^\nu=\langle\hat{T}_{\mu\nu}\rangle_{\text{c}} U^\mu U^\nu-\expval{\hat{T}_{\mu\nu}}{0_A}U^\mu U^{\nu}&=\frac{2\pi}{L^2}\sum_{n=0}^{\infty}n-\frac{1}{2\pi a} \int_{0}^{\infty}\!\dd k\;\omega_k - \frac{1}{24\pi }\frac{\dot{a}^2}{a^2}\notag\\
    &=-\frac{\pi }{6 a^2 L^2} - \frac{1}{24\pi }\frac{\dot{a}^2}{a^2}\,.
\end{align}
In arriving at \eqref{E:nbgfs}, we have introduced a cutoff to regularize the divergent series and integral. Without any cutoff scale appearing in the end results, this gives a finite well-defined expression for the energy density (see \cite{Ford75,Ford76} for an application to quantum fields in the Einstein universe). We will elaborate more on this in Sec.~\ref{S:gbeerer}.

The first expression on the righthand side of \eqref{E:nbgfs} gives the Casimir energy density for a ring of circumference $L$.  It becomes more negative when the scale factor $a(t)$ gets smaller, independent of the derivative of $a(t)$. The second term is due to the trace anomaly (`anomaly' because the classical stress tensor is traceless). It vanishes when $a(t)$ is a constant because the metric \eqref{E:rtbfsd} becomes Minkowskian. Both contributions are negative definite, so the renormalized energy density in this configuration is always negative.

We are interested in the total energy in the spacelike hypersurface orthogonal to comoving observer's four-velocity. To do so, we integrate the energy density observed by the comoving observer using the induced metric on this hypersurface. The induced metric is obtained by the projection $\gamma_{\mu\nu} = U_{\mu}U_{\nu} - g_{\mu\nu}$, such that $\gamma_{\mu\nu}U^{\nu}=0$. Integration gives the total energy of the field:
\begin{align}\label{Edkbdfgd}
    E &=  \int_{0}^{L}\!dx\sqrt{\gamma}\;\expval{T_{\mu\nu}}_{\text{reg}}U^\mu U^\nu= -\frac{1}{24\pi}\frac{\dot{a}^2}{a}-\frac{\pi}{6a}\,,
\end{align}
where by setting $L=1$, the physical length (perimeter) or volume of the ring is $a(t)$ without loss of generality. This energy is negative definite, and is unbounded from below. Since the motion of the mirror is relatively mild, the trace anomaly, dependent on the derivative of the scale factor, plays a subdominant role. The major contribution of vacuum energy in this case comes from the Casimir energy.

The pressure $P$ is given by
\begin{equation}\label{E:dfsdfs}
    P=-\frac{\partial E}{\partial V}=-\frac{\partial_t E}{\partial_t V}=-\frac{\pi}{6a^2}-\frac{\dot{a}^2}{24\pi a^2}+\frac{\ddot{a}}{12\pi a}\,,
\end{equation}
since both $E$ and $V$ are time dependent owing to the changing scale factor $a(t)$. The first term on the righthand side, devoid of any time derivative of the scale factor,  representing the Casimir energy is due to the topology of the space. The remaining terms in the square brackets of \eqref{E:dfsdfs} can be identified as contributions from the trace anomaly. Unlike the contribution associated with the Casimir energy, this component is not sign definite, depending on the higher derivatives of $a(t)$. 

With this, the vacuum compressibility $\kappa_{\textsc{vac}}$ is given by
\begin{equation}\label{E:brgbsss}
    \kappa_{\textsc{vac}} = -\frac{1}{V}\frac{\partial V}{\partial P}=\biggl[-\frac{\pi}{3a^2}-\frac{\dot{a}^2}{12\pi a^2}+\frac{\ddot{a}}{6\pi a}-\frac{\dddot{a}}{12\pi\dot{a}}\biggr]^{-1}\,.
\end{equation}
Again,  the first term inside the brackets comes from the Casimir energy, which is independent of $\dot{a}$, even though in arriving at this expression we have taken the time derivative of $P(t)$. Thus this contribution has the same form as the counterpart in the static Einstein cylinder. The trace anomaly contribution, like that in $P$, contains higher derivatives of $a(t)$. Once we have numerically obtained the time evolution of the scale factor $a(t)$, we may use  the above result to find the time evolution of $E$, $P$ and $\kappa_{\textsc{vac}}$.

It is of interest to ponder over the implications of \eqref{E:dfsdfs}. Suppose we start with a slowly varying $a(t)$ from a constant value, such that the $\ddot{a}$ term has a negligible initial contribution. The negative pressure \eqref{E:dfsdfs} generates a contracting stress, making the ring smaller. This trend is independent of the sign of $\dot{a}$. A smaller $a$ makes the inward pressure even stronger and the shrinking even faster. Once $\ddot{a}$ becomes negative, this trend becomes unstoppable and the ring spacetime is doomed to collapse. Positive initial $\dot{a}$ merely slight deters this from happening.

\begin{figure}
    \centering
    \includegraphics[width =1.0\columnwidth]{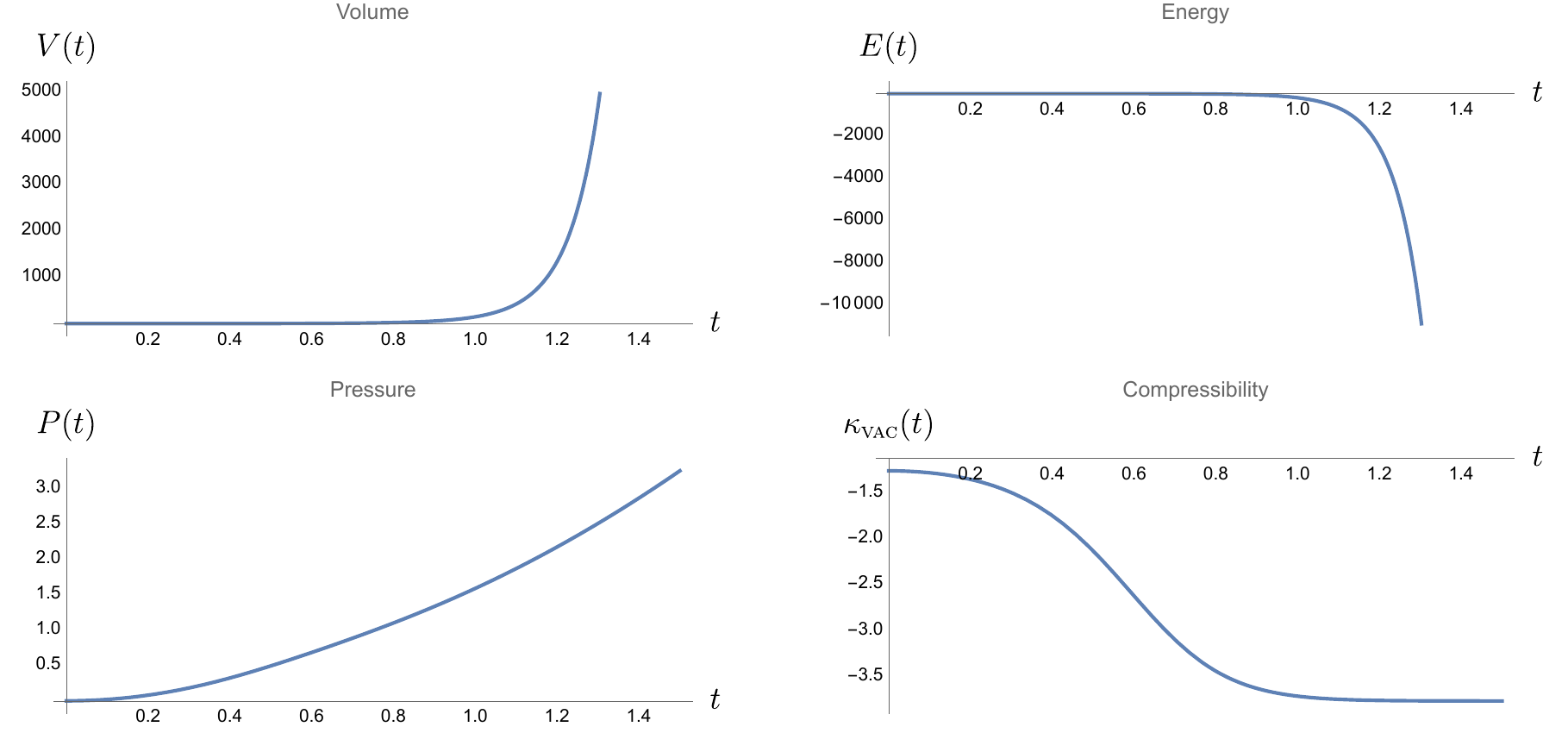}
    \caption{Here we consider a massless conformally coupled quantum field in an Einstein cylinder of topology $ \mathbf{R}^1\times S^1$ where the radius  (scale factor)  is preset as $a(t)=1+e^{t^2/\tau^2}$ with $1/\tau^2=5$. Plotted are the time dependence of the volume of the space, the vacuum energy and pressure of the field due to the Casimir effect and the trace anomaly combined, and the vacuum compressibility of the space associated with these sources.}
    \label{fig:1+1 exp2}
\end{figure}

\subsection{\texorpdfstring{An example of prescribed $a(t)$ with positive pressure}{}}
From \eqref{E:dfsdfs}, to generate a positive pressure, we need a very large $\ddot{a}$ such that the third term in \eqref{E:dfsdfs} becomes dominant. In other words, we would like the factor $a\,\ddot{a}$ to be at least as large as $\dot{a}^2$. It implies that $a(t)$ needs to be a rapidly increasing function of $t$, preferably faster than exponential growing. Then we may see the pressure taking on positive values and increase monotonically with $t$ or $a(t)$. This is pretty easy to achieve if $a(t)$ is prescribed by, say, $a(t)=1+e^{t^2/\tau^2}$. In this case, even though the volume is expanding, the total energy drops to large negative values. Apparently in this regime, the Casimir effect is negligible, so the dominant effects come from the trace anomaly, related to the derivatives of $a(t)$. The pressure in this case is positive, but it is not large enough to account for such a rapid expansion, so we expect such a scenario may not be realizable self-consistently. An external agent is needed to drive the expansion. In addition, the pressure increases with the volume, and this is the cause of a negative compressibility. It is in stark contrast with our experience with the ideal gas that when the volume expands, the pressure inside the volume decreases inversely. Even with this unusual compressibility, the spacetime seems to respond to the vacuum energy of the field in such a way that it is quite compressible at early times but becomes relatively less compressible at late times due to the larger absolute values of the vacuum compressibility.

\subsection{\texorpdfstring{An example of increasing then decreasing $a(t)$}{} }
\begin{figure}
    \centering
    \includegraphics[width =1\columnwidth]{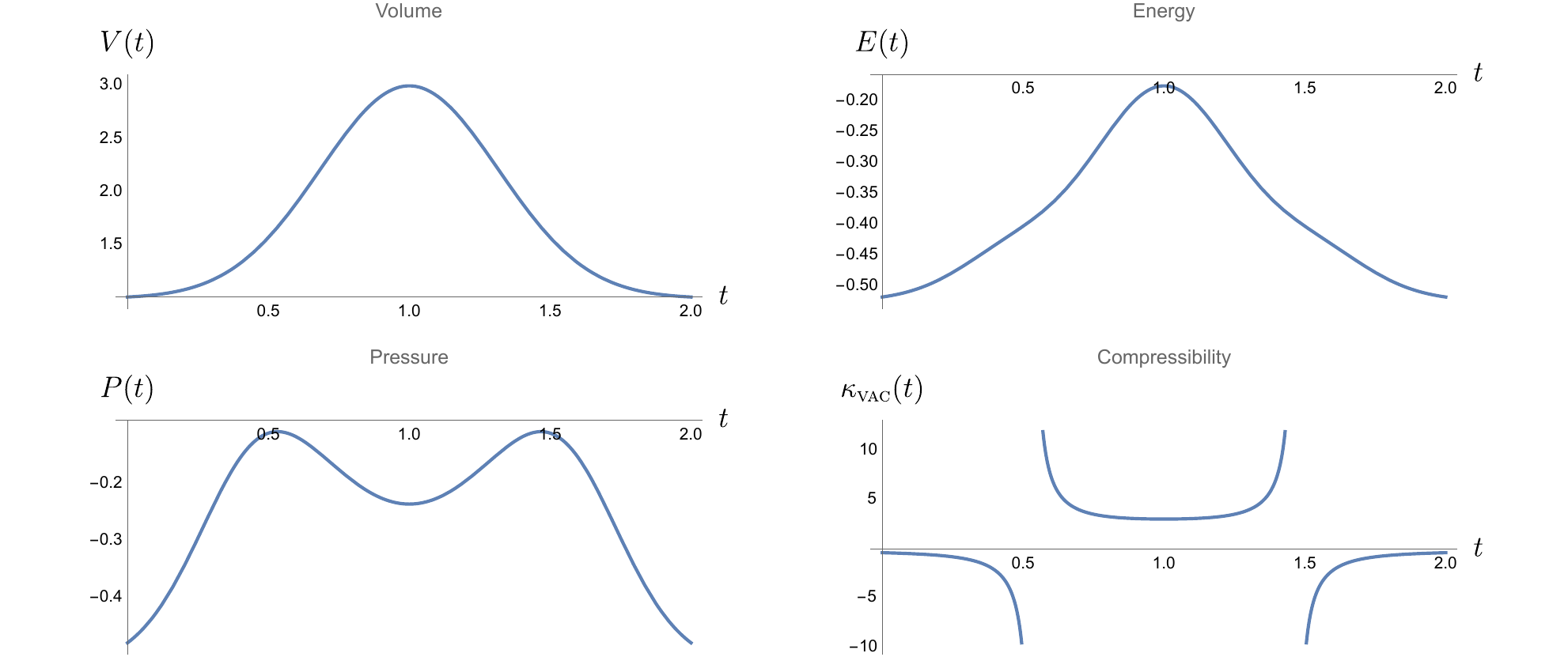}
    \caption{Time dependence of the volume, energy, pressure and vacuum compressibility in an Einstein cylinder $\mathbf{R}\times S^1$ with radius (scale factor) following a preset Gaussian time-dependence given by \eqref{E:gbirutsd}, with $a_0=1$, $\Delta=2$, $t_0=1$ and $1/\tau^2=5$. Same quantum field and vacuum processes as in Fig. \ref{fig:1+1 exp2}.}
    \label{fig:1+1 gau}
\end{figure}

In comparison, we add one more example in Fig.~\ref{fig:1+1 gau}. 
In this example, the volume of $S^1$ or the separation of two mirrors increases and then decreases, following a Gaussian function of time, 
\begin{equation}\label{E:gbirutsd}
    a(t)=a_0+\Delta\, e^{-(t-t_0)^2/\tau^2}\,,
\end{equation}
where without loss of generality $a_0$ can be identified as the scale at $t=-\infty$, $\Delta$ the maximal amount of scale change, $\tau$ the width of the Gaussian, and $t_0$ the moment of maximal separation. There are a few unusual features. The total energy reaches the least negative value at the moment $t=t_0$ where the volume is the largest. This is solely due to the Casimir effect,  the trace anomaly containing $\dot{a}$ plays no role. In contrast, for the pressure at $t=t_0$, the contributions from the Casimir effect and the trace anomaly are respectively given by
\begin{align}
    P_{\textsc{ca}}(t_0)&=-\frac{\pi}{6(a_0+\Delta)^2}\,,&P_{\textsc{ta}}(t_0)&=-\frac{\Delta/\tau^2}{6\pi(a_0+\Delta)}\,,&&\Rightarrow&\frac{P_{\textsc{ta}}(t_0)}{P_{\textsc{ca}}(t_0)}&=\frac{\Delta(a_0+\Delta)}{\pi^2\tau^2}\,.
\end{align}
\begin{figure}
    \centering
    \includegraphics[width =1.0\columnwidth]{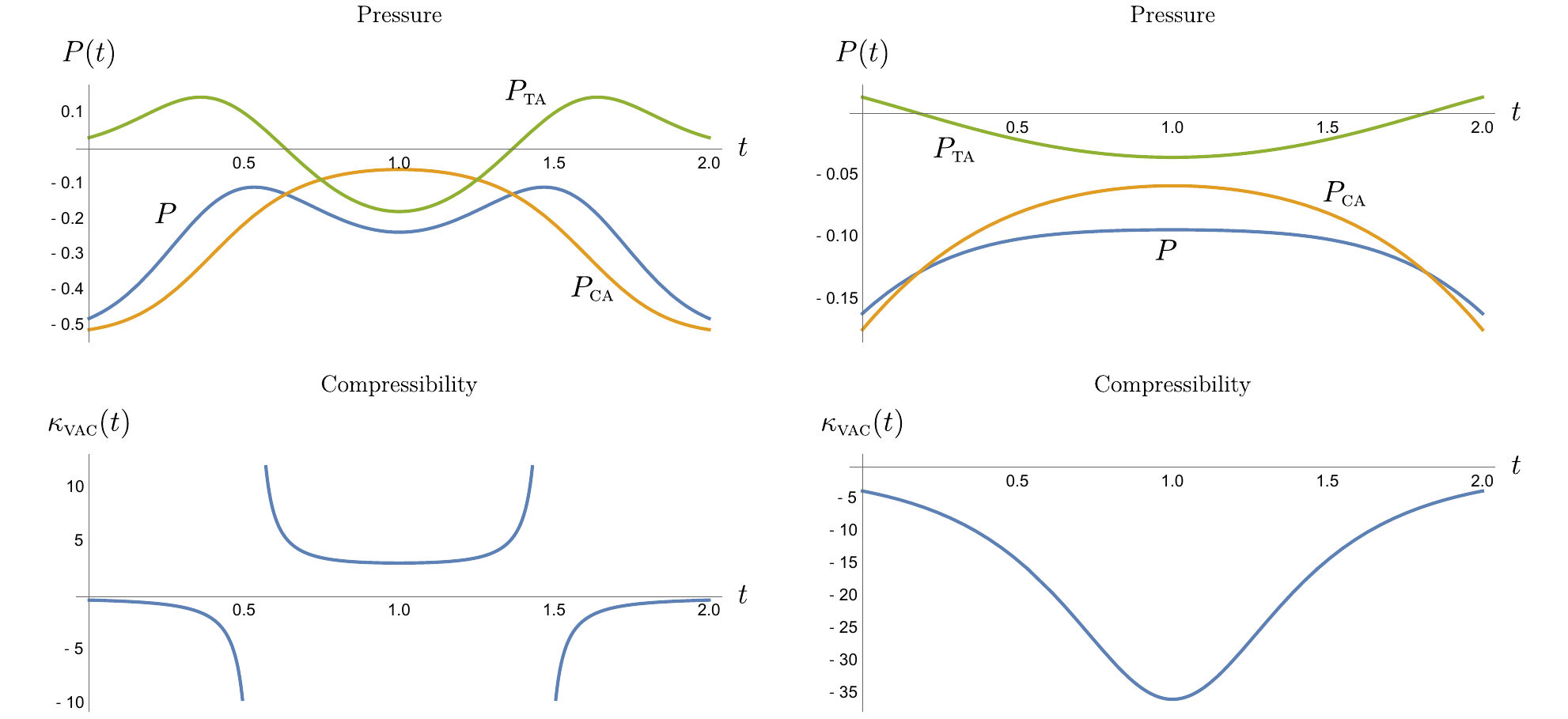}
    \caption{The time dependence of the pressure for the Gaussian evolution $a(t)=a_0+\Delta\, e^{-(t-t_0)^2/\tau^2}$ is shown in the first row for two choices of $\tau$. We decompose $P(t)$ into two contributions $P_{\textsc{ca}}$ and $P_{\textsc{ta}}$ respectively associated with the Casimir effect and trace anomaly. In particular, the distinct behavior of $P_{\textsc{ta}}$ is related to the compressibility shown in the second row. We have $1/\tau^2=5$ for the first column, and $1/\tau^2=1$ for the second column. The other parameters chosen here are $a_0=1$, $\Delta=2$, and $t_0=1$.}
    \label{fig:1+1 gauComp}
\end{figure}
Thus their relative dominance depends on the scales we choose for the scale factor. Generally speaking, since the trace anomaly is related to the time variation of $a(t)$, we expect $P_{\textsc{ta}}(t_0)$ will be more important when the Gaussian peak is very narrow, representing rapid time variation of the scale factor. The M shape of the curve of the pressure in Fig.~\ref{fig:1+1 gau} is essentially caused by the trace anomaly because it is dominant for the parameters chosen for Fig.~\ref{fig:1+1 gau}. In the case when the Casimir effect dominates, the curve monotonically increases to the least negative value, and then monotonically decreases, simply reflecting the change of scales. The most extraordinary feature of this example is that the compressibility is not well defined at certain moments. From the definition of compressibility, we see that $dV=-\kappa_\textsc{vac}V\,dP$, and such critical behavior tends to happen when $P$ has a local extremum but $a$ is still changing, so that $dP\approx0$ but $dV\neq0$. If the interpretation of compressibility still holds here, then it implies that the ring is the most deformable at this juncture. The culprit of this pathology may be traced back to the trace anomaly, as shown in Fig.~\ref{fig:1+1 gauComp}. In the left column, the parameter $1/\tau^2=5$, while $1/\tau^2=1$ for the plots in the right column. It is easier to identify the contributions from the Casimir effect and the trace anomaly in the plots of $P(t)$ in the first row. We note by comparison that the curve for the contribution from the trace anomaly tends to have more wiggles due to its dependence on various derivatives of $a(t)$. In addition, as addressed earlier, the trace-anomaly contribution in $P$ becomes more important for large $1/\tau^2$. These two factor brings in the compressibility shown in the left column of Fig.~\ref{fig:1+1 gauComp}. In comparison, the compressibility in the right column appears better behaved.

\begin{figure}
    \centering
    \includegraphics[width =1.0\columnwidth]{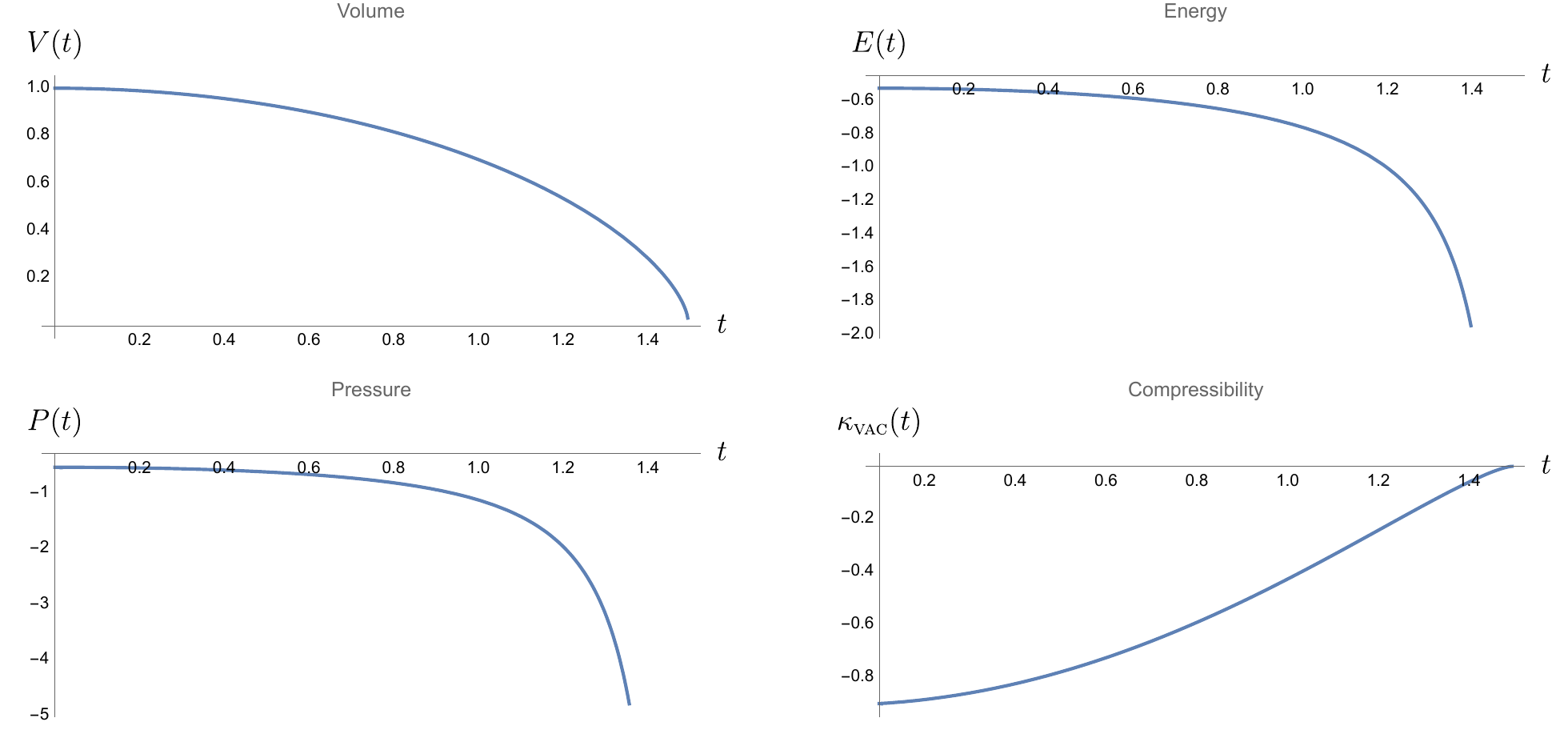}
    \caption{The time evolution of the volume, pressure, energy, and compressibility of a massless conformal field in a one-dimensional moving mirror setup, which after a periodic boundary condition is imposed, acquires a $\mathbf{R}\times S^1$ topology. The evolution of the scale factor $a(t)$ is dynamically determined by incorporating the backreactions from the vacuum energy, see \cite{XBH}. The initial conditions of $a(t)$ are $a(0)=1$ and $\dot{a}(0)=0$.}
    \label{fig:1+1 combfig}
\end{figure}

For the model outlined in Ref. \cite{XBH}, the behavior of $a(t)$ in principle can be dynamically determined if we include the backreaction considered above. The corresponding plots are shown in Fig.~\ref{fig:1+1 combfig}. The generic behavior is consistent with the above analysis. Since the increasing inward pressure causes the volume to shrink, we expect that the vacuum compressibility takes on negative values. However, it approaches zero when $S^1$ shrinks to a dot, and the space acts like a incompressible, rigid body.

\begin{figure}
    \centering
    \includegraphics[width =1.0\columnwidth]{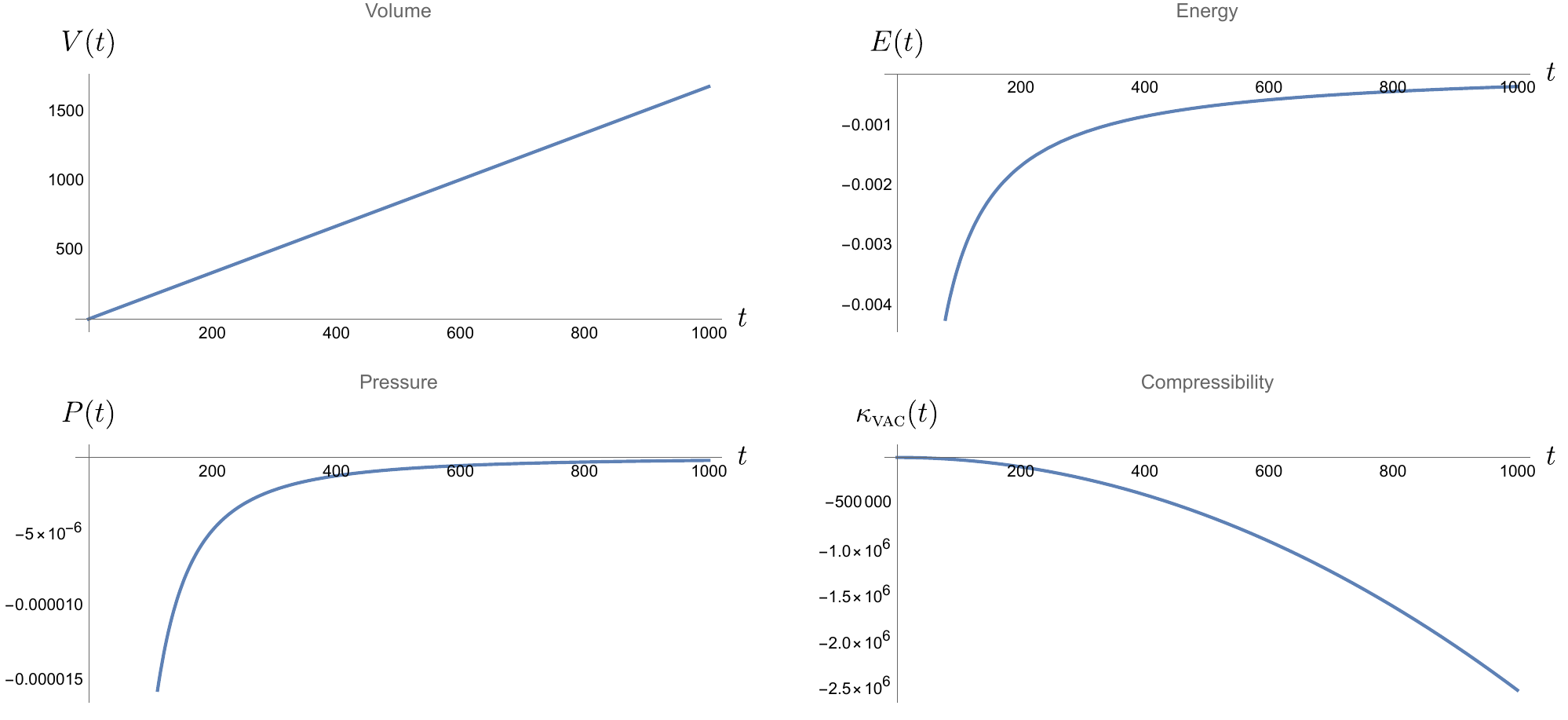}
    \caption{The time evolution of the volume, pressure, energy, and compressibility of a massless conformal field in the one-dimensional moving mirror setup in the same configuration as  in Fig.~\ref{fig:1+1 combfig} except we set $\dot{a}(0)>\mu_c$.}
    \label{fig:1+1 combfigc}
\end{figure}

It is interesting to comment on an observation that by the numerical tests, the results in Fig.~\ref{fig:1+1 combfig} is rather generic if the initial condition $\dot{a}(0)<\nu_c$, where the critical value $\nu_c$ is given by
\begin{equation}
    \nu_c=\frac{1}{\sqrt{12\pi m-1}}\,.
\end{equation}
When $\dot{a}(0)$ is greater than $\nu_c$, the dynamics of $a(t)$ has a qualitative change. It will grow indefinitely, but the energy and the pressure all approach zero. The consequence is that the compressibility grows rapidly with time, as shown in Fig.~\ref{fig:1+1 combfigc}.

\section{\texorpdfstring{Vacuum compressibility of $\mathbf{R}\times S^2$}{}}\label{S:esefsg}
Now we turn to the vacuum processes of a massive, conformally coupled quantum field in a $\mathbf{R}\times S^2$ spacetime, whose spatial sector is a two-sphere with time-dependent radius following a prescribed protocol. In this case, the vacuum processes include the Casimir effect and spontaneous particle creation. 

The line element of $\mathbf{R}\times S^2$ spacetime is     
\begin{eqnarray}
    ds^2 = a^2(\eta)\qty[d \eta^2 - d \theta^2 -\sin^2\theta\;d\varphi^2]\,,
    \label{eq: metric}
\end{eqnarray}
with $\theta \in [0,\pi]$, and $\varphi \in [0,2\pi)$. It has the same form as that of a  $(1+2)$-dimensional closed Robertson-Walker universe. 

Assume that the field has mass $M$, conformally coupled to curvature, and its action is given by
\begin{equation}
S = \frac{1}{2}\int\!d^3\vb{x} \sqrt{-g}\;\qty[g^{\mu\nu}\phi_{,\mu}\phi_{,\nu}-(M^2+\xi R)\phi^2]\,,
\end{equation} 
where the coupling strength takes on the value $\xi =1/8$, and the Ricci scalar $R$ is given by 
\begin{equation}
    R=-\frac{2}{a^2}+\frac{2a'^2}{a^4}-\frac{4a''}{a^3}\,.
\end{equation}
The field equation
\begin{equation}    
    \qty(\square +M^2+\xi R)\phi = 0\,,
    \label{eq: field equation}
\end{equation} 
permits an expansion of the form   
\begin{equation}
    \phi = \int\!\dd \tilde{u}_{\vb{k}}\;\qty[\hat{a}^{\vphantom{\dagger}}_{\vb{k}} f_{\vb{k}}^{\vphantom{*}}(x)+\hat{a}^\dagger_{\vb{k}} f_{\vb{k}}^*(x)]\,,
    \label{eq: mode expansion}.
\end{equation}
where the operator $\hat{a}_{\vb{k}}$ annihilates the vacuum state defined at the initial time $\eta_0$, and the measure $\displaystyle\int\!d \tilde{u}_{\vb{k}}$ stands for 
\begin{eqnarray}
    \int d \tilde{u}_{\vb{k}} = \sum_{l=0}^{\infty} \sum_{m=-l}^{l}\,.
\end{eqnarray}
when $\vb{k}$ takes on discrete values permitted by the indices $l$ and $m$ of the spherical harmonics $Y_{lm}(\theta,\varphi)$.

Because of the isotropy and homogeneity of space, the mode function $f_{\vb{k}}^{\vphantom{*}}(x)$ is separable and is given by $f_{\vb{k}}^{\vphantom{*}}(\eta,\theta,\varphi)=a(\eta)^{-1/2}\,Y_{lm}(\theta,\varphi)\,\chi_{l}(\eta)$ with $\chi_{l}$ satisfying 
\begin{align}
    \chi_{l}''(\eta)+\omega_{l}^2(\eta) \chi_{l}(\eta) &= 0\,,&\omega_{l}^2 = (l+1/2)^2 +a(\eta)^2M^2\,.
    \label{eq: Fourier mode}
\end{align}
The spherical harmonics $Y_{lm}(\theta,\varphi)$ are normalized to 
\begin{eqnarray}
    \int\!\dd\Omega_2\; Y_{lm}(\theta,\varphi)Y_{l'm'}^*(\theta,\varphi) =\delta_{ll'}\, \delta_{mm'}\,,
\end{eqnarray}
where $d\Omega_2$ is the solid angle element associated with $S^2$.

In this case, the expectation value of the energy density of the conformal vacuum, as measured by a comoving observer, is given by
\begin{align}
    \langle\hat{T}_{\mu\nu}\rangle U^{\mu}U^{\nu}&=\frac{1}{4\pi a^3} \sum_{l=0}^{\infty}(l+\frac{1}{2})\Bigl\{\lvert\chi'_{l}(\eta)\rvert^2+\omega_{l}^2(\eta) \lvert\chi_{l}(\eta)\rvert^2\Bigr\}\,.\label{edkfgsd}
\end{align}
Following the same procedures outlined in the previous section, we can re-write the summand $\lvert\chi'_{l}(\eta)\rvert^2+\omega_{l}^2(\eta) \lvert\chi_{l}(\eta)\rvert^2$ in terms of two time-dependent complex functions $\alpha_{l}$ and $\beta_{l}$ by the ansatz
\begin{align}
    \chi_{l}&=\frac{1}{\sqrt{2\omega_{l}(\eta)}}\,\qty[\alpha_{l}^{\vphantom{+}} e_{l}^{(-)}+\beta_{l}^{\vphantom{+}} e_{l}^{(+)}]\,,
    &\chi'_{l}&=-i\,\sqrt{\frac{\omega_{l}(\eta)}{2}}\,\qty[\alpha_{l}^{\vphantom{+}} e_{l}^{(-)}-\beta_{l}^{\vphantom{+}} e_{l}^{(+)}].
    \label{chi1d}
\end{align}
and arrive at three real first-order differential equations for $s_{l}$, $p_{l}$ and $q_{l}$
\begin{align}
    s'_{l} &= \frac{\omega'_{l}}{2\omega_{l}}\,p_{l}\,,&p'_{l} &= \frac{\omega'_{l}}{\omega_{l}}\bigl(1+2s_{l}\bigr)-2\omega_{l}\,q_{l}\,,&q'_{l} &= 2\omega_{l}\, p_{l}\,,
\end{align}
since here $Q=0$.

If we assume that no particle is present in any mode at the initial time $\eta_0$, then we have $\alpha_{l}(\eta_0)=1$, $\beta_{l}(\eta_0)=0$, which in turn implies $s_{l}(\eta_0)=p_{l}(\eta_0) =q_{l}(\eta_0)=0$. Thus we can express $\lvert\chi'_{l}(\eta)\rvert^2+\omega_{l}^2(\eta) \lvert\chi_{l}(\eta)\rvert^2$ simply by $s_{l}$, i.e., $\lvert\beta_{l}\rvert^2$, and find 
\begin{equation}
    \langle\hat{T}_{\mu\nu}\rangle U^{\mu}U^{\nu}=\frac{1}{4\pi a^3} \sum_{l=0}^{\infty}\bigl(l+\frac{1}{2}\bigr)\Bigl(2\lvert\beta_{l}\rvert^2 + 1\Bigr)\,\omega_{l}\,.
\end{equation}
What we are interested in, particle production, is contained in the $\lvert\beta_{l}\rvert^2$ term. The $\omega_{l}/2$ term contains zero-point fluctuations, and is the main culprit of UV divergence. After regularization, it will yield the Casimir energy density associated with $S^2$ topology.  The regularization is performed by subtracting out the energy density of the corresponding continuous limit, by taking $a\rightarrow \infty$
\begin{align}
    \langle\hat{T}_{\mu\nu}\rangle_{\text{phys}} U^{\mu}U^{\nu}&=\frac{1}{4\pi a^3} \sum_{l=0}^{\infty}\bigl(l+\frac{1}{2}\bigr)\Bigl(2\lvert\beta_{l}\rvert^2 + 1\Bigr)\,\omega_{l}\,-\lim_{a\rightarrow \infty}\frac{1}{4\pi a^3} \sum_{l=0}^{\infty}\bigl(l+\frac{1}{2}\bigr)\,\omega_{l}\notag\\
    &=\frac{2}{4\pi a^3} \sum_{l=0}^{\infty}\bigl(l+\frac{1}{2}\bigr)\,\lvert\beta_{l}\rvert^2\,\omega_{l}\,+\rho_{\textsc{ca}}\,,\label{E:tours}
\end{align}
where
\begin{equation}\label{E:fgvjdfg}
    \rho_{\textsc{ca}} = \frac{M^3}{2\pi}\int_0^1\!d \xi\;\frac{\xi \sqrt{1-\xi^2}}{e^{2\pi M a \xi}+1}\,, 
\end{equation}    
is the Casimir energy density~\cite{BordagBook}.

Since the space is homogeneous, the total energy of the quantum field on a spacelike hypersurface at a particular cosmic time of interest is obtained by multiplying the energy density by the volume, 
\begin{equation}\label{E:piweoru}
    E(\eta) = V\langle\hat{T}_{\mu\nu}\rangle_{\text{phys}} U^{\mu}U^{\nu} = 4\pi a^2 \langle\hat{T}_{\mu\nu}\rangle_{\text{phys}} U^{\mu}U^{\nu}.
\end{equation}
Here we will consider two illustrative examples of prescribed $a(t)$: (1) monotonic exponentially growth and (2) Gaussian, which contains a rise stage and a fall stage.

\begin{figure}
    \centering
    \includegraphics[width =1.0\columnwidth]{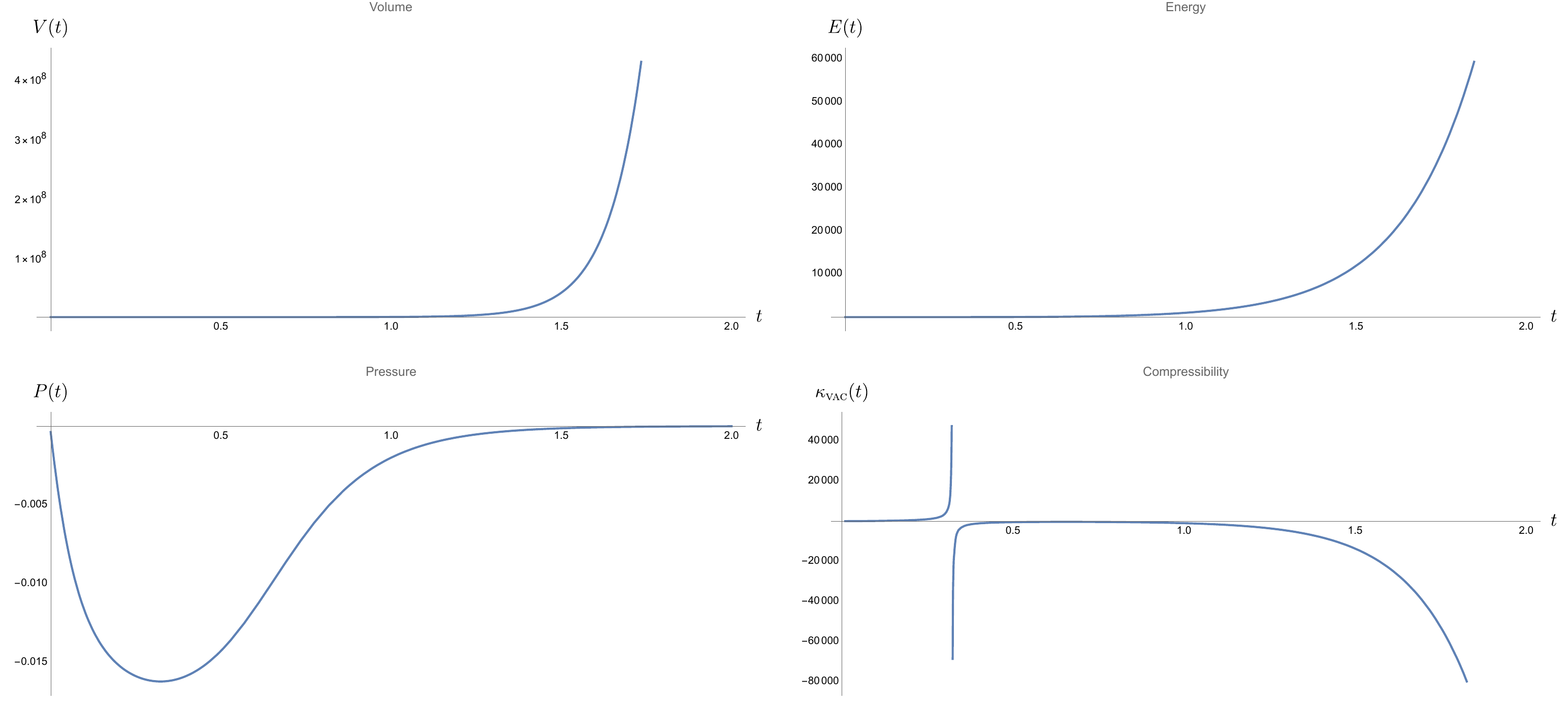}
    \caption{Time evolution of the volume, pressure, energy, and compressibility of a massive conformal field in $\mathbf{R}\times S^2$, with the scale factor taking on the exponential form, \eqref{exponential}, with $a_0=1$, $t_0=0$, and $\tau=1/5$.}
    \label{fig:exp2+1}
\end{figure}
(1)  Here we suppose the universe, which maintains a constant scale before $t=t_0$, undergoes some phase transition such that it starts exponentially expansion after $t=t_0$. The scale factor is an exponentially growing function of cosmic time $t$, 
\begin{equation}\label{exponential} 
    a(t)=
    \begin{cases}
        a_0\,,&t<t_0\,,\\
        a_0\,e^{\frac{t-t_0}{\tau}}\,,&t\geq t_0\,,
    \end{cases}
\end{equation}
where $\tau$ measures the rate of change of the scale factor, and $t_0$ represents the time theses two phases join, and, without loss of generality, we let $a_0=1$, $t_0=0$ . We can numerically solve the set of first order differential equations for $s_{\vb{k}}$, $p_{\vb{k}}$, and $q_{\vb{k}}$, and find the energy according to \eqref{E:tours} and \eqref{E:piweoru}. The pressure and compressibility then follow suit. The relevant results are shown in Fig.~\ref{fig:exp2+1}. 

The volume of $S^2$ is given by $V(t)=4\pi a^2(t)$, and the vacuum energy scales roughly proportional to the scale factor, i.e., $E(t)\propto a(t)$. Among various contributions of the vacuum energy, the Casimir energy is essentially negligible for a macroscopic universe of size greater than the Planck length, and interestingly enough, it increases to a positive constant when $t$ is greater than of the order $\tau$. Thus at late times $t>\tau$, the vacuum energy associated with spontaneous particle creation is dominant \footnote{A word of caution about the numerical calculations of the latter contribution. It is noted that for each $t$, once we have included a sufficiently large number of modes, the numerical result barely depends on the number of modes. Thus, we can set a finite upper bound of the summation in \eqref{E:tours}, instead of $\infty$, to save the computation resources. This upper bound is time dependent, and grows very rapidly with time. If we cut the corner by including an insufficient number of modes, the numerical results show an artefact, demonstrating oscillator behavior. Physically, the time-dependent cutoff is related to the fact that particles in the higher modes are  created so that the total number of created particles are expected to increase with time. Thus, we need to raise the upper bound of the summation to accommodate the contribution of newly created particles.}. It is interesting to point out that the vacuum energy density reaches a local maximum at $t\simeq 0.439$. Note that in $\mathbf{R}\times S^2$ spacetime, the vacuum energy does not contain any contribution from the trace anomaly because it is known that there is no trace anomaly in odd-dimensional spacetime~\cite{BirDav}.

The pressure $P(t)$ then takes on negative values and scales like $-a^{-1}$ at late times. That is, the spacetime experiences a very weak inward pressure. This pressure has a local extremum roughly at $t\simeq0.323$, when the Casimir energy is about to reach the saturated value. Both effects are not causally related because, as have been argued earlier, that the Casimir energy is typically small. At this moment, the universe is still expanding, so it causes the vacuum compressibility $\kappa_{\textsc{vac}}$ to be ill-defined there. This is not too unfathomable because here we stipulated the universe  to  expand in a particular way. At late times, the compressbility is then negative and decreases proportional to $-a(t)$, as a fated consequence of $E(t)$, driven by $a(t)$.

\begin{figure}
    \centering
    \includegraphics[width =1.0\columnwidth]{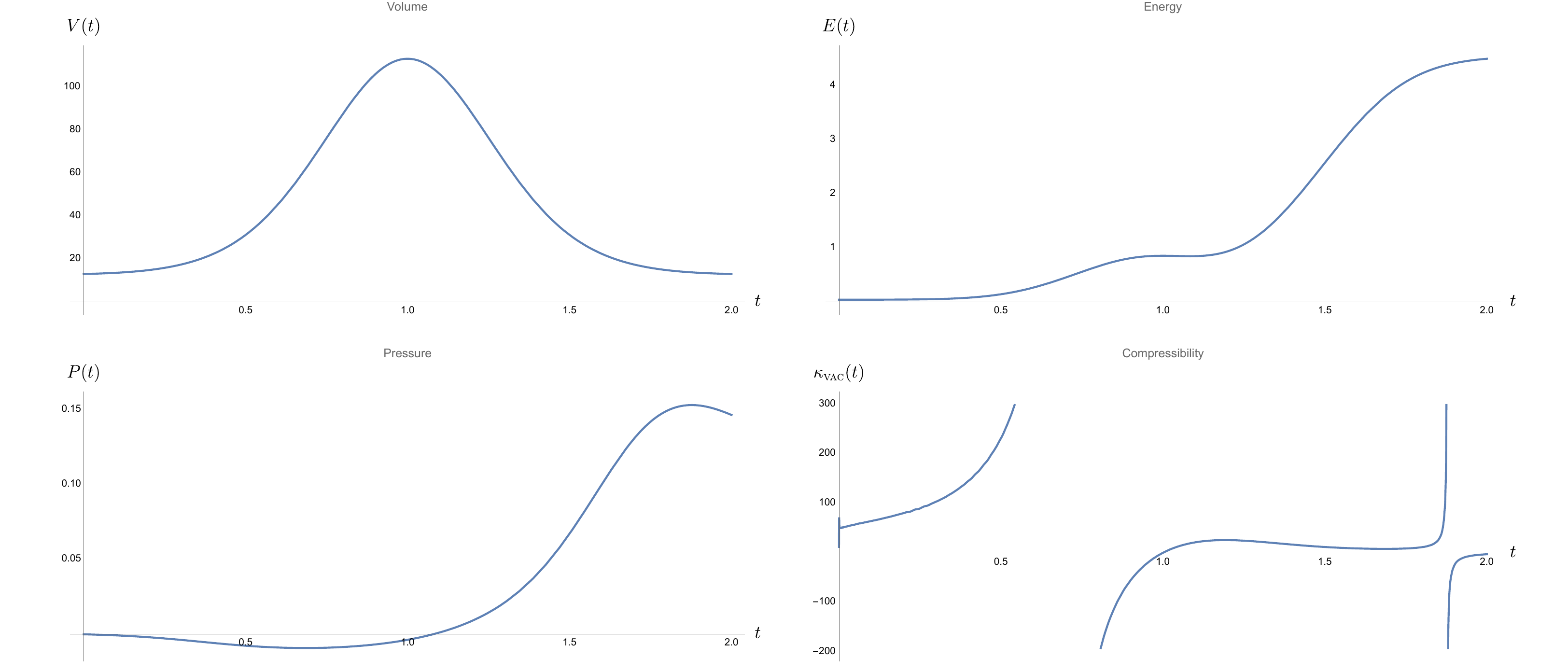}
    \caption{Time evolution of the volume, pressure, energy, and compressibility of a massive conformal field in $\mathbf{R}\times S^2$ space where the scale factor undergoes a Gaussian expansion, \eqref{gaussian} with $a_0=1$, $t_0=1$, $\Delta=2$, and $\tau^2=1/5$}
    \label{fig:2+1}
\end{figure}

(2) The scale factor is a Gaussian
\begin{equation}\label{gaussian}
   a(t) =a_0+\Delta\, e^{-\frac{(t-t_0)^2}{\tau^2}}\,.
\end{equation}
As seen in the $\mathbf{R}\times S^1$ case, it contains more scales in the expression, so the results, shown in Fig. \ref{fig:2+1}, tend  to be more complicated. In this toy model, the Universe first rapidly expands to a maximal volume at $t=t_0$ and then contracts back to the original size. For the current choice of the parameters, the curve of the vacuum energy has an inflection point at $t=t_0$ and then grows to a constant at sufficiently large time. The height of this saturated value depends on the rate of expansion and the mass of the field .

\begin{figure}
    \centering
    \includegraphics[width =1.0\columnwidth]{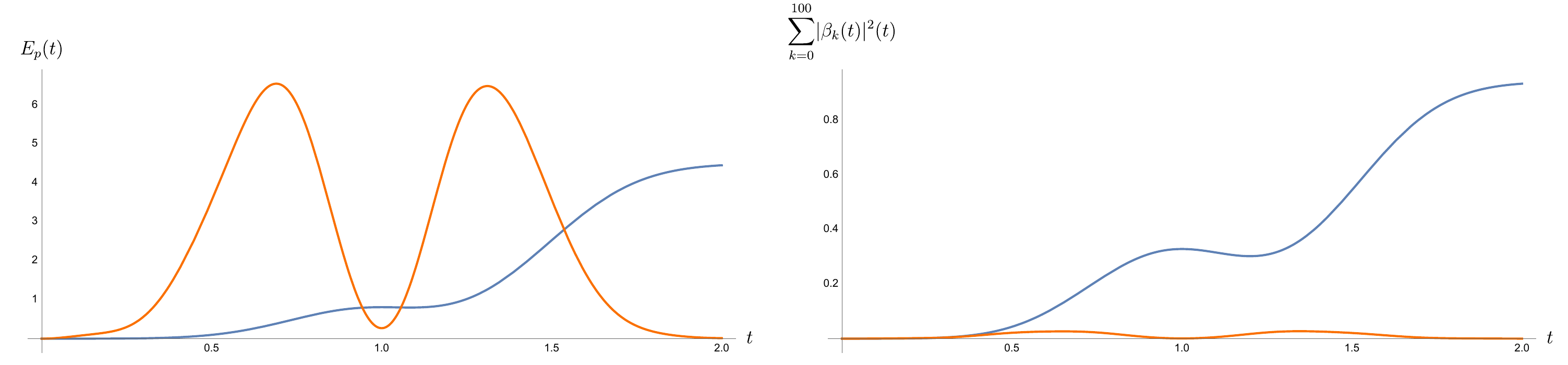}
    \caption{Comparison of the energy $E_p$ associated with spontaneous particle production and the density of created particles in the first 100 modes in $\mathbf{R}\times S^2$ when the scale factor is given by a Gaussian expansion, \eqref{gaussian} with $a_0=1$, $t_0=1$, $\Delta=2$, and $\tau^2=1/5$. The orange curve corresponds to $M=10$, while the blue curve has $M=1$.}
    \label{fig:2+1zzz}
\end{figure}

Similar to the previous case, the Casimir energy is quite small, compared to the energy $E_p$ associated with spontaneous particle production as a result of rapid expansion. However, the shape of $E_p$ is sensitive to the parameters in the configuration, in particular, the mass of the field. If we choose the values of field mass $M=10$, as shown by the orange curve in Fig.~\ref{fig:2+1zzz}, then the curve of $E_p$ shows a major double-peak feature, and then quickly decays to very small values. A larger field mass in principle can amplify the effect of the scale factor in \eqref{eq: Fourier mode}, making a given mode more energetic. Thus, the contribution to the energy attributed to particle creation can be boosted. This sounds very paradoxical because we would expect that particles of heavier mass are harder to produce. Indeed, as shown in Fig.~\ref{fig:2+1zzz}, the density of created particles is suppressed when the mass increases from $M=1$ to $M=10$, but the oscillatory frequency of a given mode increases in such a way that $E_p$ still grows with mass during the transient moments. At late times, when the universe asymptotically returns to its original size, the particle density seems to saturate to a very small value, roughly proportional to $M^{-1}$ for sufficiently large $M$. This is consistent with our expectation that it is harder to create heavier particles. 

It may seem a bit strange that the positive Casimir energy acquires the largest value when the universe is at the maximal scale. The Casimir energy density is inversely proportional to $a(t)$, and indeed it is a minimum at $t=t_0$.  However, when multiplied by the volume of the universe which  increases fast the energy takes on the greatest value. Our numerical calculations show  that both $E_p$ and the Casimir energy are actually proportional to $M$ even though \eqref{E:fgvjdfg} superficially has a overall factor of $M^3$ outside the integral, so the contribution of the Casimir energy remains small when the field mass takes on a larger value.

The temporal behavior of the pressure contains several local extrema, so the compressibility becomes ill defined there.

\section{\texorpdfstring{Vacuum compressibility of conformal scalar field in $\mathbf{R}\times S^3$}{S3}}\label{S:dbshdj}
\subsection{\texorpdfstring{Massive field:  particle creation}{}}
We first consider spontaneous particle creation of a massive conformal field in  $(1+3)$-dimensional Robertson-Walker universe with line element
\begin{eqnarray}
    ds^2 = a^2(\eta)\qty[d \eta^2 - \gamma_{ij}\,d x^i d x^j]\,,
    \label{eq: crw metric} 
\end{eqnarray}
where the spatial metric $\gamma_{ij}$ is given in hyperspherical coordinates by   
\begin{eqnarray}
    \gamma_{ij}\,d x^i d x^j = d X^2+\sin^2X\qty(d \theta^2 +\sin^2\theta\,d\varphi^2)\,, \qquad\qquad X\in [0,\pi]\,.
    \label{eq: crw space metric}
\end{eqnarray}
The wave equation governing massive conformal fields is 
\begin{eqnarray}
 \qty[\square +M^2+\frac{1}{6} R]\phi = 0\,,
    \label{eq: 31 field equation}
\end{eqnarray}
where $R = 6a^{-2}(a^{-1}a'' +1) $, and the curvature coupling constant $\xi$ is $1/6$ in $S^3$.

The mode expansion of the field operator $\hat{\phi}(x)$ has the same form as before:
\begin{eqnarray}
    \hat{\phi}(x) = \int\! d \tilde{u}\;\qty[\hat{a}^{\vphantom{\dagger}}_{\vb{k}} f^{\vphantom{\dagger}}_{\vb{k}}(x)+\hat{a}^\dagger_{\vb{k}} f_{\vb{k}}^*(x)]\,,
    \label{eq: 31 mode expansion}.
\end{eqnarray}
with the measure $\displaystyle\int\!d \tilde{u}$ defined by
\begin{eqnarray}
    \int\!d \tilde{u} = \sum_{k=1}^{\infty}\sum_{l=0}^{k-1} \sum_{m=-l}^{l}\,.
\end{eqnarray}
The mode functions $f_{\vb{k}}$, in terms of the hyperspherical harmonics, has the form 
\begin{eqnarray}
    f_{\vb{k}}(\eta,X,\theta,\varphi) &=&  a(\eta)^{-1}\chi_k(\eta)\,Y_{klm}(X,\theta,\varphi)\,,
\end{eqnarray}
with the amplitude function $\chi_k(\eta)$ obeying the wave equation
\begin{eqnarray}
    \chi_k''(\eta)+\omega_k^2(\eta) \chi_k(\eta) = 0\,,
\label{31 chi mode}
\end{eqnarray}
where $\omega_k^2 = k^2 +a(\eta)^2M^2 = k^2 +\mu(\eta)^2$.

\begin{figure}[b]
    \centering
    \includegraphics[width =1.0\columnwidth]{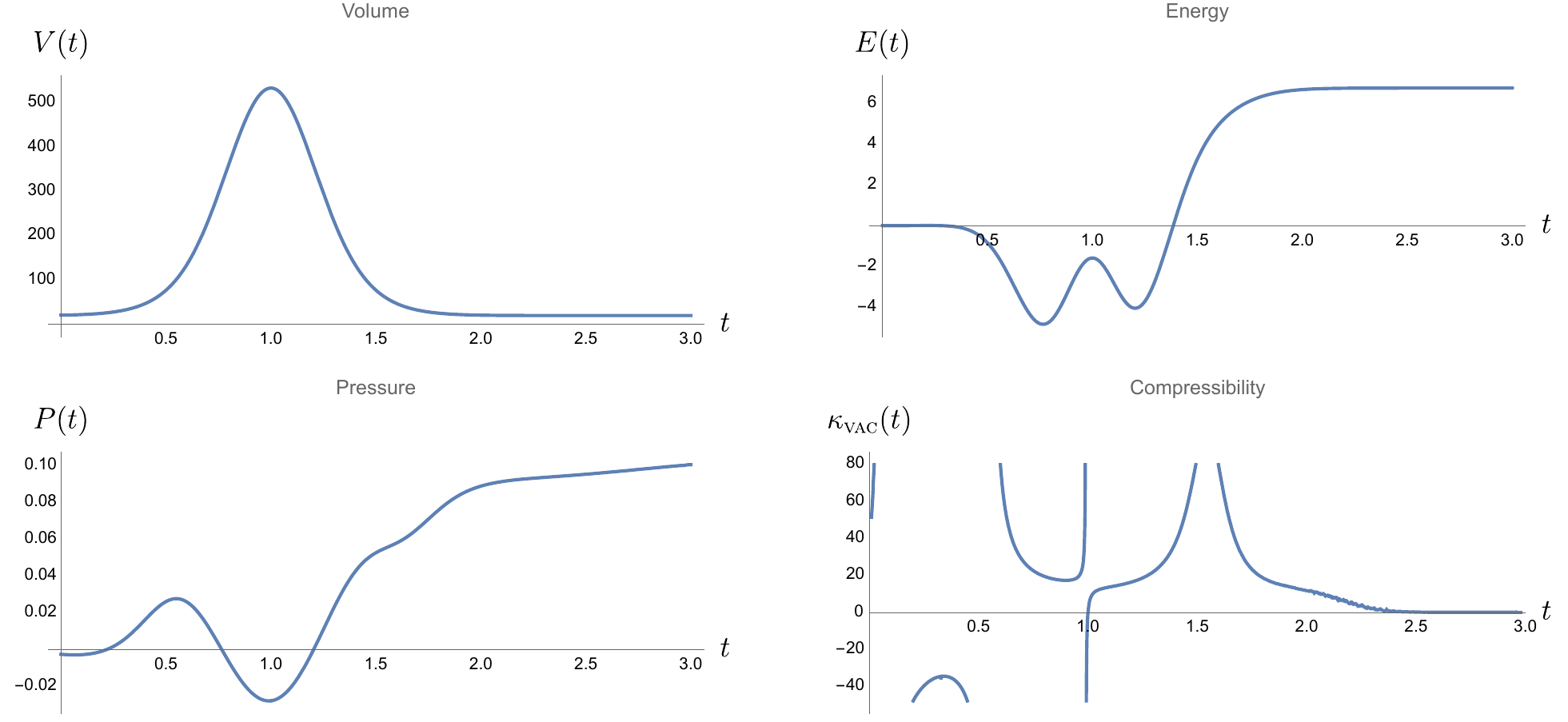}
    \caption{The time evolution of the volume, pressure, energy, and compressibility for a massive scalar field in $\mathbf{R}\times S^3$ with a Gaussian $a(t)$. We use the same set of   parameters as before.}
    \label{fig:3+1 massive comp}
\end{figure}

The procedure henceforth is similar to what was presented in the previous section for $\mathbf{R}\times S^2$ spacetime, except for the difference in the prefactors of the mode sums
\begin{eqnarray}
\expval{T_{\mu\nu}} U^{\mu}U^{\nu}=\sum_{k=1}^{\infty}\sum_{l=0}^{k-1} \sum_{m=-l}^{l}\; T_{\mu\nu}[f_{\vb{k}}^{\vphantom{*}},\,f_{\vb{k}}^*]U^{\mu}U^{\nu}=\frac{1}{4\pi^2 a^4} \sum_{k=1}^{\infty}k^2\qty[|\chi'_{k}|^2+\omega_k^2(\eta) |\chi_{k}|^2]\,.
\end{eqnarray}
The form of $\chi_k$ is the same as in the $\mathbf{R}\times S^2$ case, so is the expression of the energy density  
\begin{eqnarray}
 \expval{T_{\mu\nu}} U^{\mu}U^{\nu} &=&  \frac{1}{4\pi^2 a^4} \sum_{k=1}^{\infty}k^2\qty[\omega_k(2|\beta_k|^2 + 1)]\,.
\label{31 E summation}
\end{eqnarray}
The above expression is formally UV divergent, and  subtraction of terms of the corresponding adiabatic orders is needed to obtain a finite expression. The latter have been calculated in~\cite{Bunch80}. The finite physical energy density is given by
\begin{align}
    \expval{T_{\mu\nu}}_{\text{phys}} U^{\mu}U^{\nu} &= \frac{1}{4\pi^2 a^4} \sum_{k=1}^{\infty}k^2\qty[\omega_k(2|\beta_k|^2 + 1)] \notag\\
    &-\frac{1}{4\pi^2 a^4}\int_{0}^{\infty} \!d k\; k^2 \Biggl\{\Biggr.\omega_k + \frac{M^4a^4}{8\omega_k^5}\frac{a'^2}{a^2}- \frac{M^4a^4}{32\omega_k^7}\qty[2\frac{a'''a'}{a^2}
    -\frac{a''^2}{a^2}+\frac{4a''a'^2}{a^3}-\frac{a'^4}{a^4}] \notag\\
    &+ \frac{7M^6a^6}{16\omega_k^9}\qty[\frac{a''a'^2}{a^3}+\frac{a'^4}{a^4}]-\frac{105M^8a^8}{128\omega_k^{11}}\frac{a'^4}{a^4} \Biggl.\Biggr\}\,.
\end{align}
The Casimir energy on the other hand is given by  
\begin{equation}
    \rho_{\textsc{ca}} = \frac{1}{4\pi^2 a^4} \sum_{k=1}^{\infty}k^2\omega_k -\frac{1}{4\pi^2 a^4}\int_{0}^{\infty} \!d k\; k^2 \omega_k\,.
\end{equation}
After performing the integration, we obtain
\begin{align}
    \expval{T_{\mu\nu}}_{\text{phys}} U^{\mu}U^{\nu} &= \frac{1}{4\pi^2 a^4} \sum_{k=1}^{\infty}\qty[2|\beta_k|^2\,k^2\omega_k] -\frac{M^2 a'^2}{48\pi^2 a^4} + \rho_{\textsc{ca}}\notag\\
    &\qquad\qquad+ \frac{6}{2880\pi^2 a^4} \qty{\frac{a'''a'}{a^2}-\frac{1}{2}\frac{a''^2}{a^2}-\frac{2a''a'^2}{a^3}+\frac{1}{2}\frac{a'^4}{a^4}}\,.
\end{align}
Expressing the result in  cosmic time  
\begin{figure}
    \centering
    \includegraphics[width =1\columnwidth]{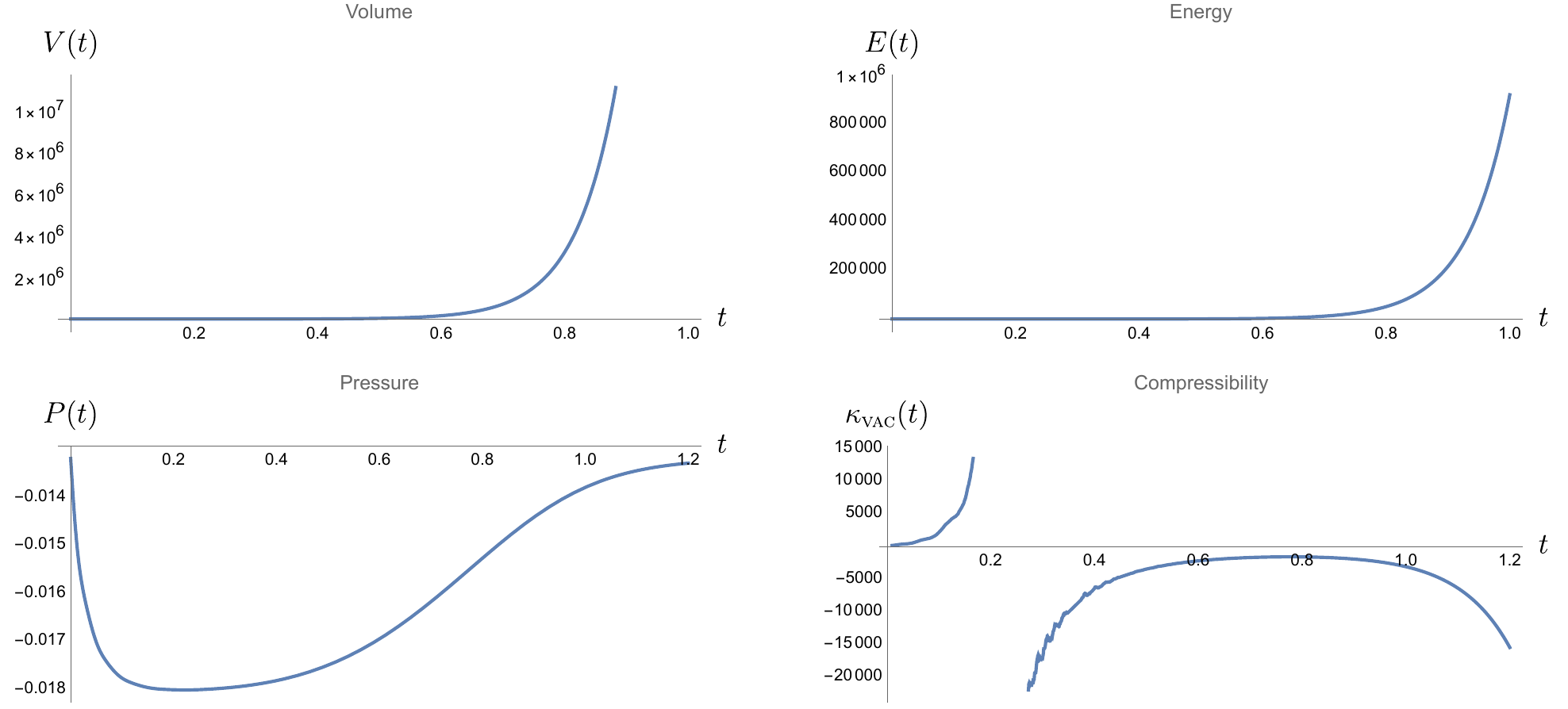}
    \caption{The time evolution of the volume, pressure, energy, and compressibility for a massive scalar field in $\mathbf{R}\times S^3$ with $a(t)$ following an exponential expansion. The parameters are the same as before, but here we choose a shorter time scale. The pressure will tend to a negative constant, not shown in the plot.}
    \label{fig:3+1 exp massive comp}
\end{figure}
\begin{align}
    \expval{T_{\mu\nu}}_{\text{phys}} U^{\mu}U^{\nu} &= \frac{1}{4\pi^2 a^4} \sum_{k=1}^{\infty}\qty[2|\beta_k|^2\,k^2\omega_k] -\frac{M^2 \dot{a}^2}{48\pi^2 a^2} + \rho_{\textsc{ca}}\notag\\
    &\qquad\qquad\qquad+ \frac{1}{480\pi^2 a^2} \qty{\frac{\dot{a}^2 \ddot{a}}{a}-\frac{\dot{a}^4}{a^2}-\frac{\ddot{a}^2}{2}+\dot{a}
   \dddot{a}}
   \label{31 cosmic total E}
\end{align}
gives the total energy. The third term is the Casimir energy and the expressions in the second line (shown in the next section) come from the trace anomaly. The second term is related to the nonzero trace of the classical stress tensor of the massive field. The numerical results for a) an exponential expansion and b) a Gaussian evolution are respectively shown in Fig.~\ref{fig:3+1 exp massive comp} and Fig.~\ref{fig:3+1 massive comp}, with the volume given by $ V(t)=2\pi^2a^3(t)$.

For the case of exponential expansion, the vacuum energy contains four components. The contribution from the Casimir energy, like before, is negligibly small for a macroscopic universe. The vacuum energy $E$ in this $\mathbf{R}\times S^3$ case scales like $a^3(t)$. At first glance it appears plausible to relate it to the fact that the number of allowed modes increases exponentially with the spatial dimensions. It turns out this is not entirely true. As will be explained later, this scaling mainly results from the trace-related contribution of the vacuum energy in a rapidly expanding universe. There are two types of contributions associated with the trace of the renormalized expectation value of the energy momentum tensor of the massive field: One is related to the nonzero trace of the classical stress tensor, and the other, independent of mass, results from the trace anomaly. For a rapidly expanding case like the current one, their contributions tend to dominate over the one due to spontaneous particle creation,  because these contributions contain   higher-order time derivatives of the scale factor. In particular, the trace anomaly contribution in a fast expanding universe typically is more important than the one associated with the classical trace as long as $m$ is smaller than the inverse of the expansion time scale. On the other hand, for a slowly (but not adiabatically slowly) expanding universe, say $\tau=1$, the contribution associated with the trace can be comparable or even smaller than the counterpart by spontaneous particle creation. 

\begin{figure}
    \centering
    \includegraphics[width =1\columnwidth]{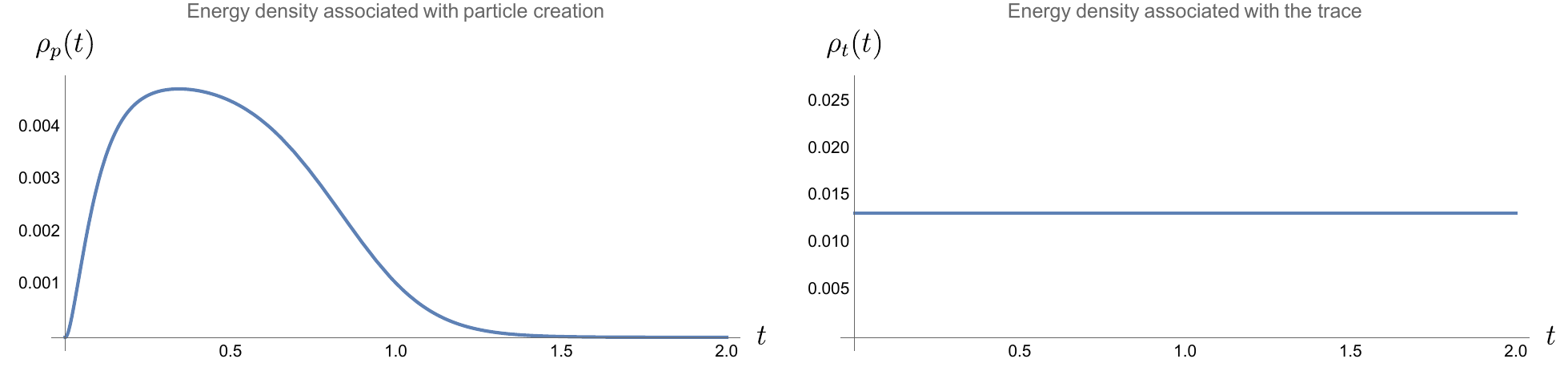}
    \caption{Comparison of the time evolution of the energy densities associated with spontaneous particle creation and the trace for a massive scalar field in an exponentially expanding 3-sphere $\mathbf{R}\times S^3$. We use the same set of  parameters as before.}
    \label{fig:3+1 exp massive ED comp}
\end{figure}

In this case the time evolution of the vacuum may reveal additional complexity due to the competition of different time scales such as the inverse mass, $\tau$, or their combinations. From the time dependence of the vacuum energy, we may roughly deduce the time evolution of the pressure and the vacuum compressibility accordingly. At late times $t\gg\tau$, $P(t)$ scales like $a^0(t)$ and is thus a negative constant. Again the pressure has a local extremum at $t\simeq 0.225$, its presence causes the compressibility to be ill defined there. 

It is interesting to compare this with the above time scale $t\simeq0.344$ when both the vacuum energy density $E/V$ and the ratio $E_p/E_t$ reach the local maximum, in which $E_t$ is the contribution associated with the trace. At first glance it is not clear why both ratios reach the local maximum at the same moment. A closer numerical inspection shows that the energy density associated with the trace quickly reaches a saturated constant for the exponential expansion case. This can be readily seen from \eqref{31 cosmic total E}
\begin{equation}
    -\frac{M^2 \dot{a}^2}{48\pi^2 a^2}+ \frac{1}{480\pi^2 a^2} \qty{\frac{\dot{a}^2 \ddot{a}}{a}-\frac{\dot{a}^4}{a^2}-\frac{\ddot{a}^2}{2}+\dot{a}
   \dddot{a}}=-\frac{M^2}{48\pi^2\tau^2}\biggl[1-\frac{1}{20M^2\tau^2}\biggr]\,.
\end{equation}
This further reveals how the magnitude and the sign of the trace-related contribution in vacuum energy can depend on the mass $M$ of the field and the expansion rate $\tau^{-1}$, and allow us to assess the importance of this contribution.

Thus, the time variation of the vacuum energy density $\rho_{\textsc{vac}}$ is mostly governed by the contribution due to spontaneous particle creation, as shown in Fig.~\ref{fig:3+1 exp massive ED comp}.

Compared with the $\mathbf{R}\times S^2$ case, there are two major distinctions: 1) The contribution of the vacuum energy scales differently among the two cases. Thus, this variation will cause the pressure and compressbility to scale in a different way. 2) The vacuum energy in $\mathbf{R}\times S^2$ spacetime is always dominated by the process of vacuum spontaneous particle production for a macroscopic universe, but in $\mathbf{R}\times S^3$ spacetime, the most important contribution depends on the expansion rate and the mass scale. For a fast expanding universe, the trace-related contribution dominates.

When the scale factor assumes the Gaussian form, \eqref{gaussian}, the time evolution of the vacuum energy gets increasingly complicated because the trace related contribution depends on various time derivatives of the scale factor, introducing multiple scales into the evolution. The behavior of the energy $E_p$ associated with spontaneous particle production is very similar to that in $\mathbf{R}\times S^2$, but the curve of vacuum energy $E$ looks somewhat different from the counterpart in the $\mathbf{R}\times S^2$ case. It can be negative before the universe reaches the maximum size. This is related to the fact that the trace related contribution $E_t$ is negative about $t=t_0$, and its absolute value is  slightly larger than $E_p$ for the current choice of parameters. It is worth mentioning that the time evolution of $E_t$ is symmetric about $t=t_0$ because each term contains an even number of derivatives of the scale factor of the Gaussian form. Thus, the contribution $E_t$ becomes negligible at late times.

It is not too surprising that the compressibility has pretty wild behavior, but it approaches zero at late times. This suggests that the space manifest itself as a rigid body. That is, the volume barely changes when the outward pressure is slightly increased.

\subsection{\texorpdfstring{Massless field: Trace anomaly}{}}\label{S:gbeerer}

As a final example we return to the effects due to the trace anomaly.  The action of a massless conformally couplws scalar field in a  closed Robertson-Walker spacetime with scalar curvature $R$ is
\begin{equation}
    S = \frac{1}{2}\int\!d^3x \sqrt{-g}\;\qty[g^{\mu\nu}\phi_{,\mu}\phi_{,\nu}-\xi R\,\phi^2]\,.
\end{equation}
It is invariant under conformal transformations, i.e., $g_{\mu\nu} \rightarrow a^2g_{\mu\nu}$ if we set $\xi = 1/6$.

The frequencies $\omega_k$ of the normal modes in Eq. \eqref{31 chi mode} become time independent and are given by $\lvert k\rvert$,
\begin{equation}
    \chi_k''(\eta)+\omega_k^2 \chi_k(\eta) = 0\,.
\end{equation} 
The energy density is formally divergent at the UV end:
\begin{eqnarray}
\langle\hat{T}_{\mu\nu}\rangle  U^{\mu}U^{\nu}&=&  \frac{1}{4\pi^2 a^4} \sum_{k=1}^{\infty}k^3\,.
\label{massless bare energy}
\end{eqnarray}
Before we carry out the renormalization of the energy density expression, we note that although the trace of the classical energy momentum tensor vanishes, the trace of the renormalized expectation values of the energy momentum tensor operator is nonzero, this anomalous trace, known as the trace anomaly, is a purely quantum effect. The renormalization procedure  breaks conformal symmetry. This nonzero trace anomaly has been calculated by various approaches, such as dimensional regularization, zeta function regularization, point-splitting regularization, and so on. Here we will take the adiabatic regularization, and the corresponding expressions for the trace anomaly for a massless conformally coupled scalar field in the Bianchi type I and the Robertson-Walker universes have been derived in \cite{Hu78,Hu79} by exploiting the difference between a massless conformal quantum field theory and a massive conformal quantum field theory after renormalization in the mass approaching  zero limit. The observation was that finite terms appear in the reduction of integrals for the energy density and the pressure  containing higher powers of mass  to integrals containing a lower power of mass when one sets the mass of a massive field to zero. These finite terms from the integrals containing higher powers of mass add up to produce the trace anomaly which would otherwise not be there if one works ab initio with a massless field throughout.

With this in mind,  we obtain the energy density in a comoving frame: 
\begin{align}
    \expval{T_{\mu\nu}}_{\text{phys}}U^{\mu}U^{\mu} &= \frac{6}{2880\pi^2 a^4}\qty[1+\frac{a'''a'}{a^2}-\frac{a''^2}{2a^2}-\frac{2a''a'^2}{a^3}+\frac{a'^4}{2a^4}]\notag\\
    &= \frac{1}{480\pi^2 a^2} \qty{\frac{1}{a^2}+\frac{\dot{a}^2 \ddot{a}}{a}-\frac{\dot{a}^4}{a^2}-\frac{\ddot{a}^2}{2}+\dot{a}
   \dddot{a}}\,.\label{E:fgfd}
\end{align}
In the second line, we express the result in terms of cosmic time. The first term in the curly brackets is the Casimir energy density, and the remaining expressions is associated with the trace anomaly of a massless conformal field in the Robertson-Walker universe~\cite{FHH79,And83}.  It is interesting to note that the same expression for the contribution associated with trace anomaly in \eqref{E:fgfd} also appears in the second line of \eqref{31 cosmic total E} in the massive field case.
\begin{figure}
    \centering    
    \includegraphics[width =1.0\columnwidth]{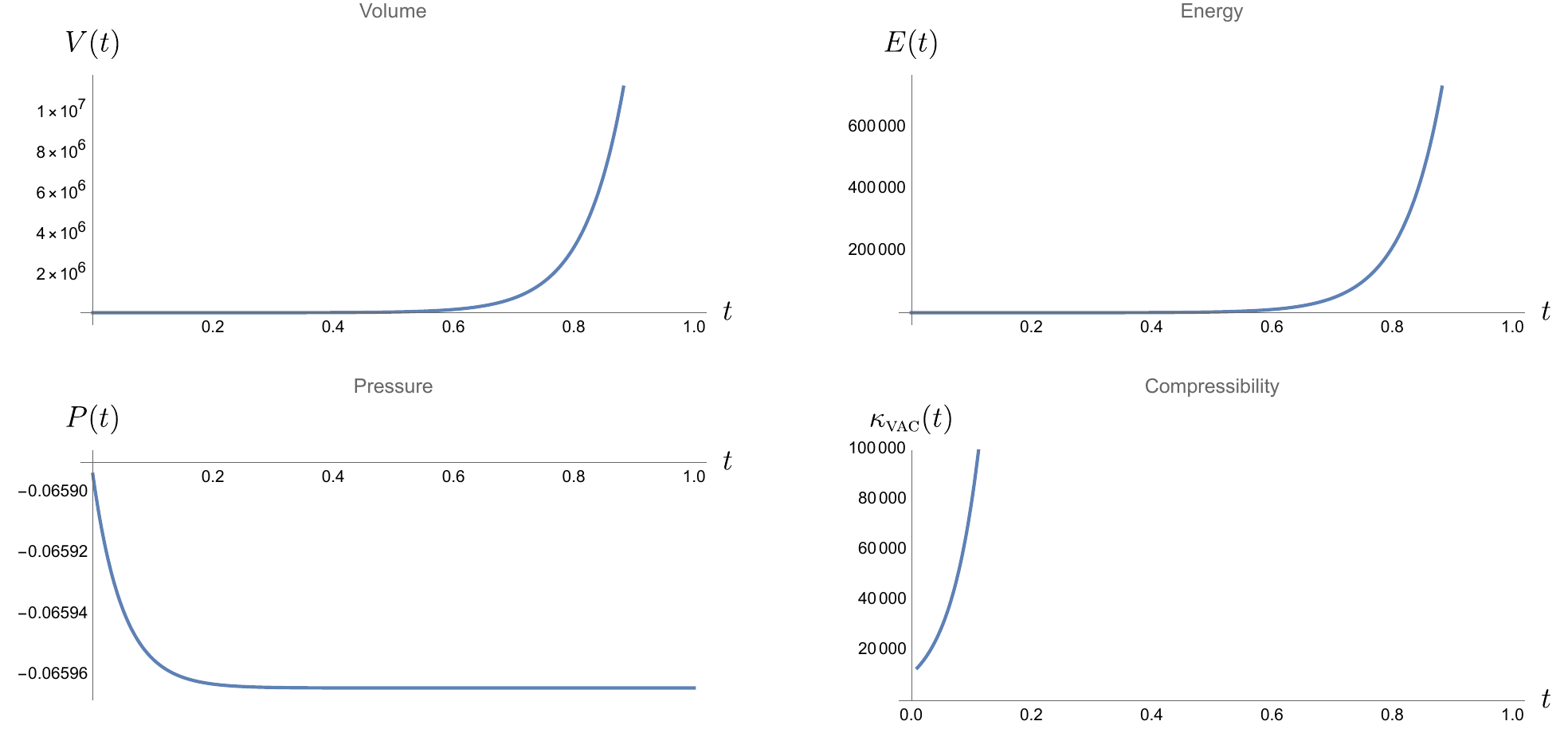}
    \caption{The time evolution of the volume, pressure, energy, and vacuum compressibility for a massless scalar field in $\mathbf{R}\times S^3$ when an exponentially expanding scale factor $a(t)$.}
    \label{fig:3+1 trace Exp}
\end{figure}

From this expression of the energy density for the massless conformal field,  retracing the steps in the massive field case, we obtain the results for the pressure, and the vacuum compressibility for a massless conformal scalar field due to the Casimir effect and the trace anomaly. There is no spontaneous particle production for a conformally coupled, massless field in a conformally static spacetime. 

\begin{figure}[htbp]
    \centering    
        \includegraphics[width =1\columnwidth]{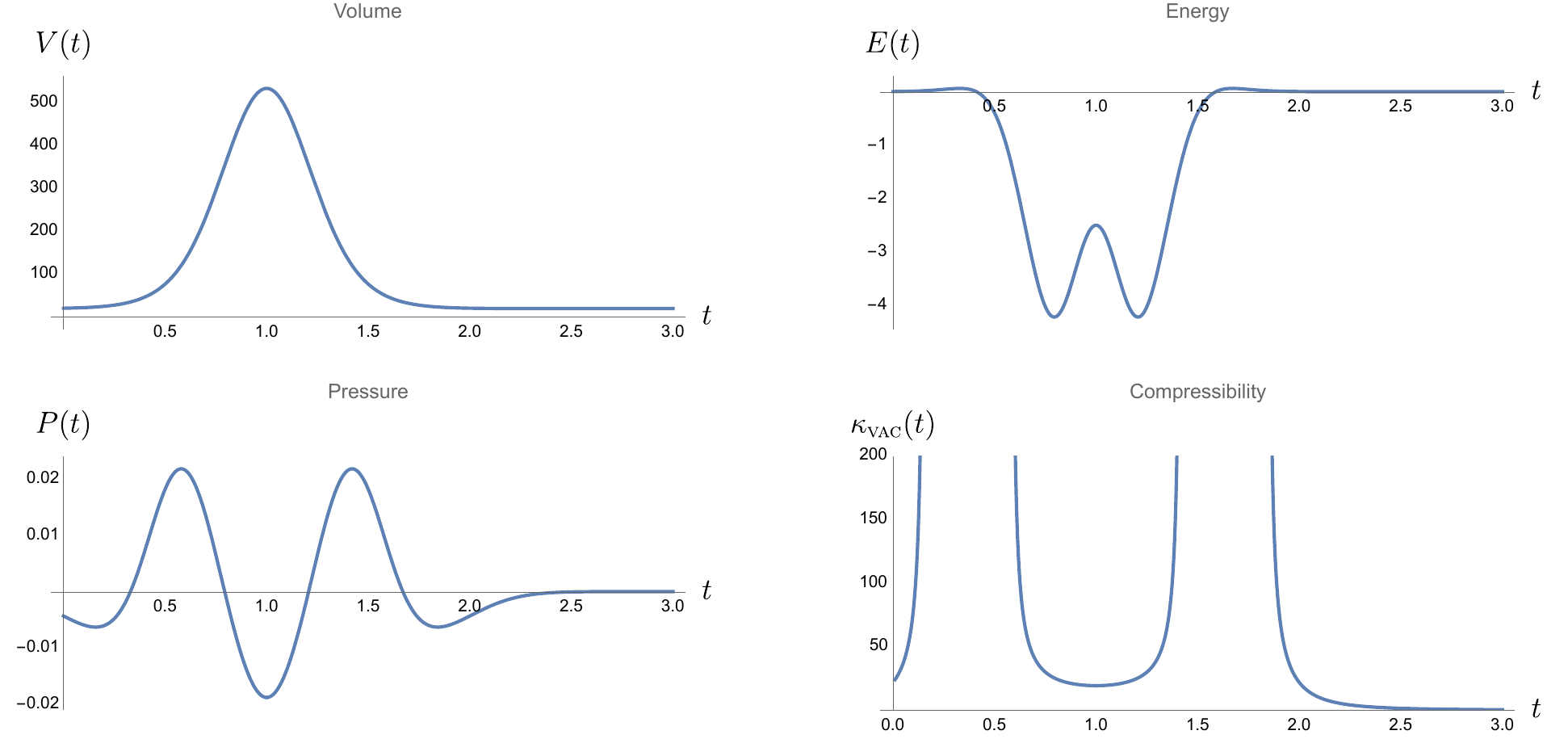}
    \caption{Time evolution of the volume, pressure, energy, and vacuum compressibility for a massless scalar field in $\mathbf{R}\times S^3$ for a scale factor $a(t)$ taking on a Gaussian dynamics}
    \label{fig:3+1 trace Gauss}
\end{figure}

Their time evolution due to these two quantum sources are shown in Fig.~\ref{fig:3+1 trace Exp} for an exponential expansion and Fig.~\ref{fig:3+1 trace Gauss} for a Gaussian evolution.  It is seen that the vacuum energy due to the Casimir effect is too small to be of our concern. The contribution from the trace anomaly of a massless field is smaller than that of a massive field because the mass dependent term is absent. As a consequence,  we do not see a local extremum in the time evolution of the pressure and thus the vacuum compressibility is well defined. However, since the pressure saturates rapidly to a negative value, the compressibility diverges accordingly.

This is a good place to comment on the comparison with the results for the $\mathbf{R}\times S^1$ case. a) Recall that in the $\mathbf{R}\times S^1$ case the scale factor $a(t)$ is dynamically determined, with the backreaction effects. That is, we solve for the dynamics of $a(t)$ with the Casimir effect and the trace anomaly incorporated, by the self-consistent interaction between the dynamics of the quantum field and the geometry. Here, the scale factor is prescribed a priori, not determined dynamically.  b) In the $\mathbf{R}\times S^1$ case, the presence of the trace anomaly accelerates the collapse of the $S^1$ space, over and above that already present due to the Casimir effect. c) The expressions for the trace anomaly in two dimensions and four dimensions are quite different: the four-dimensional case contains higher-order derivative of $a$. Therefore it is not surprising to see many qualitative differences between these two cases.

If the universe expands and contracts according to the Gaussian profile given by \eqref{E:gbirutsd}, then, as analyzed earlier,  the vacuum energy is essentially given by the contribution of the trace anomaly only. This is because the Casimir energy is negligible when $a(t)$ is large, and the contribution from the nonzero classical trace vanishes. Hence the features related to the trace-anomaly contribution we highlighted earlier can be directly applied here. Since there is no particle production, the vacuum energy is very small at late times. However, at $t\gg t_0$ the contribution from the trace anomaly is exponentially small, while that from the Casimir energy dominates.

\section{\texorpdfstring{Vacuum compressibility of $\mathbf{R}\times T^3$}{}}\label{S:t3}
% We now turn our attention to the effect of spontaneous particle creation. Creation of field quanta occurs quite generically for the quantum field in a time-dependent background. The conformally-coupled, massless scalar field in a conformally-flat spacetime is a well known exception. For the effects of particle creation of conformally-coupled massless scalar fields we need to consider anisotropic expansion, such as was done for the Bianchi Type I universe \cite{HuPar78,HarHu79}.   

We now turn our attention to the quantum effect due to anisotropic expansion. Creation of field quanta occurs quite generically for the quantum field in a time-dependent background. The conformally-coupled, massless scalar field in a conformally-flat spacetime is a well known exception. For the effects of particle creation of conformally-coupled massless scalar fields we need to consider anisotropic expansion, such as was done for the Bianchi Type I universe \cite{HuPar78,HarHu79}.   

Here we consider the effects of vacuum particle creation in a three-dimensional rectangular box. We allow the longitudinal length to change in time, but set the two transverse lengths equal and fixed. Inside the box is filled a massless, conformally-coupled quantum scalar field in its vacuum state. We assume that the field takes on periodic boundary conditions on the surfaces of the box, so for this quantum field, the box configuration is analogous to a space with three-torus $T^3$ topology. Thus, we have cast the dynamical Casimir effect of a massless, conformally-coupled scalar field into an anisotropically expanding spacetime. We are interested in the vacuum energy of created particles in this $\mathbf{R}\times T^3$ spacetime.

This spacetime has a line element of the same form as that of an anisotropically expanding,  spatially-flat  (Bianchi type-I) universe:
\begin{equation}\label{E:hdfgds}
    ds^2 = d t^2 - \qty[a_1^2(t)\,d x^2 + a_2^2(t)\,d y^2 + a_3^2(t)\,d z^2]\,,
\end{equation}
where $a_i$ are the scale factors in the three Cartesian directions. We let $a_2=a_3=1$, but $a_1=a(t)$, i.e., allowing the scale factor in the $x$ direction to change in time. 

A conformally-coupled, massless scalar field $\phi$ in this anisotropically expanding space obeys the Klein-Gordon equation: 
\begin{eqnarray}
    \partial_t^2\phi + \frac{\dot{a}}{a}\dot{\phi} - \sum_{i=1}^3\frac{\p_i^2\phi}{a_i^2}+\frac{1}{6}R\phi=0\,,
    % \label{eq: field equation}
\end{eqnarray}
where $\xi=\dfrac{n-2}{4(n-1)}=\dfrac{1}{6}$ is the conformal coupling strength, and $R=-\dfrac{2\ddot{a}}{a}$ is the scalar curvature of spacetime.

The classical energy-momentum tensor of this scalar field is given by
\begin{align}
    T_{\mu\nu} &= (\p_\mu\phi)(\p_\nu\phi)-\frac{1}{2}g_{\mu\nu}g^{\lambda\sigma}(\p_\lambda\phi)(\p_\sigma\phi) -\frac{1}{6}\left[\nabla_\mu\p_\nu(\phi^2)-g_{\mu\nu}g^{\lambda\sigma}\nabla_\lambda\p_\sigma(\phi^2)+\phi^2G_{\mu\nu}\right]\,,
\end{align}
where $G_{\mu\nu}$ is the Einstein tensor and $\nabla_{\mu}$ is the covariant derivative with respect to the metric tensor $g_{\mu\nu}$. The corresponding energy density  is
\begin{equation}\label{dbgkdf}
    T_{\mu\nu}U^{\mu}U^{\nu}=\frac{1}{2}\,\dot{\phi}^2 + \frac{1}{3}\frac{\dot{a}}{a}\,\phi\dot{\phi}+\frac{1}{6}\sum_{i=1}^3 \left(\frac{\p_i\phi}{a_i}\right)^2-\frac{1}{3}\sum_{i=1}^3 \frac{\phi\,\p_i^2\phi}{a_i^2}-\frac{1}{6}\,G_{tt}\phi^2,
\end{equation}
where an overdot indicates differentiation with respect to $t$.

It is convenient to introduce a new field variable $\chi= a^{1/3}\phi$ and a conformal time variable $\eta$ according to $\dd\eta=  a^{-1/3}{\dd t}$ with a prime indicating differentiation respect to $\eta$, and write \eqref{dbgkdf} in terms of conformal time
\begin{equation}
    T_{\mu\nu}U^{\mu}U^{\nu}= a^{-2/3}\qty[\frac{1}{2}\,a^{-2/3}\chi'^2 +\frac{1}{6}\sum_{i=1}^3\qty(\frac{\p_i\chi}{a_i})^2-\frac{1}{3}\sum_{i=1}^3\frac{\chi\,\p_i^2\chi}{a_i^2}-\frac{1}{2}a^{-2/3}Q\chi^2]\,.
\end{equation}
The positive-definite quantity $Q$ defined by 
\begin{equation}
    Q = \frac{1}{18}\sum_{i<j=1}^3\qty(\frac{a_i'}{a_i}-\frac{a_j'}{a_j})^2= \frac{1}{9}\qty(\frac{a'}{a})^2>0\,,
\end{equation}
measures the severity of anisotropic expansion. The new field variable $\chi$ satisfies the equation of motion 
\begin{equation}
    \chi'' - 2a^{2/3}\sum_{i=1}^3 \frac{\p_i^2\chi}{a_i^2}+Q\chi = 0\,.
\end{equation}
Under the periodic condition imposed at the boundaries, the corresponding field operator $\hat{\chi}$ has mode expansion of the form:
\begin{equation}
    \hat{\chi} = l^{-3/2}\sum_\mathbf{k} \qty[\hat{A}_{\vb{k}}\chi_{\vb{k}}(\eta)e^{+i\vb{k}\cdot\vb{x}}+\hat{A}_{\vb{k}}^{\dagger}\chi_{\vb{k}}^*(\eta)e^{-i\vb{k}\cdot\vb{x}}]\,,
\end{equation}
with the compact notation $$\sum_{\vb{k}}= \prod_{i=1}^{3}\sum_{n_i}$$ where $k_i=2\pi n_i/l$, $n_i\in\mathbb{Z}$, and $l$ is the coordinate length of each side of the box. The Fourier modes $\chi_{\vb{k}}$ satisfy the parametric oscillator equation:
\begin{align}
    \chi''_{\vb{k}}+[\Omega_{\vb{k}}^2(\eta)+Q(\eta)]\chi_{\vb{k}}=0\,,
    \label{xeq}
\end{align}
with $\Omega_{\vb{k}}(\eta)$ given by 
\begin{equation}
    \Omega_{\vb{k}}^2=a^{2/3}(\eta)\qty[\frac{k_x^2}{a^2(\eta)}+k_y^2+k_z^2]\,,
\end{equation}
and the normalization condition $\chi'^*_{\vb{k}}\chi^{\vphantom{*}}_{\vb{k}}-\chi_{\vb{k}}^*\chi'_{\vb{k}}=i$.

Let the initial state $\ket{0_A}$ be the vacuum state annihilated by $\hat{A}_{\vb{k}}$. Then after transforming back to the cosmic time, the vacuum expectation value of the $tt$ component of the stress-energy tensor 
\begin{equation}
    \expval{\hat{T}_{\mu\nu}}{0_A}U^{\mu}U^{\nu}= {\frac{a^{-4/3}}{2l^3}}\sum_{\vb{k}}\bqty{\vqty{\chi_{\vb{k}}'}^2+(\Omega_{\vb{k}}^2-Q)\vqty{\chi_{\vb{k}}}^2}\,,
    \label{div}
\end{equation}
gives the formal expression of the energy density. In order to obtain the explicit expressions for $\vqty{\chi_{\vb{k}}}$ and $\vqty{\chi_{\vb{k}}'}$, we solve the equation of motion \eqref{xeq} by the ansatz
\begin{align}
    \chi_{\vb{k}}&=\frac{1}{\sqrt{2\Omega_{\vb{k}}(\eta)}}\,\qty[\alpha_{\vb{k}}^{\vphantom{+}} e_{\vb{k}}^{(-)}+\beta_{\vb{k}}^{\vphantom{+}} e_{\vb{k}}^{(+)}]\,,
    &\chi'_{\vb{k}}&=-i\,\sqrt{\frac{\Omega_{\vb{k}}(\eta)}{2}}\,\qty[\alpha_{\vb{k}}^{\vphantom{+}} e_{\vb{k}}^{(-)}-\beta_{\vb{k}}^{\vphantom{+}} e_{\vb{k}}^{(+)}],
    \label{bianchi1d}
\end{align}
for two new time-dependent functions $\alpha_{k}(t)$ and $\beta_{k}(t)$, with 
\begin{equation}
    e_{\vb{k}}^{(\pm)} \equiv \exp\qty[\pm i \int_{\eta_0}^{\eta}\!\dd \eta'\;\Omega_{\vb{k}}(\eta') ]\,.
\end{equation}
The normalization condition requires $\qty|\alpha_k(\eta)|^2 - \qty|\beta_k(\eta)|^2=1$. Combining Eqs. \eqref{xeq} and \eqref{bianchi1d}, we obtain a simultaneous set of first-order differential equations for $\alpha_{\vb{k}}$ and $\beta_{\vb{k}}$,
\begin{align}
    \alpha'_{\vb{k}}&=\frac{1}{2}\qty(\frac{\Omega'_{\vb{k}}}{\Omega_{\vb{k}}}-i\,\frac{Q}{\Omega_{\vb{k}}}) (e_{\vb{k}}^{+})^2\,\beta_{\vb{k}}-i\frac{Q}{2\Omega_{\vb{k}}}\,\alpha_{\vb{k}}\,,\\
    \beta'_{\vb{k}}&=\frac{1}{2}\qty(
\frac{\Omega'_{\vb{k}}}{\Omega_{\vb{k}}}+i\,\frac{Q}{\Omega_{\vb{k}}}) (e_{\vb{k}}^{-})^2\,\alpha_{\vb{k}}+i\,\frac{Q}{2\Omega_{\vb{k}}}\,\beta_{\vb{k}}\,.
\end{align}
These are constrained complex-valued equations, so they can be re-written into three real-valued differential equations by introducing three real variables
\begin{align}
    s_{\vb{k}} &= |\beta_{\vb{k}}|^2\,,&p_{\vb{k}} &= \alpha_{\vb{k}}^{\vphantom{*}}\beta_{\vb{k}}^* e_{-}^2+\alpha_{\vb{k}}^*\beta_{\vb{k}}^{\vphantom{*}} e_{+}^2\,,&q_{\vb{k}} &=i\,(\alpha_{\vb{k}}^{\vphantom{*}}\beta_{\vb{k}}^* e_{-}^2 - \alpha_{\vb{k}}^*\beta_{\vb{k}}^{\vphantom{*}} e_{+}^2)\,.
\end{align}
Thus we obtain a simultaneous set of first-order real-valued ordinary differential equations \cite{Hu74}
\begin{align}
    s'_{\vb{k}} &= \frac{1}{2}\frac{\Omega'_{\vb{k}}}{\Omega_{\vb{k}}}p_{\vb{k}} + \frac{1}{2}\frac{Q}{\Omega_{\vb{k}}}q_{\vb{k}}\,,\\
    p'_{\vb{k}} &= \frac{\Omega'_{\vb{k}}}{\Omega_{\vb{k}}}(1+2s_{\vb{k}})-\qty(\frac{Q}{\Omega_{\vb{k}}}+2\Omega_{\vb{k}})q_{\vb{k}}\,,\\
    q'_{\vb{k}} &= \frac{Q}{\Omega_{\vb{k}}}(1+2s_{\vb{k}})+\qty(\frac{Q}{\Omega_{\vb{k}}}+2\Omega_{\vb{k}})p_{\vb{k}}\,.
\end{align}
Expressing the energy density in term of $s_{\vb{k}}$, $p_{\vb{k}}$, and $q_{\vb{k}}$, we find
\begin{align}\label{E:beirte}
    \expval{\hat{T}_{\mu\nu\;\text{reg}}}{0_A}U^{\mu}U^{\nu}&={\frac{a^{-4/3}}{2l^3}}\sum_{\vb{k}}\bqty{\Omega_{\vb{k}}(2s_{\vb{k}} + 1) -\frac{Q}{2\Omega_{\vb{k}}}(1+2s_{\vb{k}}+p_{\vb{k}})}.
\end{align}
With a suitably defined adiabatic vacuum state the parameter $s_{\vb{k}}$ can be used as a measure of particle number density at time $\eta$due to the moving boundary.

Eq.~\eqref{E:beirte} contains ultraviolet divergences which we remove by adiabatic regularization
\begin{align}
    \expval{\hat{T}_{\mu\nu\;\text{reg}}}{0_A}U^{\mu}U^{\nu}
    &={\frac{a^{-4/3}}{2l^3}}\sum_{\vb{k}}\bqty{\Omega_{\vb{k}}(2s_{\vb{k}} + 1) -\frac{Q}{2\Omega_{\vb{k}}}(1+2s_{\vb{k}}+p_{\vb{k}})}-\rho_{(0)} -\rho_{(2)} -\rho_{(4)}\,.
\end{align}
We take the zeroth- and second-order adiabatic expansion of the energy density $\rho_{(0)}$, $\rho_{(2)}$ from \cite{FulParHu74}, and subtract them from the corresponding formally divergent terms
\begin{align}\label{E:bgivta}
    \expval{\hat{T}_{\mu\nu\;\text{reg}}}{0_A}U^{\mu}U^{\nu}&={\frac{a^{-\frac{4}{3}}}{2l^3}}\sum_{\vb{k}}2s_{\vb{k}}\Omega_{\vb{k}} + \qty{{\frac{a^{-\frac{4}{3}}}{2l^3}}\sum_{\vb{k}}\Omega_{\vb{k}} -\frac{a^{-\frac{4}{3}}}{2}\int\!\frac{d^3 k}{(2\pi)^3}\;\Omega_{\vb{k}}}\notag\\
    &\qquad+\qty{{\frac{a^{-\frac{4}{3}}}{2l^3}}\sum_{\vb{k}}\qty[-\frac{Q}{2\Omega_{\vb{k}}}] -\frac{a^{-\frac{4}{3}}}{2}\int\!\frac{d^3 k}{(2\pi)^3}\;\qty[-\frac{Q}{2\Omega_{\vb{k}}}]}\\
    &\qquad+\qty{{\frac{a^{-\frac{4}{3}}}{2l^3}}\sum_{\vb{k}}\qty[-\frac{Q}{2\Omega_{\vb{k}}}(2s_{\vb{k}}+p_{\vb{k}})]}
    -{\frac{a^{-\frac{4}{3}}}{2}\int\!\frac{d^3 k}{(2\pi)^3}\;\qty[\frac{1}{8}\frac{\Omega'{}_{\vb{k}}^2}{\Omega_{\vb{k}}^3}]}-\rho_{(4)}\,.\notag
\end{align}
The first term in the first line on the righthand side is the energy density of the created particles,  the second term,  denoted by $\rho_{\textsc{ca}}$, is the Casimir energy density due to the compactness of the spatial section.

\begin{figure}
    \centering
    \includegraphics[width =1\columnwidth]{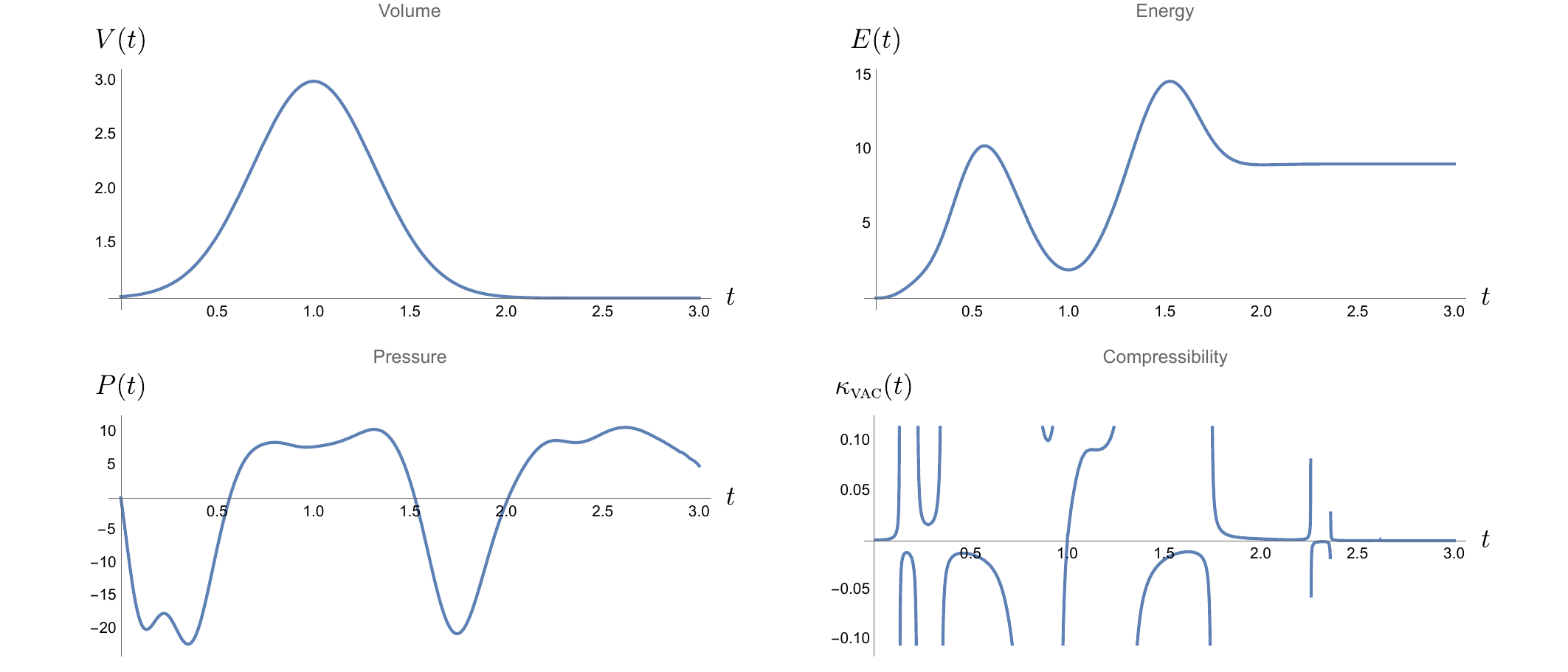}
    \caption{The volume, pressure, energy, and vacuum compressibility of a massless conformal field in a rectangular $T^3$ box when one side of the box is allowed to expand and contract in cosmic time. The scalar factor  is described by the Gaussian function, Eq.~\eqref{E:gbirutsd} with $a_0=1$, $\Delta=2$, $t_0=1$ and $1/\tau^2=5$.}
    \label{fig: expanding box}
\end{figure}

The Casimir energy in $T^3$ topology has been calculated in e.g.~\cite{MamaevTrunov79}, and is given by
\begin{equation}
    \rho_{\textsc{ca}}(\ell_1,\ell_2,\ell_3) = -\frac{\pi^2}{90\ell_1^4}-\frac{\pi^2}{6\ell_1\ell_2\ell_3^2}-\frac{\zeta_{R}(3)}{2\pi \ell_1\ell_2^3}+\frac{8\pi}{\ell_1\ell_2^2\ell_3}G\qty(\frac{\ell_3}{\ell_2})-\frac{16}{\ell_1^2\ell_2\ell_3}R\qty(\frac{\ell_2}{\ell_1},\frac{\ell_3}{\ell_1}),
\end{equation}
where $\ell_1$, $\ell_2$, $\ell_3$ are the side lengths of the rectangular box, the modified Bessel function of the second kind is given by
\begin{align}
    K_{\nu}(z) &= \frac{(z/2)^\nu \Gamma(1/2)}{\Gamma(\nu + 1/2)}\int_1^\infty\!dt\; e^{-zt} (t^2 -1 )^{(2\nu-1)/2}\,, \\
    \intertext{and}
    G(z) &= -\frac{1}{2\pi}\sum_{n=1}^{\infty}\sum_{l=1}^{\infty}\frac{n}{l}\,K_1(2\pi n l z)\,,\\
    R(z_1, z_2) &= \frac{z_1 z_2}{8}\sum_{l,p=-\infty}^{\infty}(1-\delta_{l0}\delta_{p0})\sum_{j=1}^{\infty}\qty(\frac{j}{\sqrt{l^2 z_1^2+p^2 z_2^2}})^{3/2}K_{3/2}\qty(2\pi j \sqrt{l^2 z_1^2 + p^2 z_2^2})\,.
\end{align}
For a cubic box $\rho_{\textsc{ca}}(\ell_1,\ell_1,\ell_1) = -0.838/\ell_1^4$, and for the geometry where the length of one side goes to infinity, $\rho_{\textsc{ca}}(\ell_1,\ell_1,\infty)=-0.301/\ell_1^4$. In this case, since $\rho_{\textsc{ca}}(\ell_1,\ell_1,\ell_3)$ for general $\ell_3$ is believed to be of the same order of magnitude as these two values, we expect that $\rho_{\textsc{ca}}(\ell_1,\ell_1,\ell_3)$ is a slowly varying function of $\ell_3$ such that
\begin{align}
    \rho_{\textsc{ca}}(\ell_1,\ell_1,\ell_3)&=-\frac{1}{\ell_1^4}\,f(\frac{\ell_1}{\ell_3})\,,&&\text{with}&\lim_{z\to1}f(z)&=0.838\,,&\lim_{z\to0}f(z)&=0.301\,.
\end{align}   
Thus, the dimensionless, coefficient function $f(z)$ is typically small for $0\leq z\leq1$. Then the Casimir effect contribution to the vacuum energy $E_{\textsc{ca}}$ is given  by
\begin{equation}
    E_{\textsc{ca}}=-\frac{\ell_3}{\ell_1^2}\,f(\frac{\ell_1}{\ell_3})\,,
\end{equation}
in turn, this yields the corresponding contribution to the pressure
\begin{equation}\label{E:fkdjbkfdg}
    P_{\textsc{ca}}=-\frac{1}{\ell_1^2}\frac{\partial E_{\textsc{ca}}}{\partial \ell_3}=\frac{1}{\ell_1^4}\Bigl[f(\frac{\ell_1}{\ell_3})-\frac{\ell_1}{\ell_3}\,f'(\frac{\ell_1}{\ell_3})\Bigr]\simeq\frac{1}{\ell_1^4}\,f(\frac{\ell_1}{\ell_3})\,,
\end{equation}
where the prime here represents differentiation with respect to the argument of the function. The slowly varying nature of $f(z)$ and the fact that $\ell_1/\ell_3<1$ imply that the second term in \eqref{E:fkdjbkfdg} can be ignored. Next, since the numerical values of $f(z)$ are already small, if we are interested in physically reasonable cases $\ell_3\geq \ell_1$ such that $\ell_1$ is not too small, then we can conclude the contribution of the Casimir energy to the pressure is typically negligibly small, compared with the pressure caused by particle creation, unless the latter is accidentally zero or falls under the adiabatic limit. Therefore the Casimir energy can be ignored in the calculation of the pressure. This will be further justified by Fig.~\ref{fig: expanding box} where the vacuum energy due to particle creation is numerically much larger than that from the Casimir effect.

For the term in the second line of \eqref{E:bgivta} 
\begin{align}
    -\frac{a^{-\frac{4}{3}}}{2l^3}\sum_{\vb{k}}\frac{Q}{2\Omega_{\vb{k}}}
    =-\frac{a^{-\frac{4}{3}}}{4l^3}\sum_{n_x=-\infty}^{\infty}\sum_{n_y=-\infty}^{\infty}\sum_{n_z=-\infty}^{\infty}\frac{Q}{\sqrt{a^{\frac{2}{3}}\qty[\frac{(2\pi n_x)^2}{l^2}+\frac{(2\pi n_y)^2}{l^2}+\frac{(2\pi n_z)^2}{l^2}]}}\,, \label{shear energy}
\end{align}
the zero mode in Eq.~\eqref{shear energy} causes infrared divergence, but we can exclude it since it represents the mode with infinite wavelength. The dimensions of the torus serve as an infrared cutoff of the spectrum, so the above expression can be decomposed into the following three components: (a))
\begin{align}\label{E:tgbidf}
     -\frac{a^{-\frac{4}{3}}}{2l^3}\sum_{n_x=1}^{\infty}\sum_{n_y=1}^{\infty}\sum_{n_z=1}^{\infty}\frac{Q}{\sqrt{a^{\frac{2}{3}}\qty[\frac{(2\pi n_x)^2}{l^2}+\frac{(2\pi n_y)^2}{l^2}+\frac{(2\pi n_z)^2}{l^2}]}}\,,
\end{align}
and b) 
\begin{align}\label{E:sub1tgbidf}
     -\frac{a^{-\frac{4}{3}}}{2l^3}\sum_{n_y=1}^{\infty}\sum_{n_z=1}^{\infty}\frac{Q}{\sqrt{a^{\frac{2}{3}}\qty[\frac{(2\pi n_y)^2}{l^2}+\frac{(2\pi n_z)^2}{l^2}]}}\;\Bigg|_{n_x=0}+\cdots\,,
\end{align}
where $\cdots$ are the two other terms with $n_x=0$ replaced by $n_y=0$, $n_z=0$, and (c) 
\begin{align}\label{E:sub2tgbidf}
     -\frac{a^{-\frac{4}{3}}}{2l^3}\Biggl\{\sum_{n_z=1}^{\infty}\frac{Q}{a^{\frac{1}{3}}\,\frac{2\pi n_z}{l}}\;\Bigg|_{n_x=n_y=0}+\sum_{n_y=1}^{\infty}\frac{Q}{a^{\frac{1}{3}}\,\frac{2\pi n_y}{l}}\;\Bigg|_{n_x=n_z=0}+\sum_{n_x=1}^{\infty}\frac{Q}{a^{\frac{1}{3}}\,\frac{2\pi n_x}{l}}\;\Bigg|_{n_y=n_z=0}\Biggr\}\,.
\end{align}

Eq. \eqref{E:tgbidf} can be renormalized with the help of the Abel-Plana formula~\cite{BordagBook}  
\begin{equation}
    \sum_{n=0}^\infty f(n)=\frac 1 2 f(0)+ \int_0^\infty f(x) \, dx+ i \int_0^\infty \frac{f(i t)-f(-i t)}{e^{2\pi t}-1}\,,
\end{equation}
iteratively applied to the triple sums in \eqref{E:tgbidf}. Thus, Eq.~\eqref{E:sub1tgbidf}, \eqref{E:sub2tgbidf} will be numerically renormalized  by subtract the corresponding integrals 
\begin{align}
    -\frac{a^{-\frac{4}{3}}}{2}\int_{0}^{1}\!\int_{1}^{\infty}\!\int_{1}^{\infty}\!\frac{d^3 k}{(2\pi)^3}\;\frac{Q}{\sqrt{a^{\frac{2}{3}}\qty[\frac{k_x^2}{a^2}+k_y^2+k_z^2]}} + \text{permutations of $ijk$}\,,
\end{align}
and
\begin{align}
    -\frac{a^{-\frac{4}{3}}}{2}\int_{0}^{1}\!\int_{0}^{1}\!\int_{1}^{\infty}\!\frac{d^3 k}{(2\pi)^3}\;\frac{Q}{\sqrt{a^{\frac{2}{3}}\qty[\frac{k_x^2}{a^2}+k_y^2+k_z^2]}} + \text{permutations of $ijk$}\,.
\end{align}
After renormalization, Eq.~\eqref{E:tgbidf}, \eqref{E:sub1tgbidf} and \eqref{E:sub2tgbidf} in a physically reasonable case (length scale much higher than the Planck scale) gives very small values, compared to the other contributions in the vacuum energy density, so we neglect these terms.

Now, the remaining part of energy density can be written as 
\begin{align}\label{appro ed}
    &\expval{\hat{T}_{\mu\nu\;\text{reg}}}{0_A}U^{\mu}U^{\nu}={\frac{a^{-\frac{4}{3}}}{l^3}}\sum_{\vb{k}}\bqty{s_{\vb{k}}\Omega_{\vb{k}}-\frac{Q}{4\Omega_{\vb{k}}}(2s_{\vb{k}}+p_{\vb{k}})} \\
    &\qquad-a^{-\frac{4}{3}}\int\!\frac{d^3k}{(2\pi)^3}\; \qty[ \frac{\Omega'^2_{\vb{k}}}{16\Omega_{\vb{k}}^3}-\frac{Q}{4\Omega^{\vphantom{2}}_{\vb{k}}}\qty(\frac{\Omega'^2_{\vb{k}}}{8\Omega_{\vb{k}}^4}+\frac{\Omega''_{\vb{k}}}{4\Omega^{3}_{\vb{k}}}-\frac{\Omega'_{\vb{k}}}{2\Omega_{\vb{k}}^4}-\frac{Q}{2\Omega_{\vb{k}}^2})+\Omega_{\vb{k}}^{\vphantom{2}} s_{\vb{k}}^{(4)}]\,,\notag
\end{align}
where \footnote{The notation $s_{\vb{k}}^{(4)}$ gives the fourth-adiabatic order contribution of $s_{\vb{k}}$, but it is not relevant to the current discussion.}
\begin{align}
    s_{\vb{k}}^{(4)} &= \frac{1}{16}\Biggl\{\Biggr.\frac{Q^2}{\Omega_{\vb{k}}^4}+\frac{Q'\Omega'_{\vb{k}}}{\Omega^5_{\vb{k}}}-\frac{Q}{\Omega^3_{\vb{k}}}\qty(\frac{\Omega'_{\vb{k}}}{\Omega^2_{\vb{k}}})'-\frac{3Q}{\Omega^2_{\vb{k}}}\qty(\frac{\Omega'_{\vb{k}}}{\Omega^2_{\vb{k}}})^2-\frac{1}{2\Omega^2_{\vb{k}}}\qty(\frac{\Omega'_{\vb{k}}}{\Omega^2_{\vb{k}}})''\frac{\Omega'_{\vb{k}}}{\Omega^2_{\vb{k}}}\nonumber\\
    &\qquad\qquad\qquad+\frac{1}{4\Omega^2_{\vb{k}}}\qty[\qty(\frac{\Omega'_{\vb{k}}}{\Omega^2_{\vb{k}}})']^2+\frac{1}{2\Omega^{\vphantom{2}}_{\vb{k}}}\qty(\frac{\Omega'_{\vb{k}}}{\Omega^2_{\vb{k}}})'\qty(\frac{\Omega'_{\vb{k}}}{\Omega^2_{\vb{k}}})^2+\frac{3}{16}\qty(\frac{\Omega'_{\vb{k}}}{\Omega^2_{\vb{k}}})^4\Biggl.\Biggr\}\,.
\end{align}
The integral in the second line of Eq.~\eqref{appro ed} associate with trace anomaly is divergent, and it must cancel the UV divergence in $s_{\vb{k}}$ and $p_{\vb{k}}$ of the first line. Although it is hard to further decompose $s_{\vb{k}}$ and $p_{\vb{k}}$ to find the exact divergent contribution, we could make the following observations: At large momentum, the integrand in the second line of Eq.~\eqref{appro ed} will be suppressed by $\Omega_{\vb{k}}$ in the denominators, and it goes like $1/k$ as $k\to\infty$, then it will cancel the divergent part in $s_{\vb{k}}$ and $p_{\vb{k}}$ for large momentum, and give a finite part. Therefore, if we can estimate the value of the integral in the small momentum range, and show it is small compared to the summation in the small momentum range in the first line of Eq. \eqref{appro ed}, then we can neglect the whole thing in the second line of Eq. \eqref{appro ed}. For the scale factor $a(t)$ of our interest, the integral indeed is a few orders of magnitude smaller than the rest of energy density expressions for $k_i < 20$. Therefore we can just use the first line of Eq.~\eqref{appro ed} and sum up to $k_i < 20$ to give an pretty accurate approximation of energy. The corresponding plot is shown in Fig.~\ref{fig: expanding box}. 

In the current case, the major component of the vacuum energy come from the contribution caused by spontaneous particle production induced by a mirror which moves in a specific direction and has a trajectory described by a Gaussian, or equivalently induced by anisotropic variation of the scale factor in $\mathbf{R}\times T^3$ spacetime. The result in Fig.~\ref{fig: expanding box} shows the energy does not monotonically grows with time even when the volume is expanding. A local maximum does occurs at the moment when the expansion rate is largest. 
% \textcolor{red}{In addition in the asymptotic future when the volume contracts to the initial size, the energy seems to settle down a value that is greater than its initial value. It is not yet clear whether ambiguous notion of particle numbers is the source of the non-monotonic behavior of the energy associated with particle creation?}

The accompanying time variation of the pressure in Fig.~\ref{fig: expanding box} contains more structure due to the expansion and contraction of the volume. The results show that the pressure can takes on either positive or negative values. The negative value occurs when the energy changes in the same trend as the volume does, but if the energy changes in a way opposite to how the volume changes, then the pressure can be positive. These can be seen by comparing the plots of energy and pressure, as the consequence of the definition of the pressure; however, it is not clear why a small hump shows up at early time in the figure for the time evolution of the pressure.  The zero pressure occurs when the energy take on a local maximum. From the previous discussions, we expect the compressibility can exhibit unusual behavior in the regions where $\Delta V\neq0$ but $\Delta P\approx0$. This is rather baffling indeed and needs more investigations.
% This is rather baffling indeed, and this needs more investigations, in particular, on why the energy presents such non-monotonic behavior.

\section{Summary}\label{S:reiuss}
We summarize our findings by first describing the quantum field theoretical factors and then the quantum thermodynamic features.

\subsection{Quantum Processes}
The quantum vacuum processes of interest to us include  the familiar Casimir effect, the dynamical Casimir effect which includes particle creation, and  the less familiar trace anomaly. The first depends on the way boundary conditions are imposed on the quantum field and  the topology of space it lives in. The second entails particle creation due to the parametric amplification of vacuum fluctuations of the quantum field. The third is specific to, and a manifestation of, the quantum nature of a conformal field -- `anomaly' refers to the fact that the trace of the stress energy tensor of a classical conformal field is zero, but not so for a quantum conformal field.  These three types of quantum field processes/effects have been studied by researchers in quantum field theory in curved spacetime since the 1970s. We have adopted the methodologies developed there, such as using the adiabatic regularization for the treatment of UV divergences. Here we have more extensive mathematical usages and accompanying physical discussions. For particle creation in $\mathbf{R}\times S^2$, $\mathbf{R}\times S^3$, and $\mathbf{R}\times T^3$ treated here we solve the differential equations governing the Bogoliubov coefficients without making any approximation. One new improvement over earlier treatments is to obtain the Bogoliubov coefficients (two complex or four real quantities minus one Wronskian condition) directly by solving the three first-order ODE numerically with a specified time-dependence of the scale factor. 

Going into the specifics: We have treated the effect of trace anomaly in $\mathbf{R}\times S^1$,  $\mathbf{R}\times S^3$ spacetimes and particle creation in $\mathbf{R}\times S^2$, $\mathbf{R}\times S^3$ and $\mathbf{R}\times T^3$ spacetimes. In the first case the time dependence of the scale factor is determined by self-consistently solving for the dynamics of the quantum field and the spacetime. This was done in a recent paper \cite{XBH}.  Results from there are imported into the present calculations. We use this as an example of vacuum processes where the backreaction effects are included in the consideration. In the other cases we have considered two types of evolution of the scale factor. The marked differences between the effects of the trace anomaly and particle creation have been known much earlier (compare e.g., \cite{FHH79,HarHu79}), and amply illustrated in \cite{XBH}: the backreaction on $S^1$ due to the trace anomaly accelerates the collapse of the ring while the backreaction of particle creation in $T^3$ retards the expansion of the box. 
For the other cases studied without backreaction, one distinction between the effects of particle creation and the trace anomaly is that the energy due to (bosonic) particle creation overall monotonically increases, while that due to trace anomaly can increase at one time and decrease at another time. This is because the expression for the trace anomaly contains terms of higher derivative orders with positive and negative signs. As remarked in the Introduction we need to include the energy and the pressure for all possible vacuum processes allowed before calculating the vacuum dynamical compressibility of each spacetime.  Detailed descriptions of how the vacuum energy, the vacuum pressure and the vacuum dynamical compressibility varies with time can be found in the plots given for each of the spacetimes considered.  Their overall features have a branding impact on the quantum thermodynamic functions of spacetimes.

\subsection{Quantum Thermodynamics}
The ending target from these QFTCST calculations is the energy density of the quantum field in various quantum vacuum processes. It is also the starting point for the derivation of the quantum thermodynamic functions of the system.  This can be seen from the four frames in the plots based on the results of calculations in all the cases considered: the volume, energy, pressure and compressibility. The first two from QFT calculations provide the raw materials for ensuing thermodynamic considerations. If the challenge in the QFT part is mainly technical, the QTD part is mainly conceptual. We refer the reader to the descriptions of the results following each figure for details, but  highlight two overall qualitative features below which are radically different from our traditional understanding of energy, pressure and compressibility based on classical thermodynamics:  

1) {\it Energy could be negative}: the energy due to the trace anomaly in $S^1$ and due to particle creation in $T^3$ are negative, whereas the energy due to particle creation in $S^2$ and $S^3$ spaces are positive, as is in $S^3$ due to the trace anomaly. Negative energy is a special hallmark of vacuum quantum field processes, probably known first in Casimir effect: the energy density in the space between two parallel plates is negative, resulting in an attractive Casimir force  whereas for a sphere it is positive, giving rise to an expansion of the sphere.    

2) {\it Pressure could be negative}.  Recall the definition:  $P = -\frac{\partial E}{\partial V}$.   If energy is negative as, e.g., in the $S^1$ case (see Fig.~\ref{fig:1+1 gau} \&~\ref{fig:1+1 gauComp}), then a) an increase in the volume with an increase in the (negative) energy will give a negative pressure,  while b) a decrease in the  volume with an increase of (negative) energy will give a positive pressure.  Note that the pressure calculated from the vacuum energy of a particular source, be it Casimir, trace anomaly or particle creation, is only a partial pressure. That is the reason for our earlier caution, that all partial pressures need be added up to a total vacuum pressure before one proceeds to calculate the compressibility.

3) {\it Compressibility could be negative, even approach infinity at certain times}.  Recall the definition $\kappa=-\frac{1}{V}\frac{\partial V}{\partial P}$. First off, note as stated before, that the pressure in `taking the derivative of volume with respect to the pressure' refers to the total pressure of all the partial pressures from different vacuum sources, like the three named above, whichever is present in the different cases studied. i)  In the $T^3$ example mentioned above, because of the cross-dependence both cases a) and b) end up giving the same compressibility functional dependence on time, becoming increasingly negative. ii)  A more `alarming' fact is that the {\it compressibility can  become infinite} (see Fig.~\ref{fig:exp2+1} \&~\ref{fig:2+1}) at certain points in time when the vacuum pressure is a maximum or minimum while the volume is changing.   

What we have encountered here is something very odd from the  traditional perspective of classical mechanics and thermodynamics. This is because the sources of vacuum energy in  the cases we have studied are all of quantum field origins. Our calculations are meant to bring forth something novel in the quantum thermodynamics of spacetimes, fundamentally different from spacetime dynamics described by classical thermodynamics. It behooves us to try to understand these processes from a different perspective, to expound / extend the meanings of these physical  quantities or,  even go so far as coming up with more cogent definitions for these quantum thermodynamic quantities to encompass these new phenomena.  

As extensions of the present work we shall study finite temperature quantum fields in our next paper \cite{XHH2}. We will be able to include the effects of stimulated particle creation and the considerations of heat capacity of dynamical spacetimes due to various quantum field processes.

%----------------------------
\noindent{\bf Acknowledgment} YCX thanks Bo-Hao Chng and Yushan Su for help with the Riemann sum in $T^3$, JTH and BLH thank Prof. H.-T. Cho for helpful discussions.  J.-T. Hsiang is supported by the National Science and Technology Council of Taiwan, R.O.C. under Grant No.~NSTC 112-2112-M-011-001-MY3. B.-L. Hu enjoyed the warm hospitality of Prof. Hsiang-nan Li and colleagues at the Institute of Physics, Academia Sinica, Taiwan, R.O.C. when this work began.

% \newpage

% \appendix
% \section{\texorpdfstring{Geometric Quantity in $\mathbb{R} \times S^2$}{}}
% All components are computed in the coordinate basis in which the metric was specified.

% \begin{eqnarray*}
%     \Gamma^{0}_{00} &=& \frac{a'}{a},~~~~~~~~~~~~~~~~~~\Gamma^{0}_{11} = \frac{a'}{a},\\
%     \Gamma^{0}_{22} &=& \frac{\sin^2\theta a'}{a},~~~~~~~~~~~\Gamma^{1}_{10} = \frac{a'}{a},\\
%     \Gamma^{1}_{22} &=& -\cos\theta\sin\theta ,~~~~~~\Gamma^{2}_{20} = \frac{a'}{a},\\
%     \Gamma^{2}_{21} &=& \cot\theta.
% \end{eqnarray*}

% \begin{eqnarray*}
%     R_{00} &=& \frac{2(a'^2-aa'')}{a^2}\\
%     R_{11} &=& 1+\frac{a''}{a}\\
%     R_{22} &=& \frac{(a+a'')\sin^2\theta}{a}\\
%     R &=& -\frac{2(a^2-a'^2+2aa'')}{a^4} 
% \end{eqnarray*}

% \newpage

\nocite{*}
\bibliography{refs}

\providecommand{\noopsort}[1]{}\providecommand{\singleletter}[1]{#1}%
\begin{thebibliography}{77}%
\makeatletter
\providecommand \@ifxundefined [1]{%
 \@ifx{#1\undefined}
}%
\providecommand \@ifnum [1]{%
 \ifnum #1\expandafter \@firstoftwo
 \else \expandafter \@secondoftwo
 \fi
}%
\providecommand \@ifx [1]{%
 \ifx #1\expandafter \@firstoftwo
 \else \expandafter \@secondoftwo
 \fi
}%
\providecommand \natexlab [1]{#1}%
\providecommand \enquote  [1]{``#1''}%
\providecommand \bibnamefont  [1]{#1}%
\providecommand \bibfnamefont [1]{#1}%
\providecommand \citenamefont [1]{#1}%
\providecommand \href@noop [0]{\@secondoftwo}%
\providecommand \href [0]{\begingroup \@sanitize@url \@href}%
\providecommand \@href[1]{\@@startlink{#1}\@@href}%
\providecommand \@@href[1]{\endgroup#1\@@endlink}%
\providecommand \@sanitize@url [0]{\catcode `\\12\catcode `\$12\catcode `\&12\catcode `\#12\catcode `\^12\catcode `\_12\catcode `\%12\relax}%
\providecommand \@@startlink[1]{}%
\providecommand \@@endlink[0]{}%
\providecommand \url  [0]{\begingroup\@sanitize@url \@url }%
\providecommand \@url [1]{\endgroup\@href {#1}{\urlprefix }}%
\providecommand \urlprefix  [0]{URL }%
\providecommand \Eprint [0]{\href }%
\providecommand \doibase [0]{http://dx.doi.org/}%
\providecommand \selectlanguage [0]{\@gobble}%
\providecommand \bibinfo  [0]{\@secondoftwo}%
\providecommand \bibfield  [0]{\@secondoftwo}%
\providecommand \translation [1]{[#1]}%
\providecommand \BibitemOpen [0]{}%
\providecommand \bibitemStop [0]{}%
\providecommand \bibitemNoStop [0]{.\EOS\space}%
\providecommand \EOS [0]{\spacefactor3000\relax}%
\providecommand \BibitemShut  [1]{\csname bibitem#1\endcsname}%
\let\auto@bib@innerbib\@empty
%</preamble>
\bibitem [{\citenamefont {Cho}\ \emph {et~al.}(2022)\citenamefont {Cho}, \citenamefont {Hsiang},\ and\ \citenamefont {Hu}}]{CHH1}%
  \BibitemOpen
  \bibfield  {author} {\bibinfo {author} {\bibfnamefont {H.-T.}\ \bibnamefont {Cho}}, \bibinfo {author} {\bibfnamefont {J.-T.}\ \bibnamefont {Hsiang}}, \ and\ \bibinfo {author} {\bibfnamefont {B.~L.}\ \bibnamefont {Hu}},\ }\href {\doibase 10.3390/universe8050291} {\bibfield  {journal} {\bibinfo  {journal} {Universe}\ }\textbf {\bibinfo {volume} {8}},\ \bibinfo {pages} {291} (\bibinfo {year} {2022})},\ \Eprint {http://arxiv.org/abs/2204.08634} {arXiv:2204.08634 [gr-qc]} \BibitemShut {NoStop}%
\bibitem [{\citenamefont {Zeldovich}(1970)}]{Zel70}%
  \BibitemOpen
  \bibfield  {author} {\bibinfo {author} {\bibfnamefont {Y.~B.}\ \bibnamefont {Zeldovich}},\ }\href@noop {} {\bibfield  {journal} {\bibinfo  {journal} {Pisma Zh. Eksp. Teor. Fiz.}\ }\textbf {\bibinfo {volume} {12}},\ \bibinfo {pages} {443} (\bibinfo {year} {1970})}\BibitemShut {NoStop}%
\bibitem [{\citenamefont {Jou}\ \emph {et~al.}(2010)\citenamefont {Jou}, \citenamefont {Casas-V{\'a}zquez},\ and\ \citenamefont {Lebon}}]{IrrTD}%
  \BibitemOpen
  \bibfield  {author} {\bibinfo {author} {\bibfnamefont {D.}~\bibnamefont {Jou}}, \bibinfo {author} {\bibfnamefont {J.}~\bibnamefont {Casas-V{\'a}zquez}}, \ and\ \bibinfo {author} {\bibfnamefont {G.}~\bibnamefont {Lebon}},\ }\href@noop {} {\bibfield  {journal} {\bibinfo  {journal} {Extended Irreversible Thermodynamics}\ ,\ \bibinfo {pages} {41}} (\bibinfo {year} {2010})}\BibitemShut {NoStop}%
\bibitem [{\citenamefont {Birrell}\ and\ \citenamefont {Davies}(1984)}]{BirDav}%
  \BibitemOpen
  \bibfield  {author} {\bibinfo {author} {\bibfnamefont {N.~D.}\ \bibnamefont {Birrell}}\ and\ \bibinfo {author} {\bibfnamefont {P.~C.~W.}\ \bibnamefont {Davies}},\ }\href {\doibase 10.1017/CBO9780511622632} {\emph {\bibinfo {title} {{Quantum Fields in Curved Space}}}},\ Cambridge Monographs on Mathematical Physics\ (\bibinfo  {publisher} {Cambridge Univ. Press},\ \bibinfo {address} {Cambridge, UK},\ \bibinfo {year} {1984})\BibitemShut {NoStop}%
\bibitem [{\citenamefont {Parker}\ and\ \citenamefont {Toms}(2009)}]{ParTom}%
  \BibitemOpen
  \bibfield  {author} {\bibinfo {author} {\bibfnamefont {L.}~\bibnamefont {Parker}}\ and\ \bibinfo {author} {\bibfnamefont {D.}~\bibnamefont {Toms}},\ }\href@noop {} {\emph {\bibinfo {title} {Quantum field theory in curved spacetime: quantized fields and gravity}}}\ (\bibinfo  {publisher} {Cambridge university press},\ \bibinfo {year} {2009})\BibitemShut {NoStop}%
\bibitem [{\citenamefont {Hawking}(1975)}]{Haw75}%
  \BibitemOpen
  \bibfield  {author} {\bibinfo {author} {\bibfnamefont {S.~W.}\ \bibnamefont {Hawking}},\ }\href@noop {} {\bibfield  {journal} {\bibinfo  {journal} {Communications in mathematical physics}\ }\textbf {\bibinfo {volume} {43}},\ \bibinfo {pages} {199} (\bibinfo {year} {1975})}\BibitemShut {NoStop}%
\bibitem [{\citenamefont {Bekenstein}(1972)}]{Bek72}%
  \BibitemOpen
  \bibfield  {author} {\bibinfo {author} {\bibfnamefont {J.~D.}\ \bibnamefont {Bekenstein}},\ }\href@noop {} {\bibfield  {journal} {\bibinfo  {journal} {Physical Review D}\ }\textbf {\bibinfo {volume} {5}},\ \bibinfo {pages} {2403} (\bibinfo {year} {1972})}\BibitemShut {NoStop}%
\bibitem [{\citenamefont {Bekenstein}(1973)}]{Bek73}%
  \BibitemOpen
  \bibfield  {author} {\bibinfo {author} {\bibfnamefont {J.~D.}\ \bibnamefont {Bekenstein}},\ }\href@noop {} {\bibfield  {journal} {\bibinfo  {journal} {Physical Review D}\ }\textbf {\bibinfo {volume} {7}},\ \bibinfo {pages} {2333} (\bibinfo {year} {1973})}\BibitemShut {NoStop}%
\bibitem [{\citenamefont {Hawking}(1976)}]{Haw76}%
  \BibitemOpen
  \bibfield  {author} {\bibinfo {author} {\bibfnamefont {S.~W.}\ \bibnamefont {Hawking}},\ }\href@noop {} {\bibfield  {journal} {\bibinfo  {journal} {Physical Review D}\ }\textbf {\bibinfo {volume} {13}},\ \bibinfo {pages} {191} (\bibinfo {year} {1976})}\BibitemShut {NoStop}%
\bibitem [{\citenamefont {Wald}(1994)}]{Wald}%
  \BibitemOpen
  \bibfield  {author} {\bibinfo {author} {\bibfnamefont {R.~M.}\ \bibnamefont {Wald}},\ }\href@noop {} {\emph {\bibinfo {title} {Quantum field theory in curved spacetime and black hole thermodynamics}}}\ (\bibinfo  {publisher} {University of Chicago press},\ \bibinfo {year} {1994})\BibitemShut {NoStop}%
\bibitem [{\citenamefont {Jacobson}(1995)}]{STDJac1}%
  \BibitemOpen
  \bibfield  {author} {\bibinfo {author} {\bibfnamefont {T.}~\bibnamefont {Jacobson}},\ }\href@noop {} {\bibfield  {journal} {\bibinfo  {journal} {Physical Review Letters}\ }\textbf {\bibinfo {volume} {75}},\ \bibinfo {pages} {1260} (\bibinfo {year} {1995})}\BibitemShut {NoStop}%
\bibitem [{\citenamefont {Eling}\ \emph {et~al.}(2006)\citenamefont {Eling}, \citenamefont {Guedens},\ and\ \citenamefont {Jacobson}}]{STDJac2}%
  \BibitemOpen
  \bibfield  {author} {\bibinfo {author} {\bibfnamefont {C.}~\bibnamefont {Eling}}, \bibinfo {author} {\bibfnamefont {R.}~\bibnamefont {Guedens}}, \ and\ \bibinfo {author} {\bibfnamefont {T.}~\bibnamefont {Jacobson}},\ }\href@noop {} {\bibfield  {journal} {\bibinfo  {journal} {Physical Review Letters}\ }\textbf {\bibinfo {volume} {96}},\ \bibinfo {pages} {121301} (\bibinfo {year} {2006})}\BibitemShut {NoStop}%
\bibitem [{\citenamefont {Padmanabhan}(2005)}]{STDPad}%
  \BibitemOpen
  \bibfield  {author} {\bibinfo {author} {\bibfnamefont {T.}~\bibnamefont {Padmanabhan}},\ }\href@noop {} {\bibfield  {journal} {\bibinfo  {journal} {Physics Reports}\ }\textbf {\bibinfo {volume} {406}},\ \bibinfo {pages} {49} (\bibinfo {year} {2005})}\BibitemShut {NoStop}%
\bibitem [{\citenamefont {Chirco}\ and\ \citenamefont {Liberati}(2010)}]{STDLib}%
  \BibitemOpen
  \bibfield  {author} {\bibinfo {author} {\bibfnamefont {G.}~\bibnamefont {Chirco}}\ and\ \bibinfo {author} {\bibfnamefont {S.}~\bibnamefont {Liberati}},\ }\href@noop {} {\bibfield  {journal} {\bibinfo  {journal} {Physical Review D}\ }\textbf {\bibinfo {volume} {81}},\ \bibinfo {pages} {024016} (\bibinfo {year} {2010})}\BibitemShut {NoStop}%
\bibitem [{\citenamefont {Padmanabhan}(2010)}]{PaddyRPP}%
  \BibitemOpen
  \bibfield  {author} {\bibinfo {author} {\bibfnamefont {T.}~\bibnamefont {Padmanabhan}},\ }\href@noop {} {\bibfield  {journal} {\bibinfo  {journal} {Reports on Progress in Physics}\ }\textbf {\bibinfo {volume} {73}},\ \bibinfo {pages} {046901} (\bibinfo {year} {2010})}\BibitemShut {NoStop}%
\bibitem [{\citenamefont {Verlinde}(2011)}]{Verlinde}%
  \BibitemOpen
  \bibfield  {author} {\bibinfo {author} {\bibfnamefont {E.}~\bibnamefont {Verlinde}},\ }\href@noop {} {\bibfield  {journal} {\bibinfo  {journal} {Journal of High Energy Physics}\ }\textbf {\bibinfo {volume} {2011}},\ \bibinfo {pages} {1} (\bibinfo {year} {2011})}\BibitemShut {NoStop}%
\bibitem [{\citenamefont {Hu}(2011)}]{HuGravNEqTh}%
  \BibitemOpen
  \bibfield  {author} {\bibinfo {author} {\bibfnamefont {B.~L.}\ \bibnamefont {Hu}},\ }\href@noop {} {\bibfield  {journal} {\bibinfo  {journal} {International Journal of Modern Physics D}\ }\textbf {\bibinfo {volume} {20}},\ \bibinfo {pages} {697} (\bibinfo {year} {2011})}\BibitemShut {NoStop}%
\bibitem [{\citenamefont {Hu}(1996)}]{HuGRhydro}%
  \BibitemOpen
  \bibfield  {author} {\bibinfo {author} {\bibfnamefont {B.~L.}\ \bibnamefont {Hu}},\ }\href@noop {} {\bibfield  {journal} {\bibinfo  {journal} {arXiv preprint gr-qc/9607070}\ } (\bibinfo {year} {1996})}\BibitemShut {NoStop}%
\bibitem [{\citenamefont {Volovik}(2003)}]{VolHe3}%
  \BibitemOpen
  \bibfield  {author} {\bibinfo {author} {\bibfnamefont {G.~E.}\ \bibnamefont {Volovik}},\ }\href@noop {} {\emph {\bibinfo {title} {The universe in a helium droplet}}},\ Vol.\ \bibinfo {volume} {117}\ (\bibinfo  {publisher} {OUP Oxford},\ \bibinfo {year} {2003})\BibitemShut {NoStop}%
\bibitem [{\citenamefont {Hu}(2005)}]{HuSTcond}%
  \BibitemOpen
  \bibfield  {author} {\bibinfo {author} {\bibfnamefont {B.-L.}\ \bibnamefont {Hu}},\ }\href@noop {} {\bibfield  {journal} {\bibinfo  {journal} {International journal of theoretical physics}\ }\textbf {\bibinfo {volume} {44}},\ \bibinfo {pages} {1785} (\bibinfo {year} {2005})}\BibitemShut {NoStop}%
\bibitem [{\citenamefont {Gielen}\ \emph {et~al.}(2016)\citenamefont {Gielen}, \citenamefont {Sindoni} \emph {et~al.}}]{GieSinCond}%
  \BibitemOpen
  \bibfield  {author} {\bibinfo {author} {\bibfnamefont {S.}~\bibnamefont {Gielen}}, \bibinfo {author} {\bibfnamefont {L.}~\bibnamefont {Sindoni}},  \emph {et~al.},\ }\href@noop {} {\bibfield  {journal} {\bibinfo  {journal} {SIGMA. Symmetry, Integrability and Geometry: Methods and Applications}\ }\textbf {\bibinfo {volume} {12}},\ \bibinfo {pages} {082} (\bibinfo {year} {2016})}\BibitemShut {NoStop}%
\bibitem [{\citenamefont {Oriti}(2021)}]{OritiCond}%
  \BibitemOpen
  \bibfield  {author} {\bibinfo {author} {\bibfnamefont {D.}~\bibnamefont {Oriti}}\ }(\bibinfo {year} {2021})\ \Eprint {http://arxiv.org/abs/2112.02585} {arXiv:2112.02585 [gr-qc]} \BibitemShut {NoStop}%
\bibitem [{\citenamefont {Gu}\ and\ \citenamefont {Wen}(2006)}]{GuWen06}%
  \BibitemOpen
  \bibfield  {author} {\bibinfo {author} {\bibfnamefont {Z.-C.}\ \bibnamefont {Gu}}\ and\ \bibinfo {author} {\bibfnamefont {X.-G.}\ \bibnamefont {Wen}},\ }\href@noop {} {\bibfield  {journal} {\bibinfo  {journal} {arXiv preprint gr-qc/0606100}\ } (\bibinfo {year} {2006})}\BibitemShut {NoStop}%
\bibitem [{\citenamefont {Sindoni}(2012)}]{SinEmG}%
  \BibitemOpen
  \bibfield  {author} {\bibinfo {author} {\bibfnamefont {L.}~\bibnamefont {Sindoni}},\ }\href {\doibase 10.3842/SIGMA.2012.027} {\bibfield  {journal} {\bibinfo  {journal} {SIGMA}\ }\textbf {\bibinfo {volume} {8}},\ \bibinfo {pages} {027} (\bibinfo {year} {2012})},\ \Eprint {http://arxiv.org/abs/1110.0686} {arXiv:1110.0686 [gr-qc]} \BibitemShut {NoStop}%
\bibitem [{\citenamefont {Oriti}(2018)}]{OritiQMB}%
  \BibitemOpen
  \bibfield  {author} {\bibinfo {author} {\bibfnamefont {D.}~\bibnamefont {Oriti}},\ }\href@noop {} {\bibfield  {journal} {\bibinfo  {journal} {Many-body Approaches at Different Scales: A Tribute to Norman H. March on the Occasion of his 90th Birthday}\ ,\ \bibinfo {pages} {365}} (\bibinfo {year} {2018})}\BibitemShut {NoStop}%
\bibitem [{\citenamefont {Volovik}(2023)}]{VolCMP}%
  \BibitemOpen
  \bibfield  {author} {\bibinfo {author} {\bibfnamefont {G.~E.}\ \bibnamefont {Volovik}},\ }\href {\doibase 10.31857/S1234567823190126} {\bibfield  {journal} {\bibinfo  {journal} {Pisma Zh. Eksp. Teor. Fiz.}\ }\textbf {\bibinfo {volume} {118}},\ \bibinfo {pages} {546} (\bibinfo {year} {2023})},\ \Eprint {http://arxiv.org/abs/2307.14370} {arXiv:2307.14370 [cond-mat.other]} \BibitemShut {NoStop}%
\bibitem [{\citenamefont {Hu}(2009)}]{HuE/QG}%
  \BibitemOpen
  \bibfield  {author} {\bibinfo {author} {\bibfnamefont {B.~L.}\ \bibnamefont {Hu}},\ }in\ \href@noop {} {\emph {\bibinfo {booktitle} {Journal of Physics: Conference Series}}},\ Vol.\ \bibinfo {volume} {174}\ (\bibinfo {organization} {IOP Publishing},\ \bibinfo {year} {2009})\ p.\ \bibinfo {pages} {012015}\BibitemShut {NoStop}%
\bibitem [{\citenamefont {Carlip}(2014)}]{Carlip}%
  \BibitemOpen
  \bibfield  {author} {\bibinfo {author} {\bibfnamefont {S.}~\bibnamefont {Carlip}},\ }\href {\doibase https://doi.org/10.1016/j.shpsb.2012.11.002} {\bibfield  {journal} {\bibinfo  {journal} {Studies in History and Philosophy of Science Part B: Studies in History and Philosophy of Modern Physics}\ }\textbf {\bibinfo {volume} {46}},\ \bibinfo {pages} {200} (\bibinfo {year} {2014})}\BibitemShut {NoStop}%
\bibitem [{\citenamefont {Marolf}(2015)}]{Marolf}%
  \BibitemOpen
  \bibfield  {author} {\bibinfo {author} {\bibfnamefont {D.}~\bibnamefont {Marolf}},\ }\href {\doibase 10.1103/PhysRevLett.114.031104} {\bibfield  {journal} {\bibinfo  {journal} {Phys. Rev. Lett.}\ }\textbf {\bibinfo {volume} {114}},\ \bibinfo {pages} {031104} (\bibinfo {year} {2015})}\BibitemShut {NoStop}%
\bibitem [{\citenamefont {Barceló}\ \emph {et~al.}(2021)\citenamefont {Barceló}, \citenamefont {Carballo-Rubio}, \citenamefont {Garay},\ and\ \citenamefont {García-Moreno}}]{BarGaray}%
  \BibitemOpen
  \bibfield  {author} {\bibinfo {author} {\bibfnamefont {C.}~\bibnamefont {Barceló}}, \bibinfo {author} {\bibfnamefont {R.}~\bibnamefont {Carballo-Rubio}}, \bibinfo {author} {\bibfnamefont {L.~J.}\ \bibnamefont {Garay}}, \ and\ \bibinfo {author} {\bibfnamefont {G.}~\bibnamefont {García-Moreno}},\ }\href {\doibase 10.3390/app11188763} {\bibfield  {journal} {\bibinfo  {journal} {Applied Sciences}\ }\textbf {\bibinfo {volume} {11}} (\bibinfo {year} {2021}),\ 10.3390/app11188763}\BibitemShut {NoStop}%
\bibitem [{\citenamefont {Hu}(1982)}]{HuVacVis}%
  \BibitemOpen
  \bibfield  {author} {\bibinfo {author} {\bibfnamefont {B.~L.}\ \bibnamefont {Hu}},\ }\href@noop {} {\bibfield  {journal} {\bibinfo  {journal} {Physics Letters A}\ }\textbf {\bibinfo {volume} {90}},\ \bibinfo {pages} {375} (\bibinfo {year} {1982})}\BibitemShut {NoStop}%
\bibitem [{\citenamefont {Hu}\ and\ \citenamefont {Verdaguer}(2020)}]{HuVer}%
  \BibitemOpen
  \bibfield  {author} {\bibinfo {author} {\bibfnamefont {B.-L.~B.}\ \bibnamefont {Hu}}\ and\ \bibinfo {author} {\bibfnamefont {E.}~\bibnamefont {Verdaguer}},\ }\href@noop {} {\emph {\bibinfo {title} {Semiclassical and Stochastic Gravity: Quantum Field Effects on Curved Spacetime}}}\ (\bibinfo  {publisher} {Cambridge University Press},\ \bibinfo {year} {2020})\BibitemShut {NoStop}%
\bibitem [{\citenamefont {Gr{\o}n}(1990)}]{Gron}%
  \BibitemOpen
  \bibfield  {author} {\bibinfo {author} {\bibfnamefont {{\O}.}~\bibnamefont {Gr{\o}n}},\ }\href@noop {} {\bibfield  {journal} {\bibinfo  {journal} {Astrophysics and Space Science}\ }\textbf {\bibinfo {volume} {173}},\ \bibinfo {pages} {191} (\bibinfo {year} {1990})}\BibitemShut {NoStop}%
\bibitem [{\citenamefont {Mak}\ and\ \citenamefont {Harko}(1999)}]{MakHar}%
  \BibitemOpen
  \bibfield  {author} {\bibinfo {author} {\bibfnamefont {M.}~\bibnamefont {Mak}}\ and\ \bibinfo {author} {\bibfnamefont {T.}~\bibnamefont {Harko}},\ }\href@noop {} {\bibfield  {journal} {\bibinfo  {journal} {Australian journal of physics}\ }\textbf {\bibinfo {volume} {52}},\ \bibinfo {pages} {659} (\bibinfo {year} {1999})}\BibitemShut {NoStop}%
\bibitem [{\citenamefont {Zimdahl}(2000)}]{Zimdahl}%
  \BibitemOpen
  \bibfield  {author} {\bibinfo {author} {\bibfnamefont {W.}~\bibnamefont {Zimdahl}},\ }\href@noop {} {\bibfield  {journal} {\bibinfo  {journal} {Physical Review D}\ }\textbf {\bibinfo {volume} {61}},\ \bibinfo {pages} {083511} (\bibinfo {year} {2000})}\BibitemShut {NoStop}%
\bibitem [{\citenamefont {Zimdahl}\ \emph {et~al.}(2001)\citenamefont {Zimdahl}, \citenamefont {Schwarz}, \citenamefont {Balakin},\ and\ \citenamefont {Pavon}}]{ZimPav}%
  \BibitemOpen
  \bibfield  {author} {\bibinfo {author} {\bibfnamefont {W.}~\bibnamefont {Zimdahl}}, \bibinfo {author} {\bibfnamefont {D.~J.}\ \bibnamefont {Schwarz}}, \bibinfo {author} {\bibfnamefont {A.~B.}\ \bibnamefont {Balakin}}, \ and\ \bibinfo {author} {\bibfnamefont {D.}~\bibnamefont {Pavon}},\ }\href@noop {} {\bibfield  {journal} {\bibinfo  {journal} {Physical Review D}\ }\textbf {\bibinfo {volume} {64}},\ \bibinfo {pages} {063501} (\bibinfo {year} {2001})}\BibitemShut {NoStop}%
\bibitem [{\citenamefont {Singh}\ and\ \citenamefont {Kale}(2011)}]{SinKal}%
  \BibitemOpen
  \bibfield  {author} {\bibinfo {author} {\bibfnamefont {G.}~\bibnamefont {Singh}}\ and\ \bibinfo {author} {\bibfnamefont {A.}~\bibnamefont {Kale}},\ }\href@noop {} {\bibfield  {journal} {\bibinfo  {journal} {Astrophysics and Space Science}\ }\textbf {\bibinfo {volume} {331}},\ \bibinfo {pages} {207} (\bibinfo {year} {2011})}\BibitemShut {NoStop}%
\bibitem [{\citenamefont {Bastero-Gil}\ \emph {et~al.}(2012)\citenamefont {Bastero-Gil}, \citenamefont {Berera}, \citenamefont {Cerezo}, \citenamefont {Ramos},\ and\ \citenamefont {Vicente}}]{WarmInf}%
  \BibitemOpen
  \bibfield  {author} {\bibinfo {author} {\bibfnamefont {M.}~\bibnamefont {Bastero-Gil}}, \bibinfo {author} {\bibfnamefont {A.}~\bibnamefont {Berera}}, \bibinfo {author} {\bibfnamefont {R.}~\bibnamefont {Cerezo}}, \bibinfo {author} {\bibfnamefont {R.~O.}\ \bibnamefont {Ramos}}, \ and\ \bibinfo {author} {\bibfnamefont {G.~S.}\ \bibnamefont {Vicente}},\ }\href@noop {} {\bibfield  {journal} {\bibinfo  {journal} {Journal of Cosmology and Astroparticle Physics}\ }\textbf {\bibinfo {volume} {2012}},\ \bibinfo {pages} {042} (\bibinfo {year} {2012})}\BibitemShut {NoStop}%
\bibitem [{\citenamefont {Chakraborty}\ and\ \citenamefont {Saha}(2014)}]{ChaSah}%
  \BibitemOpen
  \bibfield  {author} {\bibinfo {author} {\bibfnamefont {S.}~\bibnamefont {Chakraborty}}\ and\ \bibinfo {author} {\bibfnamefont {S.}~\bibnamefont {Saha}},\ }\href@noop {} {\bibfield  {journal} {\bibinfo  {journal} {Physical Review D}\ }\textbf {\bibinfo {volume} {90}},\ \bibinfo {pages} {123505} (\bibinfo {year} {2014})}\BibitemShut {NoStop}%
\bibitem [{\citenamefont {Paliathanasis}\ \emph {et~al.}(2017)\citenamefont {Paliathanasis}, \citenamefont {Barrow},\ and\ \citenamefont {Pan}}]{Barrow}%
  \BibitemOpen
  \bibfield  {author} {\bibinfo {author} {\bibfnamefont {A.}~\bibnamefont {Paliathanasis}}, \bibinfo {author} {\bibfnamefont {J.~D.}\ \bibnamefont {Barrow}}, \ and\ \bibinfo {author} {\bibfnamefont {S.}~\bibnamefont {Pan}},\ }\href@noop {} {\bibfield  {journal} {\bibinfo  {journal} {Physical Review D}\ }\textbf {\bibinfo {volume} {95}},\ \bibinfo {pages} {103516} (\bibinfo {year} {2017})}\BibitemShut {NoStop}%
\bibitem [{\citenamefont {Da~Silva}\ and\ \citenamefont {Silva}(2019)}]{VisDE}%
  \BibitemOpen
  \bibfield  {author} {\bibinfo {author} {\bibfnamefont {W.}~\bibnamefont {Da~Silva}}\ and\ \bibinfo {author} {\bibfnamefont {R.}~\bibnamefont {Silva}},\ }\href@noop {} {\bibfield  {journal} {\bibinfo  {journal} {Journal of Cosmology and Astroparticle Physics}\ }\textbf {\bibinfo {volume} {2019}},\ \bibinfo {pages} {036} (\bibinfo {year} {2019})}\BibitemShut {NoStop}%
\bibitem [{\citenamefont {G{\'o}mez}\ \emph {et~al.}(2023)\citenamefont {G{\'o}mez}, \citenamefont {Palma}, \citenamefont {Gonz{\'a}lez}, \citenamefont {Rinc{\'o}n},\ and\ \citenamefont {Cruz}}]{VisCos}%
  \BibitemOpen
  \bibfield  {author} {\bibinfo {author} {\bibfnamefont {G.}~\bibnamefont {G{\'o}mez}}, \bibinfo {author} {\bibfnamefont {G.}~\bibnamefont {Palma}}, \bibinfo {author} {\bibfnamefont {E.}~\bibnamefont {Gonz{\'a}lez}}, \bibinfo {author} {\bibfnamefont {{\'A}.}~\bibnamefont {Rinc{\'o}n}}, \ and\ \bibinfo {author} {\bibfnamefont {N.}~\bibnamefont {Cruz}},\ }\href@noop {} {\bibfield  {journal} {\bibinfo  {journal} {The European Physical Journal Plus}\ }\textbf {\bibinfo {volume} {138}},\ \bibinfo {pages} {738} (\bibinfo {year} {2023})}\BibitemShut {NoStop}%
\bibitem [{\citenamefont {Klinkhamer}\ and\ \citenamefont {Volovik}(2008{\natexlab{a}})}]{KliVol08a}%
  \BibitemOpen
  \bibfield  {author} {\bibinfo {author} {\bibfnamefont {F.~R.}\ \bibnamefont {Klinkhamer}}\ and\ \bibinfo {author} {\bibfnamefont {G.~E.}\ \bibnamefont {Volovik}},\ }\href {\doibase 10.1103/PhysRevD.77.085015} {\bibfield  {journal} {\bibinfo  {journal} {Phys. Rev. D}\ }\textbf {\bibinfo {volume} {77}},\ \bibinfo {pages} {085015} (\bibinfo {year} {2008}{\natexlab{a}})}\BibitemShut {NoStop}%
\bibitem [{\citenamefont {Klinkhamer}\ and\ \citenamefont {Volovik}(2008{\natexlab{b}})}]{KliVol08b}%
  \BibitemOpen
  \bibfield  {author} {\bibinfo {author} {\bibfnamefont {F.~R.}\ \bibnamefont {Klinkhamer}}\ and\ \bibinfo {author} {\bibfnamefont {G.~E.}\ \bibnamefont {Volovik}},\ }\href {\doibase 10.1103/PhysRevD.78.063528} {\bibfield  {journal} {\bibinfo  {journal} {Phys. Rev. D}\ }\textbf {\bibinfo {volume} {78}},\ \bibinfo {pages} {063528} (\bibinfo {year} {2008}{\natexlab{b}})}\BibitemShut {NoStop}%
\bibitem [{\citenamefont {Xie}\ \emph {et~al.}()\citenamefont {Xie}, \citenamefont {Hsiang},\ and\ \citenamefont {Hu}}]{XHH2}%
  \BibitemOpen
  \bibfield  {author} {\bibinfo {author} {\bibfnamefont {Y.-C.}\ \bibnamefont {Xie}}, \bibinfo {author} {\bibfnamefont {J.-T.}\ \bibnamefont {Hsiang}}, \ and\ \bibinfo {author} {\bibfnamefont {B.-L.}\ \bibnamefont {Hu}},\ }\href@noop {} {\ }\bibinfo {note} {In preparation}\BibitemShut {NoStop}%
\bibitem [{\citenamefont {Hsiang}\ \emph {et~al.}()\citenamefont {Hsiang}, \citenamefont {Cho},\ and\ \citenamefont {Hu}}]{CHH2}%
  \BibitemOpen
  \bibfield  {author} {\bibinfo {author} {\bibfnamefont {J.-T.}\ \bibnamefont {Hsiang}}, \bibinfo {author} {\bibfnamefont {H.-T.}\ \bibnamefont {Cho}}, \ and\ \bibinfo {author} {\bibfnamefont {B.-L.}\ \bibnamefont {Hu}},\ }\href@noop {} {\ }\bibinfo {note} {In preparation}\BibitemShut {NoStop}%
\bibitem [{\citenamefont {Christensen}(1978{\natexlab{a}})}]{TAevendim}%
  \BibitemOpen
  \bibfield  {author} {\bibinfo {author} {\bibfnamefont {S.~M.}\ \bibnamefont {Christensen}},\ }\href@noop {} {\bibfield  {journal} {\bibinfo  {journal} {Physical Review D}\ }\textbf {\bibinfo {volume} {17}},\ \bibinfo {pages} {946} (\bibinfo {year} {1978}{\natexlab{a}})}\BibitemShut {NoStop}%
\bibitem [{\citenamefont {Xie}\ \emph {et~al.}(2023)\citenamefont {Xie}, \citenamefont {Butera},\ and\ \citenamefont {Hu}}]{XBH}%
  \BibitemOpen
  \bibfield  {author} {\bibinfo {author} {\bibfnamefont {Y.-C.}\ \bibnamefont {Xie}}, \bibinfo {author} {\bibfnamefont {S.}~\bibnamefont {Butera}}, \ and\ \bibinfo {author} {\bibfnamefont {B.~L.}\ \bibnamefont {Hu}},\ }\href@noop {} {\  (\bibinfo {year} {2023})},\ \Eprint {http://arxiv.org/abs/2308.03129} {arXiv:2308.03129 [quant-ph]} \BibitemShut {NoStop}%
\bibitem [{\citenamefont {Hartle}(1977)}]{Har77}%
  \BibitemOpen
  \bibfield  {author} {\bibinfo {author} {\bibfnamefont {J.~B.}\ \bibnamefont {Hartle}},\ }\href@noop {} {\bibfield  {journal} {\bibinfo  {journal} {Physical Review Letters}\ }\textbf {\bibinfo {volume} {39}},\ \bibinfo {pages} {1373} (\bibinfo {year} {1977})}\BibitemShut {NoStop}%
\bibitem [{\citenamefont {Hu}\ and\ \citenamefont {Parker}(1977)}]{HuPar77}%
  \BibitemOpen
  \bibfield  {author} {\bibinfo {author} {\bibfnamefont {B.~L.}\ \bibnamefont {Hu}}\ and\ \bibinfo {author} {\bibfnamefont {L.}~\bibnamefont {Parker}},\ }\href@noop {} {\bibfield  {journal} {\bibinfo  {journal} {Phys. Lett., A;(Netherlands)}\ }\textbf {\bibinfo {volume} {63}} (\bibinfo {year} {1977})}\BibitemShut {NoStop}%
\bibitem [{\citenamefont {Hu}\ and\ \citenamefont {Parker}(1978)}]{HuPar78}%
  \BibitemOpen
  \bibfield  {author} {\bibinfo {author} {\bibfnamefont {B.~L.}\ \bibnamefont {Hu}}\ and\ \bibinfo {author} {\bibfnamefont {L.}~\bibnamefont {Parker}},\ }\href {\doibase 10.1103/PhysRevD.17.933} {\bibfield  {journal} {\bibinfo  {journal} {Phys. Rev. D}\ }\textbf {\bibinfo {volume} {17}},\ \bibinfo {pages} {933} (\bibinfo {year} {1978})}\BibitemShut {NoStop}%
\bibitem [{\citenamefont {Fischetti}\ \emph {et~al.}(1979)\citenamefont {Fischetti}, \citenamefont {Hartle},\ and\ \citenamefont {Hu}}]{FHH79}%
  \BibitemOpen
  \bibfield  {author} {\bibinfo {author} {\bibfnamefont {M.~V.}\ \bibnamefont {Fischetti}}, \bibinfo {author} {\bibfnamefont {J.~B.}\ \bibnamefont {Hartle}}, \ and\ \bibinfo {author} {\bibfnamefont {B.~L.}\ \bibnamefont {Hu}},\ }\href {\doibase 10.1103/PhysRevD.20.1757} {\bibfield  {journal} {\bibinfo  {journal} {Phys. Rev. D}\ }\textbf {\bibinfo {volume} {20}},\ \bibinfo {pages} {1757} (\bibinfo {year} {1979})}\BibitemShut {NoStop}%
\bibitem [{\citenamefont {Hartle}\ and\ \citenamefont {Hu}(1979)}]{HarHu79}%
  \BibitemOpen
  \bibfield  {author} {\bibinfo {author} {\bibfnamefont {J.~B.}\ \bibnamefont {Hartle}}\ and\ \bibinfo {author} {\bibfnamefont {B.~L.}\ \bibnamefont {Hu}},\ }\href@noop {} {\bibfield  {journal} {\bibinfo  {journal} {Physical Review D}\ }\textbf {\bibinfo {volume} {20}},\ \bibinfo {pages} {1772} (\bibinfo {year} {1979})}\BibitemShut {NoStop}%
\bibitem [{\citenamefont {Hartle}\ and\ \citenamefont {Hu}(1980)}]{HarHu80}%
  \BibitemOpen
  \bibfield  {author} {\bibinfo {author} {\bibfnamefont {J.~B.}\ \bibnamefont {Hartle}}\ and\ \bibinfo {author} {\bibfnamefont {B.~L.}\ \bibnamefont {Hu}},\ }\href@noop {} {\bibfield  {journal} {\bibinfo  {journal} {Physical Review D}\ }\textbf {\bibinfo {volume} {21}},\ \bibinfo {pages} {2756} (\bibinfo {year} {1980})}\BibitemShut {NoStop}%
\bibitem [{\citenamefont {Hartle}(1981)}]{HarV}%
  \BibitemOpen
  \bibfield  {author} {\bibinfo {author} {\bibfnamefont {J.~B.}\ \bibnamefont {Hartle}},\ }\href@noop {} {\bibfield  {journal} {\bibinfo  {journal} {Physical Review D}\ }\textbf {\bibinfo {volume} {23}},\ \bibinfo {pages} {2121} (\bibinfo {year} {1981})}\BibitemShut {NoStop}%
\bibitem [{\citenamefont {Anderson}(1985)}]{And85}%
  \BibitemOpen
  \bibfield  {author} {\bibinfo {author} {\bibfnamefont {P.~R.}\ \bibnamefont {Anderson}},\ }\href@noop {} {\bibfield  {journal} {\bibinfo  {journal} {Physical Review D}\ }\textbf {\bibinfo {volume} {32}},\ \bibinfo {pages} {1302} (\bibinfo {year} {1985})}\BibitemShut {NoStop}%
\bibitem [{\citenamefont {Calzetta}\ and\ \citenamefont {Hu}(1987)}]{CalHu87}%
  \BibitemOpen
  \bibfield  {author} {\bibinfo {author} {\bibfnamefont {E.}~\bibnamefont {Calzetta}}\ and\ \bibinfo {author} {\bibfnamefont {B.-L.}\ \bibnamefont {Hu}},\ }\href@noop {} {\bibfield  {journal} {\bibinfo  {journal} {Physical Review D}\ }\textbf {\bibinfo {volume} {35}},\ \bibinfo {pages} {495} (\bibinfo {year} {1987})}\BibitemShut {NoStop}%
\bibitem [{\citenamefont {Hu}(1974)}]{Hu74}%
  \BibitemOpen
  \bibfield  {author} {\bibinfo {author} {\bibfnamefont {B.~L.}\ \bibnamefont {Hu}},\ }\href@noop {} {\bibfield  {journal} {\bibinfo  {journal} {Physical Review D}\ }\textbf {\bibinfo {volume} {9}},\ \bibinfo {pages} {3263} (\bibinfo {year} {1974})}\BibitemShut {NoStop}%
\bibitem [{\citenamefont {Nagatani}\ and\ \citenamefont {Shigetomi}(2000)}]{2J}%
  \BibitemOpen
  \bibfield  {author} {\bibinfo {author} {\bibfnamefont {Y.}~\bibnamefont {Nagatani}}\ and\ \bibinfo {author} {\bibfnamefont {K.}~\bibnamefont {Shigetomi}},\ }\href@noop {} {\bibfield  {journal} {\bibinfo  {journal} {Physical Review A}\ }\textbf {\bibinfo {volume} {62}},\ \bibinfo {pages} {022117} (\bibinfo {year} {2000})}\BibitemShut {NoStop}%
\bibitem [{\citenamefont {Lin}\ \emph {et~al.}(2016)\citenamefont {Lin}, \citenamefont {Chou},\ and\ \citenamefont {Hu}}]{LCH_EinCyl}%
  \BibitemOpen
  \bibfield  {author} {\bibinfo {author} {\bibfnamefont {S.-Y.}\ \bibnamefont {Lin}}, \bibinfo {author} {\bibfnamefont {C.-H.}\ \bibnamefont {Chou}}, \ and\ \bibinfo {author} {\bibfnamefont {B.-L.}\ \bibnamefont {Hu}},\ }\href@noop {} {\bibfield  {journal} {\bibinfo  {journal} {Journal of High Energy Physics}\ }\textbf {\bibinfo {volume} {2016}},\ \bibinfo {pages} {1} (\bibinfo {year} {2016})}\BibitemShut {NoStop}%
\bibitem [{\citenamefont {Ford}(1975)}]{Ford75}%
  \BibitemOpen
  \bibfield  {author} {\bibinfo {author} {\bibfnamefont {L.~H.}\ \bibnamefont {Ford}},\ }\href {\doibase 10.1103/PhysRevD.11.3370} {\bibfield  {journal} {\bibinfo  {journal} {Phys. Rev. D}\ }\textbf {\bibinfo {volume} {11}},\ \bibinfo {pages} {3370} (\bibinfo {year} {1975})}\BibitemShut {NoStop}%
\bibitem [{\citenamefont {Ford}(1976)}]{Ford76}%
  \BibitemOpen
  \bibfield  {author} {\bibinfo {author} {\bibfnamefont {L.~H.}\ \bibnamefont {Ford}},\ }\href {\doibase 10.1103/PhysRevD.14.3304} {\bibfield  {journal} {\bibinfo  {journal} {Phys. Rev. D}\ }\textbf {\bibinfo {volume} {14}},\ \bibinfo {pages} {3304} (\bibinfo {year} {1976})}\BibitemShut {NoStop}%
\bibitem [{\citenamefont {Bordag}\ \emph {et~al.}(2009)\citenamefont {Bordag}, \citenamefont {Klimchitskaya}, \citenamefont {Mohideen},\ and\ \citenamefont {Mostepanenko}}]{BordagBook}%
  \BibitemOpen
  \bibfield  {author} {\bibinfo {author} {\bibfnamefont {M.}~\bibnamefont {Bordag}}, \bibinfo {author} {\bibfnamefont {G.~L.}\ \bibnamefont {Klimchitskaya}}, \bibinfo {author} {\bibfnamefont {U.}~\bibnamefont {Mohideen}}, \ and\ \bibinfo {author} {\bibfnamefont {V.~M.}\ \bibnamefont {Mostepanenko}},\ }\href@noop {} {\emph {\bibinfo {title} {Advances in the Casimir effect}}},\ Vol.\ \bibinfo {volume} {145}\ (\bibinfo  {publisher} {OUP Oxford},\ \bibinfo {year} {2009})\BibitemShut {NoStop}%
\bibitem [{\citenamefont {Bunch}(1980)}]{Bunch80}%
  \BibitemOpen
  \bibfield  {author} {\bibinfo {author} {\bibfnamefont {T.}~\bibnamefont {Bunch}},\ }\href@noop {} {\bibfield  {journal} {\bibinfo  {journal} {Journal of Physics A: Mathematical and General}\ }\textbf {\bibinfo {volume} {13}},\ \bibinfo {pages} {1297} (\bibinfo {year} {1980})}\BibitemShut {NoStop}%
\bibitem [{\citenamefont {Hu}(1978)}]{Hu78}%
  \BibitemOpen
  \bibfield  {author} {\bibinfo {author} {\bibfnamefont {B.~L.}\ \bibnamefont {Hu}},\ }\href@noop {} {\bibfield  {journal} {\bibinfo  {journal} {Physical Review D}\ }\textbf {\bibinfo {volume} {18}},\ \bibinfo {pages} {4460} (\bibinfo {year} {1978})}\BibitemShut {NoStop}%
\bibitem [{\citenamefont {Hu}(1979)}]{Hu79}%
  \BibitemOpen
  \bibfield  {author} {\bibinfo {author} {\bibfnamefont {B.~L.}\ \bibnamefont {Hu}},\ }\href@noop {} {\bibfield  {journal} {\bibinfo  {journal} {Physics Letters A}\ }\textbf {\bibinfo {volume} {71}},\ \bibinfo {pages} {169} (\bibinfo {year} {1979})}\BibitemShut {NoStop}%
\bibitem [{\citenamefont {Anderson}(1983)}]{And83}%
  \BibitemOpen
  \bibfield  {author} {\bibinfo {author} {\bibfnamefont {P.}~\bibnamefont {Anderson}},\ }\href@noop {} {\bibfield  {journal} {\bibinfo  {journal} {Physical Review D}\ }\textbf {\bibinfo {volume} {28}},\ \bibinfo {pages} {271} (\bibinfo {year} {1983})}\BibitemShut {NoStop}%
\bibitem [{\citenamefont {Fulling}\ \emph {et~al.}(1974)\citenamefont {Fulling}, \citenamefont {Parker},\ and\ \citenamefont {Hu}}]{FulParHu74}%
  \BibitemOpen
  \bibfield  {author} {\bibinfo {author} {\bibfnamefont {S.~A.}\ \bibnamefont {Fulling}}, \bibinfo {author} {\bibfnamefont {L.}~\bibnamefont {Parker}}, \ and\ \bibinfo {author} {\bibfnamefont {B.~L.}\ \bibnamefont {Hu}},\ }\href@noop {} {\bibfield  {journal} {\bibinfo  {journal} {Physical Review D}\ }\textbf {\bibinfo {volume} {10}},\ \bibinfo {pages} {3905} (\bibinfo {year} {1974})}\BibitemShut {NoStop}%
\bibitem [{\citenamefont {Mamaev}\ and\ \citenamefont {Trunov}(1979)}]{MamaevTrunov79}%
  \BibitemOpen
  \bibfield  {author} {\bibinfo {author} {\bibfnamefont {S.~G.}\ \bibnamefont {Mamaev}}\ and\ \bibinfo {author} {\bibfnamefont {N.~N.}\ \bibnamefont {Trunov}},\ }\href@noop {} {\bibfield  {journal} {\bibinfo  {journal} {Teoreticheskaya i Matematicheskaya Fizika}\ }\textbf {\bibinfo {volume} {38}},\ \bibinfo {pages} {345} (\bibinfo {year} {1979})}\BibitemShut {NoStop}%
\bibitem [{\citenamefont {Bander}\ and\ \citenamefont {Itzykson}(1966)}]{bander1966group}%
  \BibitemOpen
  \bibfield  {author} {\bibinfo {author} {\bibfnamefont {M.}~\bibnamefont {Bander}}\ and\ \bibinfo {author} {\bibfnamefont {C.}~\bibnamefont {Itzykson}},\ }\href@noop {} {\bibfield  {journal} {\bibinfo  {journal} {Reviews of Modern Physics}\ }\textbf {\bibinfo {volume} {38}},\ \bibinfo {pages} {346} (\bibinfo {year} {1966})}\BibitemShut {NoStop}%
\bibitem [{\citenamefont {Christensen}(1978{\natexlab{b}})}]{christensen1978regularization}%
  \BibitemOpen
  \bibfield  {author} {\bibinfo {author} {\bibfnamefont {S.~M.}\ \bibnamefont {Christensen}},\ }\href {\doibase 10.1103/PhysRevD.17.946} {\bibfield  {journal} {\bibinfo  {journal} {Phys. Rev. D}\ }\textbf {\bibinfo {volume} {17}},\ \bibinfo {pages} {946} (\bibinfo {year} {1978}{\natexlab{b}})}\BibitemShut {NoStop}%
\bibitem [{\citenamefont {Anderson}\ and\ \citenamefont {Parker}(1987)}]{AndPar}%
  \BibitemOpen
  \bibfield  {author} {\bibinfo {author} {\bibfnamefont {P.~R.}\ \bibnamefont {Anderson}}\ and\ \bibinfo {author} {\bibfnamefont {L.}~\bibnamefont {Parker}},\ }\href@noop {} {\bibfield  {journal} {\bibinfo  {journal} {Physical Review D}\ }\textbf {\bibinfo {volume} {36}},\ \bibinfo {pages} {2963} (\bibinfo {year} {1987})}\BibitemShut {NoStop}%
\bibitem [{\citenamefont {Hu}\ \emph {et~al.}(1994)\citenamefont {Hu}, \citenamefont {Kang},\ and\ \citenamefont {Matacz}}]{HKM94}%
  \BibitemOpen
  \bibfield  {author} {\bibinfo {author} {\bibfnamefont {B.~L.}\ \bibnamefont {Hu}}, \bibinfo {author} {\bibfnamefont {G.}~\bibnamefont {Kang}}, \ and\ \bibinfo {author} {\bibfnamefont {A.}~\bibnamefont {Matacz}},\ }\href@noop {} {\bibfield  {journal} {\bibinfo  {journal} {International Journal of Modern Physics A}\ }\textbf {\bibinfo {volume} {9}},\ \bibinfo {pages} {991} (\bibinfo {year} {1994})}\BibitemShut {NoStop}%
\bibitem [{\citenamefont {Hu}(1983)}]{Hu83}%
  \BibitemOpen
  \bibfield  {author} {\bibinfo {author} {\bibfnamefont {B.~L.}\ \bibnamefont {Hu}},\ }\href@noop {} {\bibfield  {journal} {\bibinfo  {journal} {Physics Letters A}\ }\textbf {\bibinfo {volume} {97}},\ \bibinfo {pages} {368} (\bibinfo {year} {1983})}\BibitemShut {NoStop}%
\bibitem [{\citenamefont {Parker}(1983)}]{Par83}%
  \BibitemOpen
  \bibfield  {author} {\bibinfo {author} {\bibfnamefont {L.}~\bibnamefont {Parker}},\ }\href@noop {} {\bibfield  {journal} {\bibinfo  {journal} {Physical Review Letters}\ }\textbf {\bibinfo {volume} {50}},\ \bibinfo {pages} {1009} (\bibinfo {year} {1983})}\BibitemShut {NoStop}%
\bibitem [{\citenamefont {Parker}\ and\ \citenamefont {Fulling}(1974)}]{ParFul74}%
  \BibitemOpen
  \bibfield  {author} {\bibinfo {author} {\bibfnamefont {L.}~\bibnamefont {Parker}}\ and\ \bibinfo {author} {\bibfnamefont {S.~A.}\ \bibnamefont {Fulling}},\ }\href@noop {} {\bibfield  {journal} {\bibinfo  {journal} {Physical Review D}\ }\textbf {\bibinfo {volume} {9}},\ \bibinfo {pages} {341} (\bibinfo {year} {1974})}\BibitemShut {NoStop}%
\bibitem [{\citenamefont {Fulling}\ and\ \citenamefont {Parker}(1974)}]{FulPar74}%
  \BibitemOpen
  \bibfield  {author} {\bibinfo {author} {\bibfnamefont {S.}~\bibnamefont {Fulling}}\ and\ \bibinfo {author} {\bibfnamefont {L.}~\bibnamefont {Parker}},\ }\href@noop {} {\bibfield  {journal} {\bibinfo  {journal} {Annals of Physics}\ }\textbf {\bibinfo {volume} {87}},\ \bibinfo {pages} {176} (\bibinfo {year} {1974})}\BibitemShut {NoStop}%
\end{thebibliography}%
\end{document}